\documentstyle[epsfig]{article}

\newcommand{\qed}{{$\Box$}}

\newcommand{\ba}{\begin{array}}
\newcommand{\ea}{\end{array}}

\newtheorem{lemma}{Lemma}
\newtheorem{hypo}{Assumption}
\newcommand{\bhypo}{\begin{hypo}}
\newcommand{\ehypo}{\end{hypo}}
\newtheorem{defi}{Definition}
\newcommand{\ble}{\begin{lemma}}
\newcommand{\ele}{\end{lemma}}
\newcommand{\bde}{\begin{defi}}
\newcommand{\ede}{\end{defi}}

\newtheorem{prop}{Proposition}
\newcommand{\epr}{\end{prop}}
\newcommand{\bpr}{\begin{prop}}
\newtheorem{teo}{Theorem}
\newcommand{\bth}{\begin{teo}}
\newcommand{\eth}{\end{teo}}
\newtheorem{rema}{Remark}
\newcommand{\bre}{\begin{rema}}
\newcommand{\ere}{\end{rema}}
\newcommand{\ee}{\end{equation}}
\newcommand{\be}{\begin{equation}}

\begin{document}

\centerline{ \huge \bf On the Logarithmic Asymptotics  of the}
\vskip 0.3 cm
\centerline{ \huge \bf   Sixth Painlev\'e  Equation (Summer 2007)}

\vskip 0.3 cm
\centerline{\Large Davide Guzzetti}

\vskip 0.3 cm

\begin{abstract}   
 We compute the monodromy 
group associated to the solutions of the sixth Painlev\'e equation with a logarithmic
 asymptotic  behavior at a critical point and we characterize the asymptotic behavior in terms of the
 monodromy itself.  
\vskip 0.3 cm

\end{abstract}

\vskip 1 cm

\section{Introduction}

This paper appeared as a preprint in August 2007. It is published in
J. Phys. A: Math. Theor. {\bf 41},  (2008), 205201-205247. It was on the
archive in January 2008 (arXiv:0801.1157). This version does not differ from the
published one, except for two facts: 1)the addition of subsection \ref{cioco},
which proves that tr$(M_0M_x)=-2$ for solutions $y(x)\sim {a\over( \ln 
  x)^n}$,
$n=1,2$, $x\to 0$. 2) The explicit parametrization of the critical behavior of the general log-solutions  in terms of monodromy data, in the last part of  Proposition 2.
\vskip 0.3 cm

We consider the sixth Painlev\'e equation: $$
{d^2y \over dx^2}={1\over 2}\left[ 
{1\over y}+{1\over y-1}+{1\over y-x}
\right]
           \left({dy\over dx}\right)^2
-\left[
{1\over x}+{1\over x-1}+{1\over y-x}
\right]{dy \over dx}
$$
$$
+
{y(y-1)(y-x)\over x^2 (x-1)^2}
\left[
\alpha+\beta {x\over y^2} + \gamma {x-1\over (y-1)^2} +\delta
{x(x-1)\over (y-x)^2}
\right]
,~~~~~\hbox{(PVI)}.
$$ 
The generic solution  has essential 
singularities and/or branch points in 0,1,$\infty$. It's behavior at
these points  is called {\it critical}. 
   Other singularities which may appear are poles and depend on the initial conditions. 
 A solution of (PVI)    
 can be analytically continued to a meromorphic function on the universal
   covering of ${\bf P}^1\backslash \{ 0,1 ,\infty \}$. 
 For generic values of the integration constants and of the parameters $\alpha$,$\beta$,$\gamma$,$\delta$,
 it cannot be expressed via elementary or classical
   transcendental functions. For this reason, it  is called a {\it
   Painlev\'e transcendent}.  Solving (PVI) means: ~i) Determine 
the critical behavior of the transcendents 
at the {\it critical points} $x=0,1,\infty$. Such a behavior must depend on 
two integration constants. ~ii) Solve the {\it connection
  problem}, namely: find the relation between couples of integration
constants  at $x=0,1,\infty$. 

 In this paper we compute the monodromy 
group associated to the solutions of the sixth Painlev\'e equation with a logarithmic
 asymptotic  behavior,  and we parametrize 
 the asymptotic behavior in terms of the
 monodromy itself.  

\subsection{ Associated Fuchsian System} 

(PVI) is the isomonodromy deformation equation of  a Fuchsian system of
 differential equations \cite{JMU}: 
\be
   {d\Psi\over d\lambda}=A(x,\lambda)~\Psi,~~~~~
A(x,\lambda):=\left[ {A_0(x)\over \lambda}+{A_x(x) \over \lambda-x}+{A_1(x)
\over
\lambda-1}\right],~~~\lambda\in{\bf C}.
\label{SYSTEM}
\ee
The  $2\times 2$ matrices  $A_i(x)$  depend 
 on $x$ in such a way that there exists a fundamental matrix solution $\Psi(\lambda,x)$ such that its monodromy  
 does not change for  small deformations of $x$. They also depend on the 
 parameters $\alpha,\beta,\gamma,\delta$ of (PVI) through more elementary 
parameters $\theta_0,\theta_x,\theta_1,\theta_{\infty}$, according to 
the following relations:  
\begin{equation}
 -A_\infty:=A_0+A_1+A_x = -{\theta_{\infty}\over 2}
 \sigma_3,~~\theta_\infty\neq 0.~~~~~
\hbox{ Eigenvalues}~( A_i) =\pm {1\over 2} \theta_i, ~~~i=0,1,x;
\label{caffe0}
\end{equation}
 \begin{equation}
    \alpha= {1\over 2} (\theta_{\infty} -1)^2,
~~~-\beta={1\over 2} \theta_0^2, 
~~~ \gamma={1\over 2} \theta_1^2,
~~~ \left({1\over 2} -\delta \right)={1\over 2} \theta_x^2 
\label{caffe1}
\end{equation}
Here $\sigma_3:=\pmatrix{1&0\cr 0&-1}$ is the Pauli matrix. The condition $\theta_\infty\neq 0$ is not restrictive, because $\theta_\infty=0$ is equivalent to $\theta_\infty=2$.   
 The equations of monodromy preserving deformation (Schlesinger equations), can be written in Hamiltonian form and reduce
 to (PVI), being the transcendent $y(x)$ the solution $\lambda$  of
 $A(x,\lambda)_{1,2}=0$. Namely:
\be
y(x)= 
{x~(A_0)_{12} \over x~\left[
(A_0)_{12}+(A_1)_{12}
\right]- (A_1)_{12}},
\label{leadingtermaprile}
\ee
The matrices $A_i(x)$, $i=0,x,1$, 
 depend on $y(x)$, ${d y(x)\over dx}$ and $\int y(x)$ 
through rational functions, which are given in \cite{JMU} and in subsection 
\ref{monodinsimM}.

\vskip 0.2 cm 
 This paper is devoted to the computation of the 
 monodromy group of (\ref{SYSTEM}) associated to the solutions with a logarithmic 
critical behavior. 
This is  part of a our project to classify 
 the critical behaviors in terms of the monodromy data of the system 
(\ref{SYSTEM}). 
This project has been carried on through our papers  \cite{guz1} \cite{guz3} \cite{guz5}.   

\subsection{Background} 
 
 In \cite{guz5} we developed a {\it matching technique}. It enabled us to compute the first leading terms of the 
critical behavior of a transcendent $y(x)$ in a constructive way.   
 Originally, such an approach  
was suggested by Its and Novokshenov in \cite{its}, 
 for the second and third Painlev\'e equations. 
 The method of  Jimbo \cite{Jimbo} can be regarded as a matching procedure.   
 This approach was further developed and
used by Kapaev, Kitaev,  Andreev, and Vartanian (see for example the case of 
 the fifth
Painlev\'e equation,  in \cite{KitaevAndreev}). 

Our approach of  \cite{guz5} 
is new, because we introduced non-fuchsian systems (systems with irregular singularities) associated to (PVI). This allowed us to obtained  new asymptotic behaviors. 

 Denote by $M_0$, $M_x$, $M_1$ a monodromy representation of (\ref{SYSTEM}). 
The critical behaviors associated to monodromy matrices satisfying the 
relation tr$(M_iM_j)\neq \pm 2$, $i\neq j\in\{0,x,1\}$, is known from the 
work  \cite{Jimbo}. The matching procedure was 
developed in \cite{guz5}, as a general method to study the cases 
tr$(M_iM_j)=\pm 2$ and the non generic cases  of $\alpha,\beta,\gamma,\delta$.
 
 The   values of the traces $\hbox{\rm tr}(M_0M_x)$, 
 $\hbox{\rm tr}(M_1M_x)$, $\hbox{\rm tr}(M_0M_1)$ characterize 
the  critical behaviors at $x=0,1,\infty$ respectively.  
For example, in the generic case studied in \cite{Jimbo} we find 
 the following behaviors at the critical points 
\cite{Jimbo}\cite{DM}\cite{guz1}\cite{guz3}\cite{guz4}\cite{Boalch}\cite{Sh}:   
   $$
y(x)=
\left\{
\matrix{
 a x^{1-\sigma}(1+O(|x|^{\epsilon})),&~~~~x\to 0,
\cr
\cr
y(x)= 1-a^{(1)}(1-x)^{1-\sigma^{(1)}} (1+O(|1-x|^{\epsilon})),&~~~~x\to 1,
\cr
\cr
y(x)= a^{(\infty)}
 x^{\sigma^{(\infty)}}(1+O(|x|^{-\epsilon})),~~~~x\to \infty,
}
\right.
$$
where $\epsilon$ is a small positive number, $a$, $\sigma$, $a^{(1)}$ , 
 $\sigma^{(1)}$,  $a^{(\infty)}$ , 
 $\sigma^{(\infty)}$ are complex numbers such that $a$, $a^{(i)}\neq 0$ and 
  $ 0< \Re \sigma<1$, $0< \Re \sigma^{(1)}<1$, $0< \Re \sigma^{(\infty)}<1$.
The connection problem among the three sets of parameters $(a,\sigma)$, $(a^{(1)},\sigma^{(1)})$, $(a^{(\infty)}\sigma^{(\infty)})$  was first solved 
in \cite{Jimbo} and its solution implies that: 
$$ 
 2 \cos(\pi \sigma)=\hbox{tr}(M_0M_x), ~~~
2 \cos(\pi \sigma^{(1)})=\hbox{tr}(M_1M_x),~~~
2 \cos(\pi \sigma^{(\infty)})=\hbox{tr}(M_0M_1);
$$
while $a$, $a^{(1)}$, $ a^{(\infty)}$ are rational functions of the  
tr$(M_iM_j)$'s ($i\neq j=0,x,1$)  and depend on the $\theta_\nu$'s 
($\nu=0,x,1,\infty$) through 
trigonometric functions and $\Gamma$-functions rationally combined. 
In this sense, the three traces determine the critical behavior 
at the three critical points.

\subsection{Critical Behaviors} 
 
We summarize  the 
results obtained by the matching procedure in \cite{guz5}. 
We consider only the point $x=0$, because the critical behaviors at $x=1,\infty$ can be obtained by the action of Backlund transformations of (PVI) ( see subsection \ref{CCPR}). 
 Let $\sigma$ be a complex number defined by:
$$
\hbox{tr} (M_0M_x)=2\cos(\pi\sigma),~~~~~0\leq \Re \sigma\leq 1. 
$$
 The 
matching procedure yields the following behaviors for $x\to 0$: 
\be
\label{intrJimbo}
y(x)\sim ~
a~x^{1-\sigma},~~~~~~~~~~~~ ~~~\hbox{\rm if $\Re \sigma>0$};
\ee
$$
y(x)\sim
x\left\{
i{A}~\sin\bigl(i\sigma\ln
x+\phi\bigr)
+{\theta_0^2-\theta_x^2+\sigma^2\over 2\sigma^2}
\right\}
,~~~\hbox{\rm if $\Re\sigma=0$},~~~\sigma\neq 0.
$$
In the above formulae, $\sigma$ is one of the integration constants, 
 while $a$, or $\phi$,  is the other.  $A $ is: 
$${A}:=\left[{\theta_0^2\over
    \sigma^2}-\left({\theta_0^2-\theta_x^2+\sigma^2\over
    2\sigma^2}\right)^2\right]^{1\over 2}. 
$$  
 For special values of $\sigma\neq 0$, the first leading term above is zero 
and we need to consider the next leading terms: 
$$
y(x)\sim{\theta_0\over \theta_0+\theta_x}~x ~\mp~{r\over\theta_0+\theta_x}
~x^{1+\sigma},~~~\sigma=\pm(\theta_0+\theta_x)\neq 0,
$$
$$
y(x)\sim {\theta_0\over \theta_0-\theta_x}~x~\mp
~{r\over\theta_0-\theta_x}~x^{1+\sigma},
~~~
\sigma=\pm(\theta_0-\theta_x)\neq 0.
$$
\vskip 0.2 cm
\noindent
When $\sigma = 0$: 
\be
\label{intrlog1}
y(x)\sim
x\left\{
{\theta_x^2-\theta_0^2\over 4} 
\left[
\ln x +{4r+2\theta_0\over \theta_0^2-\theta_x^2}
\right]^2 +{\theta_0^2\over \theta_0^2-\theta_x^2}
\right\}
,~~~~~\hbox{ if } \theta_0^2\neq\theta_x^2,
\ee
\be
\label{intrlog2}
y(x)\sim x~(r~\pm~\theta_0~\ln x),~~~~~~~\hbox{ if } \theta_0^2=\theta_x^2.
\ee
Here $r$ is an integration constant. 

 In \cite{guz5} we also computed all  the solutions with 
Taylor expansions at a critical point. 

Taylor solutions are studied also in \cite{kaneko}, by the isomonodromy 
deformation method; and in  \cite{Bruno} \cite{Bruno1} \cite{Bruno2} 
\cite{Bruno3}  
by a power geometry technique. 
 In \cite{Bruno} \cite{Bruno1} \cite{Bruno2} 
\cite{Bruno3}  
\cite{Bruno6}, A.D.Bruno and I.V.Goryuchkina constructed  the asymptotic
 expansions, including logarithmic ones, by a power geometry technique 
\cite{Bruno5}. All the asymptotic expansions obtained by this 
technique are summarized in \cite{Bruno6}, and include -- to use the terminology of \cite{Bruno6} -- power-logarithmic expansions and 
logarithmic complicated expansions.     
  The logarithmic asymptotics for real solutions of (PVI) is 
studied in \cite{qlu}. Our approach, being based on the method of 
isomonodromy deformations, allows to solve the connection problem, while the 
results  \cite{Bruno}--
\cite{Bruno6} and \cite{qlu} are local.

The monodromy data for the solution (\ref{intrJimbo}) are computed in 
\cite{Jimbo}\cite{DM}\cite{guz1}\cite{guz3}\cite{guz4}\cite{Boalch}. The monodromy data for the Taylor expansions are computed in \cite{guz5} and \cite{kaneko}. 

The monodromy associated to the logarithmic behaviors are not known. Their computation is the main result of  the present paper. We are going to  
show that logarithmic critical behaviors at $x=0$ are associated to tr$(M_0M_x)=\pm 2$, at $x=1$ to tr$(M_1M_x)=\pm 2$, and at $x=\infty$ to tr$(M_0M_1)=\pm 2$.

\subsection{Results of this Paper} 

In this paper: 

\vskip 0.2 cm 
1) In Section \ref{ClassificationTerms}  we justify  the  
project of classifying the transcendents in terms of monodromy
 data of (\ref{SYSTEM}). We establish the necessary and sufficient conditions such 
that 
 there exist a one to one correspondence between a set of monodromy data 
of  system (\ref{SYSTEM}) and a transcendent of (PVI). The result is 
Proposition \ref{pro1}. The definition of {\it monodromy data} itself 
is given in Section  \ref{ClassificationTerms}.

\vskip 0.2 cm
2) We compute the monodromy data associated to the logarithmic solutions (\ref{intrlog1}) in the generic case $\theta_0,\theta_x,\theta_1,\theta_\infty\not \in{\bf Z}$. The result is Proposition \ref{monodromiagen}, Section \ref{sectionmonodromiagen}. In particular, tr$(M_0M_x)=2$.

\vskip 0.2 cm 
3) In Proposition \ref{propmonodromiagen1} of Section \ref{monodromiagen1}, we compute the monodromy group associated to the solution (\ref{intrlog2}). In  particular,  tr$(M_0M_x)=2$.  The parameter $r$ will be computed as a function of the $\theta_\nu$'s, $\nu=0,x,1,\infty$ and of  $\hbox{\rm tr}(M_0M_1)$.

\vskip 0.2 cm

\vskip 0.2 cm
4) We consider a  non generic case of (\ref{intrlog1}), which occurs  when:
\be
\label{valuetheta}
\theta_x=\theta_1=0,~~~~~\theta_\infty=1,~~~~~\theta_0=2p\neq 0,~~~p\in{\bf Z}.
\ee
Therefore:
\be
\label{nongenintro}
y(x)\sim x~\left[1 -p^2\left(\ln x +{r+p\over p^2}\right)^2\right],~~~~~x \rightarrow 0.
\ee
 The monodromy of the associated system (\ref{SYSTEM}) is  computed  in 
Proposition \ref{monodromiaCHAZYmy}, Section \ref{sectionmonodromiaCHAZYmy}. 
It is important to observe that the monodromy is independent of $r$.  This 
means that the parameter $r$ cannot be determined in terms of the monodromy 
data. Therefore, (\ref{nongenintro}) is a {\it one parameter class of 
solutions ({\rm parameter $r$}) associated to the same monodromy data}.
 We prove  in Proposition \ref{monodromiaCHAZYmy} 
that the solution (\ref{nongenintro}) is associated to:  
$$
\hbox{\rm tr}(M_0M_x)=2,~~~\hbox{\rm tr}(M_0M_1)=2,~~~\hbox{\rm tr}(M_1M_x)=-2.
$$
This special values of the traces imply that the behavior at $x=\infty$ and 
$x=1$ is also logarithmic. $\hbox{tr}(M_0M_x)=2$ is associated to  the 
logarithmic behavior of type $\ln^2x$ at $x=0$. 
 $\hbox{tr}(M_0M_1)=2$ is associated to the  
logarithmic behavior of type $\ln^2(1/x)$ at $x=\infty$.  
$\hbox{tr}(M_1M_x)=-2$ is associated 
to the  
logarithmic behavior of type $1/\ln^2(1-x)$ at $x=1$. Actually a solution (\ref{nongenintro}) 
has the following behaviors at the three critical points:
\be
\label{imbroglio}
  y(x)\sim
\left\{
\matrix{
x\left[ 1-p^2(\ln x +\rho_0)^2\right],& ~~~x\to 0,
\cr
\cr
1-p^2\left(\ln{1\over x} +\rho_\infty\right)^2,& ~~~x\to\infty,
\cr
\cr
1-{1\over p^2 (\ln(1-x)+\rho_1)^2},&~~~x\to 1.
}
\right.
\ee
where: 
$$\rho_0={(r+p)\over p^2},~~~
\rho_\infty= {\pi(4\ln 2-1+\rho_0)\over \pi -i (4\ln 2-1+\rho_0)}-2\ln 2+1,
~~~\rho_1={\pi^2\over 4\ln 2 -1+\rho_0}-\ln 2 +1.
$$ 
The behavior at $x=1$ differs from those at $x=0,\infty$ for the inverse of  
$\ln (1-x) $ appears. This is actually due to the fact that tr$(M_1M_x)=-2$. 
 We will prove the above behaviors in section \ref{caseimbro},  and  in the second paper by a different method.

\vskip 0.2 cm 
 In general, the logarithmic behaviors of ``type (\ref{intrlog1})'' at the critical points are as follows: 
\be
\left\{\matrix{ y(x)\sim
x\left\{
{\theta_x^2-\theta_0^2\over 4} 
\left[
\ln x +{4r+2\theta_0\over \theta_0^2-\theta_x^2}
\right]^2 +{\theta_0^2\over \theta_0^2-\theta_x^2}
\right\}
,&~~~x\to 0.
\cr
\cr
y(x)\sim
 {\theta_0^2\over \theta_0^2-\theta_1^2}+
{\theta_1^2-\theta_0^2\over 4} 
\left[
\ln {1\over x} + {4 r + 2\theta_0\over \theta_0^2-\theta_1^2}
\right]^2,&~~~x\to\infty.
\cr
\cr
y(x)\sim
 1 -(1-x) \left\{{\theta_1^2\over \theta_1^2-\theta_x^2} +
{\theta_x^2-\theta_1^2\over 4}
\left[
\ln (1-x) + {4 r + 2\theta_1 \over \theta_1^2-\theta_x^2}
\right]^2
\right\},&~~~x\to 1.
}
\right.
\label{ass1}
\ee
\vskip 0.2 cm 
\be
\label{ass2}
\left\{
\matrix{y(x)=
{4\over [\theta_1^2-(\theta_\infty-1)^2]\ln^2 x } 
\left[
1+{8r+4\theta_\infty-4\over \theta_1^2-(\theta_\infty-1)^2}{1\over \ln x} 
+
O\left({1\over \ln^2x}\right)
\right],&~~~x\to 0.
\cr
\cr
y(x)={4~x\over [(\theta_\infty-1)^2-\theta_x^2]\ln^2x}
\left[
1-{8r+4(\theta_\infty-1)\over \theta_x^2-(\theta_\infty-1)^2}{1\over \ln x}+
O\left(1\over \ln^2 x
\right)
\right],&~~~x\to\infty.
\cr
\cr
y(x)=
 1+{4\over (\theta_1^2-\theta_0^2)\ln^2(x-1)}
\left[
1-{8r+4\theta_0\over \theta_0^2-\theta_1^2}{1\over \ln(x-1)}+O\left({1\over \ln^2(x-1)}\right)
\right],&~~~x\to 1.
}
\right.
\ee
\vskip 0.3 cm 
We stress that a solution with a log-behavior at  $x=0$ (for example) does not necessarily have a log-behavior at $x=1$ or $x=\infty$.   
The case (\ref{nongenintro}) is special, in that  the log-behavior appears at the three critical points. 
\vskip 0.2 cm 
In general, the log-behaviors of ``type (\ref{intrlog2})'' are:
\be
\label{Ass1}
\left\{
\matrix{
y(x)\sim 
x~(r~\pm~\theta_0~\ln x),&~~~~~x\to 0,&~~~~~\theta_0^2=\theta_x^2.
\cr
\cr
y(x)\sim
 r\pm\theta_0 \ln x,&
~~~~~x\to\infty,&~~~~~
\theta_0^2= \theta_1^2.
\cr
\cr
y(x)\sim
1-(1-x)
\bigl(
r\pm \theta_1\ln(1-x)
\bigr),&~~~~~x\to 1,&~~~~~\theta_1^2= \theta_x^2.
}
\right.
\ee
\vskip 0.1 cm 
\be
\label{Ass2}
\left\{
\matrix{
y(x)=
{1\over r\pm(\theta_\infty-1)\ln x}\left[
1+O\left(
{1\over \ln x}
\right)
\right],&~~~~~
x\to 0,&~~~~~(\theta_\infty-1)^2= \theta_1^2.
\cr
\cr
y(x)=
\pm{x\over (\theta_\infty-1)\ln x}\left[
1\mp{r\over (\theta_\infty-1)\ln x}+O\left(
{1\over \ln^2 x}
\right)
\right]
,&~~~~~x\to \infty,&~~~~~(\theta_\infty-1)^2=\theta_x^2.
\cr
\cr
y(x)=
 1\pm{1\over \theta_0\ln(x-1)}
\left[
1\mp {r\over \theta_0\ln(x-1)}+O\left({1\over \ln^2(x-1)}\right)
\right],&~~~~~x\to 1,&~~~~~(\theta_\infty-1)^2=\theta_0^2.
}
\right.
\ee

The above are  proved in Section \ref{monodinsim}, 
 making use of the Backlund transformations of (PVI). 
The behaviors in  (\ref{ass1}), (\ref{Ass1}) 
 are associated to $\hbox{\rm tr}(M_0M_x)=2$ when $x\to 0$; to $\hbox{\rm tr}(M_0M_1)=2$ when $x\to \infty$;  
to $\hbox{\rm tr}(M_1M_x)=2$ when $x\to 1$. This fact is proved in Section \ref{CCPR}. The behaviors in (\ref{ass2}), (\ref{Ass2}) are 
associated to $\hbox{\rm tr}(M_0M_x)=-2$ when $x\to 0$; 
to $\hbox{\rm tr}(M_0M_1)=-2$ when $x\to \infty$; 
to  $\hbox{\rm tr}(M_1M_x)=-2$ when $x\to 1$. This fact is proved in
 section \ref{cioco}.

\vskip 0.3 cm
{\bf Acknowledgments:} The paper was written at RIMS, Kyoto
University,  supported by the Kyoto Mathematics COE.



\section{Matching Procedure}
\label{MatchingProcedure}

 This section is a review of the matching procedure of \cite{guz5}. We explain how the asymptotic behavior of a transcendent is derived, and how the associated monodromy is computed. 

\subsection{Leading Terms of $y(x)$}
\label{matchleadingaprile}

We consider $x\to 0$. We divide the $\lambda$-plane into
two domains. The ``outside'' domain  is defined for $\lambda$ sufficiently big:
\be
|\lambda|\geq |x|^{\delta_{OUT}},~~~~~\delta_{OUT}>0. 
\label{dominioOUTbasta}
\ee
Therefore, (\ref{SYSTEM}) can be written as: 
\be
{d\Psi\over d \lambda}=
\left[
{A_0+A_x\over \lambda}+{A_x\over \lambda}~\sum_{n=1}^{\infty}\left({x\over \lambda}\right)^n+ {A_1\over \lambda-1}
\right]~\Psi.
\label{SYSTEM1aprile}
\ee
The ``inside'' domain  is defined for $\lambda$ comparable with $x$, namely:
\be
|\lambda|\leq |x|^{\delta_{IN}},~~~~~\delta_{IN}>0. 
\label{dominioINbasta}
\ee
Therefore, $\lambda\to 0$ as $x\to 0$, and we rewrite ({\ref{SYSTEM})
  as: 
\be
{d \Psi \over d\lambda}
=
\left[
{A_0\over \lambda} + {A_x \over \lambda -x} - A_1 \sum_{n=0}^\infty 
\lambda^n
\right]~
\Psi.
\label{SYSTEM0aprile}
\ee

\vskip 0.2 cm 
If the behavior of $A_0(x)$, $A_1(x)$ and $A_x(x)$ is sufficiently
good, we expect that the higher order terms in the series of
(\ref{SYSTEM1aprile}) and (\ref{SYSTEM0aprile}) are small
corrections which can be neglected when $x\to 0$. If this is the case,
(\ref{SYSTEM1aprile}) and (\ref{SYSTEM0aprile}) reduce respectively  to:
\be
{d\Psi_{OUT}\over d \lambda}=
\left[
{A_0+A_x\over \lambda}+{A_x\over \lambda}~\sum_{n=1}^{N_{OUT}}\left({x\over \lambda}\right)^n+ {A_1\over \lambda-1}
\right]~\Psi_{OUT},
\label{nonfuchsianSYSTEMOUT}
\ee
\be
{d \Psi_{IN} \over d\lambda}
=
\left[
{A_0\over \lambda} + {A_x \over \lambda -x} - A_1 \sum_{n=0}^{N_{IN}} 
\lambda^n
\right]~
\Psi_{IN},
\label{nonfuchsianSYSTEMIN}
\ee
where $N_{IN}$, $N_{OUT}$ are suitable integers. 
The simplest  reduction  is to Fuchsian systems:
\be
\label{fuchsianSYSTEMOUT}
{d\Psi_{OUT}\over d \lambda}=
\left[
{A_0+A_x\over \lambda}+ {A_1\over \lambda-1}
\right]~\Psi_{OUT},
\ee
\be
{d \Psi_{IN} \over d\lambda}
=
\left[
{A_0\over \lambda} + {A_x \over \lambda -x}
\right]~
\Psi_{IN}.
\label{fuchsianSYSTEMIN}
\ee

In \cite{guz5} we  considered reduced
non-fuchsian systems for the first time in the literature, where  
 the fuchsian reduction has
been privileged. We showed that in some relevant cases it cannot be
used, being the non-fuchsian reduction necessary. 

\vskip 0.2 cm

 Generally speaking, we can parameterize the elements of
 $A_0+A_x$ and $A_1$ of (\ref{fuchsianSYSTEMOUT}) in terms of
 $\theta_1$, the eigenvalues of $A_0+A_x$ and  the eigenvalues
 $\theta_\infty$ of $A_0+A_x+A_1$. We also need an additional unknown
 function of $x$.   In the same way, we 
can explicitly parameterize the elements of
 $A_0$ and $A_x$  in (\ref{fuchsianSYSTEMIN}) in terms of
 $\theta_0$, $\theta_x$, the   eigenvalues of $A_0+A_x$  and  another
 additional unknown
 function of $x$.  Cases when the reductions   
 (\ref{nonfuchsianSYSTEMOUT}) and (\ref{nonfuchsianSYSTEMIN}) are
 non-fuchsian deserve   
    particular care, as it has been done in \cite{guz5}.
 Our purpose is to find the leading terms  of the unknown functions 
 when $x\to 0$, in order to determine  the  critical behavior of 
 $A_0(x)$, $A_1(x)$, $A_x(x)$ and of  
 (\ref{leadingtermaprile}).

 The leading term can be  obtained as a result of two facts: 

\noindent
 i) Systems (\ref{nonfuchsianSYSTEMOUT})
 and (\ref{nonfuchsianSYSTEMIN}) are isomonodromic. This imposes 
  constraints on the form of the unknown functions. Typically, one of
 them must be constant. 

\noindent
ii) Two fundamental matrix 
 solutions $\Psi_{OUT}(\lambda,x)$, $\Psi_{IN}(\lambda,x)$ must
match in the region of overlap, provided this is not empty: 
\be\Psi_{OUT}(\lambda,x)
\sim 
\Psi_{IN}(\lambda,x), ~~~~~
|x|^{\delta_{OUT}} \leq |\lambda|\leq |x|^{\delta_{IN}},~~~x\to 0  
\label{overlapINOUT}
\ee  
This relation is to be intended in the sense that the leading terms 
of the local behavior of $\Psi_{OUT}$ and $\Psi_{IN}$ for $x\to
0$ must be equal. 
This  determines a simple relation between the two functions of $x$
appearing in $A_0$, $A_x$, $A_1$, $A_0+A_x$.    (\ref{overlapINOUT})
also implies that $\delta_{IN}\leq\delta_{OUT}$.

\vskip 0.2 cm 
Practically, to fulfill point ii), we  match  a fundamental solution of 
(\ref{nonfuchsianSYSTEMOUT}) for $\lambda\to 0$,  with 
 a fundamental solution of the system obtained from (\ref{nonfuchsianSYSTEMIN}) 
by the change of variables  $\mu:=\lambda/x$, namely with a solution of: 
\be
{d \Psi_{IN} \over d\mu}
=
\left[
{A_0\over \mu} + {A_x \over \mu -1} - xA_1~ \sum_{n=0}^{N_{IN}} 
x^n\mu^n
\right]~
\Psi_{IN},~~~~~\mu:={\lambda\over x}.
\label{nonfuchsianSYSTEMINmu}
\ee

\vskip 0.2 cm

To summarize,  {\it matching} two fundamental 
 solutions of the  reduced   {\it 
isomonodromic} systems (\ref{nonfuchsianSYSTEMOUT}) and
(\ref{nonfuchsianSYSTEMIN}), we  obtain the leading term(s), for $x\to 0$,  
 of the entries of the matrices of the original system
 (\ref{SYSTEM}). The procedure is algorithmic, the only  assumption being 
(\ref{overlapINOUT}). 

\vskip 0.3 cm
 This method is sometimes called {\it coalescence of singularities},
 because the singularity $\lambda=0$ and $\lambda=x$ coalesce to
 produce  system (\ref{nonfuchsianSYSTEMOUT}), while the singularity
 $\mu={1\over x}$ and $\mu=\infty$ coalesce to produce system
 (\ref{nonfuchsianSYSTEMINmu}). 
Coalescence of singularities was first used by M. Jimbo in 
\cite{Jimbo} to compute the monodromy matrices of (\ref{SYSTEM}) for the 
class of solutions of (PVI) with leading term $y(x)\sim ~a~x^{1-\sigma}$,
$0<\Re\sigma<1$.

\subsection{Computation of the Monodromy Data}
\label{MonodromyPasqua}

In  the ``$\lambda$-plane'' 
${\bf C}\backslash\{0,x,1\}$ we fix a base point $\lambda_0$ 
and  three loops, which are numbered in order 1, 2, 3 according to a 
counter-clockwise order referred to $\lambda_0$.  We choose $0,x,1$ to be the order $1,2,3$. 
We denote the loops by $\gamma_0$, $\gamma_x$, $\gamma_1$. See figure \ref{figure1}.   
\begin{figure}
\epsfxsize=12cm
\centerline{\epsffile{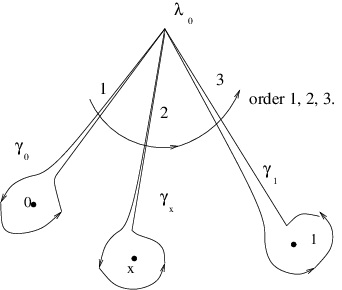}}
\caption{The ordered basis of loops}
\label{figure1}
\end{figure}
 The monodromy matrices of a fundamental solution $\Psi(\lambda)$  w.r.t. this base of loops are denoted  $M_0$, $M_x$, $M_1$. The loop at infinity will be $\gamma_\infty=\gamma_0\gamma_x\gamma_1$, so $M_\infty=M_1M_xM_0$. 
 As a consequence, the following relation
 holds:   
$$
 \cos(\pi \theta_0) \hbox{tr}(M_1 M_x) + \cos(\pi \theta_1) \hbox{tr}(M_0 M_x) + \cos(\pi \theta_x) \hbox{tr}(M_1 M_0)
$$
$$
= 2\cos(\pi \theta_\infty)+ 4 \cos(\pi \theta_1)\cos(\pi \theta_0)\cos(\pi \theta_x). 
$$ 
The monodromy matrices are determined by tr$(M_\nu)$, tr$(M_\nu M_\mu)$ , $\nu,\mu=0,x,1,\infty$ \cite{Boalch}.

As a consequence of isomonodromicity, 
there exists a fundamental solution $\Psi_{OUT}$ of
(\ref{nonfuchsianSYSTEMOUT}) such that
 $$
M^{OUT}_1=M_1,~~~~~M^{OUT}_\infty=M_\infty,
$$
where $M^{OUT}_1$ and $M^{OUT}_\infty$ are the monodromy matrices of  
$\Psi_{OUT}$ at
$\lambda=1,\infty$. Moreover,  $M^{OUT}_0= M_xM_0$.  
 There also exists a fundamental solution $\Psi_{IN}$ of
(\ref{nonfuchsianSYSTEMIN}) such that:
$$
M^{IN}_0=M_0,~~~~~M^{IN}_x=M_x,
$$
where $M^{IN}_0 $ and $M^{IN}_x $  are the monodromy matrices of  
$\Psi_{IN}$ at $\lambda=0,x$. 

\vskip 0.2 cm 
The method of coalescence of singularities is useful when 
the monodromy of the reduced systems
(\ref{nonfuchsianSYSTEMOUT}), (\ref{nonfuchsianSYSTEMIN}) can be
explicitly computed. This is the case  when the reduction
is fuchsian (namely (\ref{fuchsianSYSTEMOUT}),
(\ref{fuchsianSYSTEMIN})), because fuchsian systems with three
singular points are equivalent  to a Gauss hyper-geometric equation
(see Appendix 1). For the non-fuchsian reduction, 
in  general  we can compute the monodromy 
when (\ref{nonfuchsianSYSTEMOUT}),
(\ref{nonfuchsianSYSTEMIN}) are solvable in terms of special or
elementary functions.  

\vskip 0.2 cm 

In order for this procedure to work, not only $\Psi_{OUT}$ and $\Psi_{IN}$ 
 must match with each other, as in subsection
 \ref{matchleadingaprile}, 
but also $\Psi_{OUT}$ 
 must match with a fundamental matrix solution  $\Psi$ of
 (\ref{SYSTEM}) in a domain of the $\lambda$ plane, and   $\Psi_{IN}$ 
 must match with {\it the same}
 $\Psi$ in another domain of the $\lambda$ plane. 
  
\vskip 0.2 cm 
The standard choice of $\Psi$ is as 
follows: 
\be
\Psi(\lambda)= 
\left\{ 
\matrix{
\left[
I+O\left({1\over \lambda}\right)
\right]~\lambda^{-{\theta_\infty\over 2}\sigma_3} \lambda^{R_\infty},&~~~\lambda\to\infty;
\cr
\cr
\psi_0(x) \bigl[I+O(\lambda)\bigr]~\lambda^{{\theta_0\over
    2}\sigma_3}\lambda^{R_0}C_0,&~~~\lambda\to 0;
\cr
\cr
\psi_x(x)\bigl[I+O(\lambda-x)\bigr]~(\lambda-x)^{{\theta_x\over
    2}\sigma_3}(\lambda-x)^{R_x}C_x,&~~~\lambda\to x;
\cr
\cr
\psi_1(x)\bigl[I+O(\lambda-1)\bigr]~(\lambda-1)^{{\theta_1\over
    2}\sigma_3}(\lambda-1)^{R_1}C_1,&~~~\lambda\to 1;
}\right.
\label{PSIlocalelo}
\ee
Here $\psi_0(x)$,  $\psi_x(x)$, $\psi_1(x)$ are the diagonalizing
matrices of $A_0(x)$, $A_1(x)$, $A_x(x)$ respectively. They are
defined by multiplication to the right by arbitrary diagonal matrices,
possibly depending on $x$. $C_\nu$, $\nu=\infty,0,x,1$,
 are invertible {\it connection matrices}, independent of $x$
\cite{JMU}. Each $R_\nu$,  
$\nu=\infty,0,x,1$, is also independent of $x$, and:
$$
R_\nu=0 \hbox{ if } \theta_\nu\not\in {\bf Z},~~~~~
R_\nu=\left\{
\matrix{
\pmatrix{0 & *\cr 0 & 0},~~~ \hbox{ if } \theta_\nu>0 \hbox{
  integer}
 \cr
\cr
\pmatrix{0 & 0\cr * & 0},~~~ \hbox{ if } \theta_\nu<0 \hbox{
  integer}
}
\right.
$$
If $\theta_i=0$, $i=0,x,1$, then 
 $R_i$ is to be considered the Jordan form $\pmatrix{0 & 1 \cr 0 & 0}$
of $A_i$. 
Note that for the loop $\lambda \mapsto \lambda e^{2\pi i}$,
$|\lambda|>\max\{1,|x|\}$, we immediately compute the monodromy at infinity:  
$$
M_\infty=\exp\{-i\pi\theta_\infty\}~\exp\{ 2\pi i R_\infty\}. 
$$ 

\vskip 0.5 cm
Let  
$\Psi_{OUT}$ and $\Psi_{IN}$ be the solutions of 
 (\ref{nonfuchsianSYSTEMOUT}) and (\ref{nonfuchsianSYSTEMIN}) 
matching as in (\ref{overlapINOUT}). We explain how they are matched
with (\ref{PSIlocalelo}).

\vskip 0.3 cm
\noindent
{\bf (*) Matching $\Psi~\leftrightarrow~ \Psi_{OUT}$:}~~

$\lambda=\infty$ is a fuchsian singularity of
(\ref{nonfuchsianSYSTEMOUT}), with residue
  $-A_\infty/\lambda$. Therefore, we can always find a
  fundamental matrix solution with
  behavior:   
$$
\Psi_{OUT}^{Match}~=\left[
I+O\left({1\over \lambda}\right)
\right]~\lambda^{-{\theta_\infty\over 2}\sigma_3} \lambda^{R_\infty},
~~~\lambda\to\infty.
$$
This solution matches with $\Psi$. 
Also $\lambda=1$ is a fuchsian singularity of
(\ref{nonfuchsianSYSTEMOUT}). Therefore, we have: 
$$
 \Psi_{OUT}^{Match}~
=\psi_1^{OUT}(x)\bigl[I+O(\lambda-1)\bigr]~(\lambda-1)^{{\theta_1\over
    2}\sigma_3}(\lambda-1)^{R_1}C^{OUT}_1,~~~\lambda\to 1;
$$
Here $C^{OUT}_1$ is a suitable connection matrix. $\psi_1^{OUT}(x)$ is
the matrix that diagonalizes the leading 
terms of $A_1(x)$. 
Therefore, $\psi_1(x)\sim \psi_1^{OUT}(x)$ for
$x\to 0$. As a consequence of  isomonodromicity, $R_1$ is the same of 
$\Psi$.  

As a consequence of the matching  $\Psi~\leftrightarrow~
\Psi_{OUT}^{Match}$, the monodromy of $\Psi$ at $\lambda=1$ is:  
$$
M_1={C_1}^{-1}\exp\{i\pi \theta_1\sigma_3\} \exp\{2\pi i R_1\} C_1,
~~\hbox{ with } 
C_1\equiv C^{OUT}_1.
$$
\vskip 0.2 cm 
We finally need an invertible connection matrix  $C_{OUT}$ to connect  
$\Psi_{OUT}^{Match}$ with the solution  $\Psi_{OUT}$ appearing in 
 (\ref{overlapINOUT}). Namely,   $
\Psi_{OUT}^{Match}= \Psi_{OUT} C_{OUT}.
$

\vskip 0.3 cm
\noindent
{\bf (*) Matching $\Psi~\leftrightarrow~ \Psi_{IN}$:}~~ 

\vskip 0.2 cm
As a consequence of the matching $\Psi ~\leftrightarrow~
\Psi_{OUT}^{Match}$, we have to choose  the  IN-solution which
 matches with $\Psi_{OUT}^{Match}$. This is $
\Psi_{IN}^{Match}:=\Psi_{IN} C_{OUT}$.  
 
\vskip 0.18 cm
Now,  $\lambda =0, x$ are fuchsian singularities of
 (\ref{nonfuchsianSYSTEMIN}). Therefore: 
$$
\Psi_{IN}^{Match}=
\left\{
\matrix{
\psi_0^{IN}(x) \bigl[I+O(\lambda)\bigr]~\lambda^{{\theta_0\over
    2}\sigma_3}\lambda^{R_0}C^{IN}_0,&~~~\lambda\to 0;
\cr
\cr
\psi_x^{IN}(x)\bigl[I+O(\lambda-x)\bigr]~(\lambda-x)^{{\theta_x\over
    2}\sigma_3}(\lambda-x)^{R_x}C^{IN}_x,&~~~\lambda\to x;
}
\right.
$$
The above hold for fixed small $x\neq 0$.  
Here $C^{IN}_0$ and  $C^{IN}_x$ are suitable connection matrices. 
$\psi_0^{IN}(x)$ and $ \psi_x(x)^{IN}$ are diagonalizing matrices of 
the leading terms of  
$A_0(x)$ and $A_x(x)$. For $x\to 0$ they match with  $\psi_0(x)$ and 
$ \psi_x(x)$ of $\Psi$ in (\ref{PSIlocale}).  
On the other hand, as a consequence of isomonodromicity, the matrices
$R_0$ and $R_x$ are the same of
$\Psi$. The above $\Psi_{IN}^{Match}$ has the same behavior of $\Psi$ at $\lambda\to 0$ and $\lambda\to x$; moreover, it is an approximation of $\Psi$ for $x$ small. The 
matrices $C_0^{IN}$, $C_x^{IN}$ are independent of $x$.  So, the matching 
$\Psi~\leftrightarrow~ \Psi_{IN}$ is realized and 
the connection matrices  $C_0$ and $C_x$ coincide with 
   $C^{IN}_0$,
$C^{IN}_x$  respectively. As a result, we obtain
the monodromy matrices for $\Psi$:
$$
M_0= {C_0}^{-1} \exp\{i\pi\theta_0\sigma_3\} \exp\{ 2\pi i R_0\} C_0,
~~~~~ C_0\equiv C_0^{IN},
$$ 
$$
 M_x= {C_x}^{-1} \exp\{i\pi\theta_x\sigma_3\} \exp\{ 2\pi i R_x\}
C_x,~~~~~ C_x\equiv C_x^{IN}. 
$$

\vskip 0.2 cm 
Our reduction is useful if 
 the connection matrices $C_1^{OUT}$, $C^{IN}_0$, $C^{IN}_x$ can be computed
explicitly. This is  possible  for the fuchsian reduced systems   
(\ref{fuchsianSYSTEMOUT}), (\ref{fuchsianSYSTEMIN}). 
For  non-fuchsian reduced systems, we  discussed the 
computability in \cite{guz5}.


\section{Classification in Terms of Monodromy Data}
\label{ClassificationTerms}
Two  conjugated systems: 
$$
 {d\Psi\over d\lambda}=A(x,\lambda)~\Psi,~~~{d\tilde{\Psi}\over d\lambda}=\tilde{A}(x,\lambda)~\tilde{\Psi},
$$
$$
\tilde{\Psi}= W \Psi, ~~~~~ \det(W)\neq 0,~~~~~\tilde{A}=WAW^{-1},
$$
admit fundamental matrix solutions with the same monodromy matrices (w.r.t. the same basis of loops). The matrix $\tilde{A}(x,\lambda)$ 
defines the same solution of (PVI) associated to $A(x,\lambda)$  only 
if  the following condition holds:  
$$
\tilde{A}_0+\tilde{A}_1+\tilde{A}_x = -{\theta_{\infty}\over 2}
 \sigma_3,~~~~~\hbox{ where } \tilde{A}_i =WA_iW^{-1},~~~i=0,x,1.
$$ 
Namely, $W\sigma_3W^{-1}=\sigma_3$. This occurs if and only if $W$ is 
{\it diagonal}. 
 The transformation of $A(x,\lambda)$ is therefore: 
$$
 WA(x,\lambda)W^{-1}= \pmatrix{ A_{11}(x,\lambda) & {w_2\over w_1} A_{12}(x,\lambda) 
\cr
              {w_1\over w_2} A_{21}(x,\lambda)  &  A_{22}(x,\lambda)}, ~~~~~\hbox{ where } 
W=\pmatrix{w_1 & 0 \cr 0 & w_2}.
$$
 We conclude that the equation $A_{12}(x,\lambda)=0$ is the same and then:

\vskip 0.2 cm
{\it Two conjugate fuchsian systems, satisfying (\ref{caffe0}) (\ref{caffe1}),
 define the same solution of PVI 
if and only if the conjugation is diagonal.}
\vskip 0.3 cm 

Note that $\theta_\infty\neq 0$ is a necessary condition, otherwise any $W$ 
would be acceptable and then $A_{12}(x,\lambda)=0$ would not define $y(x)$ uniquely. 

\vskip 0.2 cm 

The problem of finding 
 a (branch of a) 
 transcendent associated to a  monodromy representation 
 is the problem of finding a fuchsian system (\ref{SYSTEM})
 having the given monodromy. This problem is called {\it
 Riemann-Hilbert problem}, or { \it $21^{th}$ Hilbert problem}. For a
 given PVI  there
 is a one-to-one correspondence between a monodromy representation and a
 branch of a transcendent if and only if 
the Riemann-Hilbert problem has a unique
 solution $A(x,\lambda)$, defined up to diagonal conjugation.  

\vskip 0.2 cm 
\noindent
{\bf $\bullet$ Riemann-Hilbert problem (R.H.)}: find the 
 coefficients $A_i(x)$, $i=0,x,1$ from the following {\it monodromy data}:

a) A fixed order of the poles $0,x,1$. Namely, we choose a base of loops.  
Here we choose the order (1,2,3)=(0,$x$,1). 
See figure \ref{figure1}.


b) The exponents  $\theta_0,\theta_x,\theta_1,\theta_{\infty}$, with  
$\theta_\infty \neq 0$. The restriction $\theta_\infty \neq 0$ is just technical, due to the present definition of $\theta_\infty$, and is not a limitation, because by definition $\theta_\infty=0$ is equivalent to $\theta_\infty=2$.

c) Matrices  $R_0,R_x,R_1,
R_\infty$, such that:  
$$
R_\nu=0 \hbox{ if } \theta_\nu\not\in {\bf Z},~~~~~
R_\nu=\left\{
\matrix{
\pmatrix{0 & *\cr 0 & 0},~~~ \hbox{ if } \theta_\nu>0 \hbox{
  integer}
 \cr
\cr
\pmatrix{0 & 0\cr * & 0},~~~ \hbox{ if } \theta_\nu<0 \hbox{
  integer}
}
\right.
$$
$$
R_j=\pmatrix{0 & 1 \cr 0 & 0},~~~\hbox{ if } \theta_j=0,~~~j=0,x,1.
$$

c) three  monodromy matrices $M_0$, $M_x$, $M_1$ relative to the 
loops, similar to the matrices $\exp\{ i\pi\theta_i\sigma_3\}
\exp\{ 2\pi i R_i\}$, $i=0,x,1$, satisfying  (for the chosen order of loops 
$\gamma_0\gamma_x\gamma_1=\gamma_\infty$):  
$$ 
  M_1~M_x~M_0 = e^{-i\pi \theta_{\infty}\sigma_3 }e^{2 \pi i  R_{\infty}}
$$

\vskip 0.2 cm
Solving the Riemann-Hilbert problem means that we have to find invertible {\it connection matrices}, $C_\nu$, $\nu=\infty,0,x,1$, such that. 
\be
C_j^{-1} e^{i\pi\theta_j\sigma_3} e^{2\pi i R_j} C_j =M_j,~~~~~j=0,x,1;
\label{conn1}
\ee
\be
C_{\infty}^{-1}  e^{-i\pi\theta_\infty\sigma_3} e^{2\pi i R_\infty}C_{\infty}
=
 e^{-i\pi\theta_\infty\sigma_3} e^{2\pi i R_\infty}.
\label{conn2}
\ee
 and a matrix valued meromorphic
function $\Psi(x,\lambda)$  such that:  
\be
\Psi(x,\lambda)= 
\left\{ 
\matrix{
\left[
I+O\left({1\over \lambda}\right)
\right]~\lambda^{-{\theta_\infty\over 2}\sigma_3} \lambda^{R_\infty}C_{\infty},
&~~~\lambda\to\infty;
\cr
\cr
\psi_0(x) \bigl[I+O(\lambda)\bigr]~\lambda^{{\theta_0\over
    2}\sigma_3}\lambda^{R_0}C_0,&~~~\lambda\to 0;
\cr
\cr
\psi_x(x)\bigl[I+O(\lambda-x)\bigr]~(\lambda-x)^{{\theta_x\over
    2}\sigma_3}(\lambda-x)^{R_x}C_x,&~~~\lambda\to x;
\cr
\cr
\psi_1(x)\bigl[I+O(\lambda-1)\bigr]~(\lambda-1)^{{\theta_1\over
    2}\sigma_3}(\lambda-1)^{R_1}C_1,&~~~\lambda\to 1;
}\right.
\label{PSIlocale}
\ee
Here $\psi_0$,  $\psi_x$, $\psi_1$ are invertible
matrices depending on $x$.  The
coefficient of the fuchsian system are then given by 
$$
 A(x;\lambda):= {d \Psi(x,\lambda) \over d\lambda} \Psi(x;\lambda)^{-1}
.$$

\vskip 0.2 cm
A $2 \times 2$ R.H. is always solvable at a fixed $x$ \cite{AB}. 
As a 
function of $x$, the solution $A(x;\lambda)$ 
extends to a meromorphic function on the universal covering of 
$\bar{\bf C}\backslash \{0,1,\infty\}$. 
Now we prove the following fact:
\vskip 0.2 cm 

{\it  
 The R.H. admits diagonally conjugated solutions (fuchsian systems), 
except  when at least one $\theta_\nu\in{\bf Z}\backslash\{0\}$ 
and simultaneously  $R_\nu=0$. 
}
\vskip 0.2 cm 
This can be equivalently stated in the form of the following:
\bpr
There is a one to one correspondence between the {\rm monodromy data}  
$\theta_0,\theta_x,\theta_1$, $R_0,R_x,R_1$, $\theta_\infty\neq 0, ~R_\infty$, $M_0,M_x,M_1$ (defined up to conjugation), satisfying a), b), c) above, 
 and a (branch of a) transcendent $y(x)$,  except  when 
at least one $\theta_\nu\in{\bf Z}\backslash\{0\}$ and simultaneously  
$R_\nu=0$.
\label{pro1}
\epr

To say in other words, the one to one correspondence is realized 
if and only if one of the following conditions is satisfied: 
\vskip 0.2 cm 
(1) $\theta_\nu \not\in {\bf Z}$,  for every  $\nu=0,x,1,\infty$;
\vskip 0.2 cm 
(2) if some $\theta_\nu\in {\bf Z}$ and  $R_\nu\neq 0$, $\theta_\nu\neq 0$
\vskip 0.2 cm 
(3) if some $\theta_j=0$ ($j=0,x,1$) and simultaneously $\theta_\infty\not \in {\bf Z}$, or $\theta_\infty\in{\bf Z}$ and $R_\infty\neq 0$. 

\vskip 0.2 cm
Note that for $\theta_j=0$, $M_j$ can be put in Jordan form $\pmatrix{1 & 2\pi i \cr
0 & 1}$. Therefore Proposition \ref{pro1} says that: 

\vskip 0.2 cm
{\it There is one to one correspondence except when one of the 
matrices $M_i$ ($i=0,x,1$), or $M_\infty=M_1M_xM_0$, is equal to $\pm I$. 
} 

\vskip 0.2 cm
\noindent
{\it Proof:}
The proof is based on the observation that a triple of monodromy matrices $M_0$, $M_x$, $M_1$
may be realized by two fuchsian systems which are not conjugated. 
    The crucial point is that  the solutions 
of (\ref{conn1}), (\ref{conn2})
are not unique. Two sets of  particular solutions  $C_{\nu}$ and $\tilde{C}_{\nu}$($\nu=0,x,1,\infty$)  
give  to   fuchsian systems:   
$$
{d \Psi(x,\lambda)\over d\lambda} \Psi(x,\lambda)^{-1}
=A(x,\lambda),~~~~~
{d \tilde{\Psi}(x,\lambda)\over d\lambda}\tilde{\Psi}(x,\lambda)^{-1} =
\tilde{A}(x,\lambda).
$$
These may be  not diagonally conjugated. If this happens, 
there is no one-to-one correspondence between a set of monodromy data and a solutions of PVI.

\vskip 0.2 cm 
We study the structure of the solutions of (\ref{conn1}), (\ref{conn2}).  
 Equation (\ref{conn2}) has the following solutions:

\vskip 0.2 cm 
i) If $\theta_\infty\not\in{\bf Z}$ (and then $R_\infty=0$), 
 $$
C_\infty= \pmatrix{p_\infty & 0 \cr 0 & q_\infty}, ~~~p_\infty,q_\infty\in{\bf C}\backslash\{0\}
$$

ii) If $\theta_\infty\in{\bf Z}$ and $R_\infty\neq 0$, 
$$
C_{\infty}=\pmatrix{p_\infty & q_\infty \cr 0 & p_\infty}, ~~~\hbox{ if } 
R_{\infty}= \pmatrix{0 & * \cr 0 & 0}
$$
$$
C_{\infty}=\pmatrix{p_\infty &0 \cr  q_\infty  & p_\infty}, ~~~\hbox{ if } 
R_{\infty}= \pmatrix{0 & 0 \cr * & 0}.
$$
where $p_\infty,q_\infty\in{\bf C}$, $p_\infty\neq 0$.
\vskip 0.2 cm 
iii) If $\theta_\infty\in{\bf Z}$ and $R_\infty= 0$, then $C_\infty$ is {\it any} invertible matrix.

\vskip 0.2 cm 

 Equation (\ref{conn1}),  may have different solutions $C_j$ and $\tilde{C}_j$. Therefore $C_j\tilde{C}_j^{-1}$ is a solution of: 
$$
\bigl(C_j\tilde{C}_j^{-1}\bigr)^{-1} ~e^{i \pi \theta_j \sigma_3} e^{2\pi i R_j} 
~C_j\tilde{C}_j^{-1}
=
 e^{i \pi \theta_j \sigma_3} e^{2\pi i R_j}.  
$$

\vskip 0.2 cm 
i) If $\theta_j\not\in{\bf Z}$ (and then $R_j=0$), we have: 
 $$
C_j\tilde{C}_j^{-1}= \pmatrix{a_j & 0 \cr 0 & b_j},
~~~~~a_j,b_j\in{\bf C}\backslash\{0\}
$$

ii) If $\theta_j\in{\bf Z}$ and $R_j\neq 0$, we have:  
$$
C_j\tilde{C}_j^{-1}=\pmatrix{a_j & b_j \cr 0 & a_j}, 
~~~a_j,b_j\in{\bf C},~~a_j\neq 0;~~~
~~~\hbox{ if } 
R_j= \pmatrix{0 & * \cr 0 & 0}
$$
$$
C_j\tilde{C}_j^{-1}=\pmatrix{a_j &0 \cr  b_j  & a_j},~~~a_j,b_j\in{\bf C},~~a_j\neq 0;~~~ ~~~\hbox{ if } 
R_j= \pmatrix{0 & 0 \cr * & 0}.
$$
In particular, for $\theta_j=0$ , $R_j$ is the Jordan form  $\pmatrix{0 & 1 \cr 0 & 0}$. 
\vskip 0.2 cm
iii) If $\theta_j\in{\bf Z}$ and $R_j= 0$, then $
C_j\tilde{C}_j^{-1}$ is any invertible matrix $\pmatrix{a&b\cr c&d}$.

\vskip 0.2 cm 

 Let $C_\nu$  and $\tilde{C}_\nu$ ($\nu=0,x,1,\infty$) be two sets of solutions 
of (\ref{conn1}) (\ref{conn2}) and let us denote by $\Psi$ and $\tilde{\Psi}$ the corresponding solutions of the R.H. 
We observe that:

i) for $\theta_j\not\in{\bf Z}$ $(j=0,x,1$): 
$$
    (\lambda-j)^{{\theta_j\over 2} \sigma_3}\pmatrix{a_j & 0 \cr 0 & b_j}= 
\pmatrix{a_j & 0 \cr 0 & b_j}
 (\lambda-j)^{{\theta_j\over 2} \sigma_3}.
$$

ii) For $\theta_j\in {\bf Z}$ and $R_j\neq 0$: 
$$
 (\lambda-j)^{{\theta_j\over 2} \sigma_3}(\lambda-j)^{R_j}
\pmatrix{a_j & b_j \cr 0 & a_j}= 
\left[a_j I+ (\lambda-j)^{|\theta_j|}\pmatrix{ 0 & b_j \cr 0 & 0}\right]
(\lambda-j)^{{\theta_j\over 2} \sigma_3}(\lambda-j)^{R_j},$$
or 
$$
 (\lambda-j)^{{\theta_j\over 2} \sigma_3}(\lambda-j)^{R_j}
\pmatrix{a_j &0 \cr  b_j  & a_j}= 
\left[a_j I+ (\lambda-j)^{|\theta_j|}\pmatrix{ 0 & 0 \cr b_j  & 0}\right]
(\lambda-j)^{{\theta_j\over 2} \sigma_3}(\lambda-j)^{R_j},
$$
for $R_j$ upper or lower triangular respectively. 

\vskip 0.2 cm 
iii) For $\theta_j\in {\bf Z}$ and $R_j= 0$:
$$
 (\lambda-j)^{{\theta_j\over 2} \sigma_3}\pmatrix{a & b \cr c & d}=
\pmatrix{a & b \lambda^{\theta_j} \cr c \lambda^{-\theta_j} & d}~
(\lambda-j)^{{\theta_j\over 2} \sigma_3}
$$

\vskip 0.2 cm 
We conclude that, for $\lambda \to j$:
$$
\Psi\tilde{\Psi}^{-1} \sim 
\left\{ \matrix{
\pmatrix{a_j & 0 \cr 0 & b_j},
~~~~~~~~~~ & \hbox{ if } \theta_j \not\in{\bf Z};
\cr
\cr
\cr
\left\{\matrix{
a_j I, \hbox{ if } \theta_j\neq 0,
\cr
\cr
 \pmatrix{ a_j & b_j \cr 0 & a_j},
 \hbox{ if } \theta_j=0,
}\right.
~~~~~~~ &\hbox{ if } \theta_j\in{\bf Z}, ~R_j\neq 0
\cr
\cr
\cr
\left\{\matrix{
\hbox{ Arbitrary invert. matrix, if } \theta_j=0,
\cr
\cr
{\cal C}~(\lambda-j)^{-|\theta_j|}\to \infty, \hbox{ otherwise}, 
}\right. 
& 
\hbox{ if } \theta_j\in{\bf Z},~R_j=0
}\right.
$$
The matrix ${\cal C}$ above is ${\cal C}=\pmatrix{0 & * \cr 0 & 0 }$ or   
${\cal C}=\pmatrix{0 & 0 \cr * & 0 }$. 

\vskip 0.2 cm 
Let $C_\infty$ and $\tilde{C}_\infty$ be two solutions of (\ref{conn2}).

\vskip 0.2 cm 
i) If $\theta_\infty\not\in{\bf Z}$ (and then $R_\infty=0$), we have
 $$
C_\infty\tilde{C}_\infty^{-1}= \pmatrix{a_\infty & 0 \cr 0 & b_\infty}, ~~~a_\infty,
b_\infty\in{\bf C}\backslash \{0\}.
$$

ii) If $\theta_\infty\in{\bf Z}$ and $R_\infty\neq 0$, we have 
$$
C_{\infty}\tilde{C}_\infty^{-1}=\pmatrix{a_\infty & b_\infty \cr 0 & a_\infty},
 ~~~a_\infty,
b_\infty\in{\bf C},~~a_\infty\neq 0;~~~
 ~~~\hbox{ if } 
R_{\infty}= \pmatrix{0 & * \cr 0 & 0}
$$
$$
C_{\infty}\tilde{C}_\infty^{-1}=\pmatrix{a_\infty &0 \cr  b_\infty  & a_\infty},~~~a_\infty,
b_\infty\in{\bf C},~~a_\infty\neq 0;~~~ ~~~\hbox{ if } 
R_{\infty}= \pmatrix{0 & 0 \cr * & 0}.
$$

iii) If $\theta_\infty\in{\bf Z}$ and $R_\infty= 0$, then $C_\infty\tilde{C}_\infty^{-1}$ is {\it any} invertible matrix. 

\vskip 0.2 cm 
 
Therefore, for $\lambda\to\infty$ we have: 
$$
\Psi\tilde{\Psi}^{-1} \sim \left\{ 
\matrix{
\pmatrix{a_\infty & 0 \cr 0 & b_\infty}~~~~~~~~~~ & \hbox{ if } \theta_\infty \not\in{\bf Z};
\cr
\cr
\left(
I + O\left({1\over \lambda}\right)
\right)
\left(
a_\infty I+ {b_\infty\over \lambda^{|\theta_\infty|}}
\right)\to a_\infty I,
~~~~~~~ &\hbox{ if } \theta_\infty\in{\bf Z}\backslash\{0\}, ~R_\infty\neq 0
\cr
\cr
{\cal C}_\infty \lambda^{|\theta_\infty|} 
\to \infty, ~~~~~~~~~~~~
& ~~~~~~~~~~~
\hbox{ if } \theta_\infty\in{\bf Z}\backslash\{0\},~R_\infty=0
}
\right.
$$
The matrix ${\cal C}_\infty$ above is 
${\cal C}_\infty=\pmatrix{0 & * \cr 0 & 0 }$ or   
${\cal C}_\infty=\pmatrix{0 & 0 \cr * & 0 }$.

\vskip 0.2 cm
From the above result we conclude that 
$\Psi \tilde{\Psi}^{-1}$ is analytic on $\bar{\bf C}$ and then it is a 
 constant matrix $W$, except  when 
at least one $\theta_\nu\in{\bf Z}\backslash\{0\}$ and simultaneously  
$R_\nu=0$. Except for this case, we have: 
$$
\Psi=W \tilde{\Psi} 
~~
\Longrightarrow 
~~
\tilde{A}(x,\lambda)=W A(x,\lambda) W^{-1}.
$$ 
We observe that:  
$W=\lim_{\lambda\to\infty}\Psi\tilde{\Psi}^{-1}$ 
(in the cases  $\theta_\infty\not\in{\bf Z}$, or for  
$\theta_{\infty}\in{\bf Z}$ ($\theta_\infty\neq 0$) and 
$R_\infty\neq 0$). Therefore $W$ is diagonal.

\vskip 0.2 cm

 Proposition \ref{pro1} is proved. 
\qed


\section{ Logarithmic asymptotics  (\ref{intrlog1}) and (\ref{intrlog2})}
\vskip 0.2 cm 

\noindent
 We consider cases when (\ref{SYSTEM}) can be reduced to the 
fuchsian systems (\ref{fuchsianSYSTEMOUT}) and
 (\ref{fuchsianSYSTEMIN}).  Let $\sigma$ be a complex number defined, up
to sign, by:
$$
\hbox{tr}~ (M_0M_x)=2\cos(\pi\sigma),~~~~~| \Re \sigma|\leq 1. 
$$ 

In our paper \cite{guz5},  
 we computed all the asymptotic behaviors for $0\leq \Re 
\sigma <1$, as they can be obtained from the matching procedure when 
(\ref{fuchsianSYSTEMOUT}) and
 (\ref{fuchsianSYSTEMIN}) are fuchsian.  
Among them, we obtained  (\ref{intrlog1}) and (\ref{intrlog2}).

\vskip 0.2 cm 
\noindent
{\it Note:} 
 For solutions with expansion:   
{\small
$$
y(x)= x (A_1 + B_1 \ln x + C_1 \ln^2 x + D_1 \ln^3 x + ...)+
x^2(A_2+B_2\ln x +...)+...,~~~~~x\to 0.  
$$
}
only the following cases are possible:      
{\small
\be
y(x)=
\left\{
\matrix{ 
 {\theta_0\over \theta_0\pm\theta_x} x + O(x^2) ~~\hbox{ [Taylor
     expansion]},
\cr\cr 
x~\left(
{\theta_0^2-B_1^2\over \theta_0^2-\theta_x^2} + B_1\ln x + 
{\theta_x^2-\theta_0^2\over 4} \ln^2 x 
\right)
+x^2(...)  +...,
\cr\cr
x~(A_1\pm \theta_0 \ln x)+x^2(...)+...,~~~\hbox{ and }\theta_0=\pm \theta_x.
}
\right.
\label{fivecases}
\ee
}
$A_1$ and $B_1$ are parameters. We see that the higher orders in (\ref{intrlog1}) and (\ref{intrlog2})  are $O(x^2\ln^m x)$, for some integer $m>0$.


\subsection{Review of the Derivation of  (\ref{intrlog1}) and (\ref{intrlog2})}

Let $x\to 0$. The reduction to the fuchsian systems (\ref{fuchsianSYSTEMOUT}) is
possible if  in the domain (\ref{dominioOUTbasta}) we have: 
\be
 |(A_0+A_x)_{ij}| \gg \left|(A_x)_{ij}~{x\over \lambda}\right|,~~~\hbox{
   namely: }~
|(A_0+A_x)_{ij}| \gg \left|(A_x)_{ij}~x^{1-\delta_{OUT}}\right|.
\label{condition1}
\ee
Let us denote with $\hat{A}_i$ the leading term of the matrix $A_i$,
$i=0,x,1$. We can  substitute 
(\ref{fuchsianSYSTEMOUT}) with: 
\be
{d\Psi_{OUT}\over d\lambda} = \left[{\hat{A}_0+\hat{A}_x\over
    \lambda}+ 
{\hat{A}_1\over \lambda-1}\right]~\Psi_{OUT}
\label{system1}
\ee

\ble
   If the approximation 
(\ref{fuchsianSYSTEMOUT}) is possible, then $ 
  \hat{A}_0+\hat{A}_x $ has eigenvalues $\pm {\sigma\over 2}\in {\bf
  C}$  independent of $x$, defined (up to sign and addition 
of an integer) by $
\hbox{\rm tr}(M_xM_0)= 2\cos(\pi \sigma)
$. Let $r_1\in{\bf C}$, $r_1\neq 0$. 
For $\theta_\infty\neq 0$, the leading terms are:
\be
\hat{A_1}= 
\pmatrix{ {\sigma^2-\theta_\infty^2-\theta_1^2\over 4 \theta_\infty} & 
-r_1
\cr
{[\sigma^2-(\theta_1-\theta_\infty)^2][\sigma^2-(\theta_1+\theta_\infty)^2]\over
  16 \theta_\infty^2 }~{1\over r_1} 
&
- 
 {\sigma^2-\theta_\infty^2-\theta_1^2\over 4 \theta_\infty} 
},
\label{hatA1}
\ee
and
\be
\hat{A_0}+\hat{A_x}=
\pmatrix{
{\theta_1^2-\sigma^2-\theta_\infty^2\over 4 \theta_\infty}
&
r_1
\cr
-{[\sigma^2-(\theta_1-\theta_\infty)^2][\sigma^2-(\theta_1+\theta_\infty)^2]\over
  16 \theta_\infty^2 }~{1\over r_1}
& 
-{\theta_1^2-\sigma^2-\theta_\infty^2\over 4 \theta_\infty}
}.
\label{hatA0Ax}
\ee
\label{elicopter1}
\ele

\vskip 0.2 cm
\noindent
{\it Proof:}  Observe that $
\hbox{tr}(\hat{A}_0+\hat{A}_x)=~\hbox{tr}(A_0+A_x)=0$, 
thus, for any $x$,  
$\hat{A}_0+\hat{A}_x$ has eigenvalues of opposite sign, that we
denote $\pm \tilde{\sigma}(x)/2$.    
Then, we recall that 
 $x$ is a monodromy preserving deformation, therefore the monodromy 
matrices 
of (\ref{system1}) are  
independent of $x$.  At $\lambda=0,1,\infty$ they are:
$$
   M^{OUT}_0=\left\{\matrix{M_xM_0
                                     \cr 
                            M_0M_x}
\right.
, ~~~M^{OUT}_1=M_1,~~~M^{OUT}_{\infty}=M_{\infty}.
$$
Thus,  det$(M^{OUT}_0)=1$, because
det$(M_x)$=det$(M_0)=1$. Therefore, there exists a constant matrix $D$
and a complex constant number $\sigma$  
such that:  
$$ 
D^{-1}~M^{OUT}_0~D~=\left\{ \matrix{
 \hbox{diag}(\exp\{-i\pi \sigma\}, \exp\{i\pi \sigma\})
,\cr
\cr
 \pmatrix{\pm 1 & * \cr 0 & \pm 1} ,\hbox{ or } \pmatrix{\pm 1 & 0 \cr * & \pm 1},~~~\sigma\in{\bf Z}
}
\right. 
$$ 
We conclude that $\tilde{\sigma}(x)\equiv \sigma$. We also have 
 $\hbox{tr}(M_0^{OUT})= 2\cos(\pi\sigma)$.

\vskip 0.2 cm 
 Now consider the gauge:  
\be
\Phi_1 := \lambda^{-{\sigma\over 2}} (\lambda-1)^{-{\theta_1\over 2}} 
~\Psi_{OUT}.~~~~~
{d\Phi_1\over d\lambda} = \left[{\hat{A}_0+\hat{A}_x-{\sigma\over 2}\over \lambda}+ {\hat{A}_1-{\theta_1\over 2}\over \lambda-1}\right]~\Phi_1
\label{systemPhi1}
\ee
We can identify $\hat{A}_0+\hat{A}_x-{\sigma\over 2}$ and
  $\hat{A}_1-{\theta_1\over 2}$ with $B_0$ and $B_1$  of  Proposition
 \ref{matrices} in Appendix 1, case  (\ref{1}), with $
a=  {\theta_\infty\over 2} +{\theta_1\over 2}+{\sigma\over 2}$, 
        $ b= - {\theta_\infty\over 2} +{\theta_1\over 2}+{\sigma\over
  2}$, $
         c= \sigma$. 
 \qed

\vskip 0.3 cm
 
In principle, $r_1$ may be a function of $x$. If the monodromy of system (\ref{system1}) 
depends on $r_1$, then $r_1$ is a constant independent of $x$. This is
the case here. 

 For all the computations which  follow, 
 involving system (\ref{system1}) or (\ref{systemPhi1}), we note that 
the hypothesis $\theta_\infty\neq 0$ excludes  cases (\ref{6}),  (\ref{7}) and the Jordan cases (\ref{8})--(\ref{10}).

\vskip 0.5 cm


The reduction to the fuchsian system (\ref{fuchsianSYSTEMIN}) is
possible for $x\to 0$ in the domain (\ref{dominioINbasta}) if: 
\be
\left|{(A_0)_{ij}\over \lambda}+{(A_x)_{ij}\over \lambda-x}
\right|
\gg
\left|(A_1)_{ij}\right|,
~~~\hbox{ namely:}~
\left|
{(A_0+A_x)_{ij}\over x^{\delta_{IN}}} 
\right| \gg 
\left|(A_1)_{ij}\right|.
\label{condition0}
\ee
We can rewrite   (\ref{fuchsianSYSTEMIN}) using just   the leading terms of
the matrices: 
\be
{d\Psi_{IN}\over d\lambda}=\left[
{\hat{A_0}\over \lambda}+{\hat{A_x}\over \lambda-x}\right] \Psi_{IN},
\label{systemapprox0}
\ee
Then, we re-scale $\lambda$ and consider the following system:  
$$
{d \Psi_{IN} \over d\mu} = \left( 
         {\hat{A}_0\over \mu} +{\hat{A}_x \over \mu-1} 
\right) \Psi_{IN},~~~~~\mu:={\lambda\over x}
$$
We know that there exists  a    
matrix $K_0(x)$ such that: 
$$
  {K_0}^{-1}(x)~(\hat{A_0}+\hat{A}_x)~K_0(x)= \pmatrix{{\sigma\over 2}& 0 \cr 
                                                0 & -{\sigma\over 2}}
,~~\hbox{ or }~\pmatrix{0 & 1 \cr 0 & 0 }.
$$
Let $
\hat{\hat{A_i}} := {K_0}^{-1} \hat{A}_i K_0$, $i=0,x$. By a gauge
transformation, we get the system: 
\be
\Psi_{IN}=: K_0(x)~\Psi_0,
~~~~~~~
{d\Psi_0\over d \mu}=\left[{\hat{\hat{A_0}}\over
    \mu}+{\hat{\hat{A_x}}\over \mu-1} 
 \right] \Psi_0,
\label{system0}
\ee 

\vskip 0.2 cm
\noindent
{\it Important Remark} (see \cite{guz5}):  Conditions (\ref{condition1}), (\ref{condition0}) are 
satisfied if and only if $|\Re \sigma|<1$, $0<\delta_{IN}\leq \delta_{OUT}<1$.


\subsection{Matching for $\sigma=0$. Proof of  (\ref{intrlog1}) and (\ref{intrlog2})}

We suppose now $\sigma=0$. 

\subsubsection{ Case $\theta_0\pm \theta_x\neq 0$. Proof of (\ref{intrlog1})}
\label{secsigma0}
\ble
Let $r_1\in{\bf C}$, $r_1\neq 0$. 
The matrices of system (\ref{system1}) are:  
$$
\hat{A_1}= 
\pmatrix{ -{{\theta_\infty}^2+{\theta_1}^2\over 4 \theta_\infty} & 
-r_1
\cr
{[\theta_1^2-\theta_\infty^2]^2\over 16 \theta_\infty^2 r_1} 
&
 {{\theta_\infty}^2+{\theta_1}^2\over 4 \theta_\infty} 
},~~~
\hat{A_0}+\hat{A_x}=
\pmatrix{
{\theta_1^2-\theta_\infty^2\over 4 \theta_\infty}
&
r_1
\cr
-{[\theta_\infty^2-\theta_1^2]^2\over 16 \theta_\infty^2 r_1} 
& 
{\theta_\infty^2-\theta_1^2\over 4 \theta_\infty}
}, ~~~~\forall r_1\neq 0.
$$
A fundamental matrix solution can be chosen with the following
behavior at $\lambda=0$: 
$$
\Psi_{OUT}(\lambda)=[G_0 + O(\lambda)]~\pmatrix{1 & \log \lambda \cr
0 & 1},~~~~~~~
G_0= \pmatrix{ 1 & 0 \cr
                     {{\theta_\infty}^2-{\theta_1}^2\over 4 \theta_\infty~r_1}
&
{1\over r_1}
}.
$$
\ele

\noindent
{\it Proof:}  The system (\ref{systemPhi1}) is: 
 $$
{d\Phi_1\over d\lambda} = \left[{\hat{A}_0+\hat{A}_x\over \lambda}+ {\hat{A}_1-{\theta_1\over 2}\over \lambda-1}\right]~\Phi_1, 
$$
 We identify 
 $\hat{A}_0+\hat{A}_x$ and $\hat{A}_1-{\theta_1\over 2}$ with $B_0$
 and $B_1$ of   proposition
 \ref{matrices} in Appendix 1, diagonalizable case (\ref{1})  
(we recall that (\ref{6})--(\ref{10}) never occur when $\theta_\infty\neq 0$) 
with $
a=  {\theta_\infty\over 2} +{\theta_1\over 2}$, 
      $   b= - {\theta_\infty\over 2} +{\theta_1\over 2}$, 
          $c= 0$. 

The behavior of a fundamental solution 
is a standard result in the theory of Fuchsian
systems. The matrix $G_0$ is defined by $
{G_0}^{-1} \left(\hat{A_0}+\hat{A_x}\right)G_0=
\pmatrix{ 0 & 1 \cr 0 & 0}
$. 
\qed

\ble
Let $r\in{\bf C}$. The matrices of system (\ref{system0}) are:
\be
\hat{\hat{A_0}}= \pmatrix{ r +{\theta_0\over 2} &
{4~r~(r+\theta_0) \over \theta_x^2-\theta_0^2} \cr
{\theta_0^2-\theta_x^2\over 4} & -r-{\theta_0\over 2}
},~~~~~
\hat{\hat{A_x}}=\pmatrix{
-r-{\theta_0\over 2}  & 1- {4~r~(r+\theta_0) \over \theta_x^2-\theta_0^2}
\cr
{\theta_x^2-\theta_0^2\over 4} 
&
r+{\theta_0\over 2}.
}.
\label{saravero?0}
\ee
There exist a fundamental solution of (\ref{system0}) with the
following behavior at $\mu=\infty$:
$$
\Psi_0(\mu)=
\left[I+O\left({1\over \mu}\right)\right]~
\pmatrix{ 1 & \log \mu \cr 0 &1 }, ~~~\mu\to\infty.
$$
\ele

\noindent
{\it Proof:} We do a gauge transformation:  
\be
\Phi_0:= \mu^{-{\theta_0\over 2}} (\mu-1)^{-{\theta_x\over 2}} ~\Psi_0,~~~~~
{d \Phi_0\over d\mu}= \left[ {\hat{\hat{A_0}}-{\theta_0\over 2}\over 
\mu} + {\hat{\hat{A_x}}-{\theta_x\over 2}\over \mu-1} \right] ~\Phi_0. 
\label{systemPhi00}
\ee
We 
identify $\hat{\hat{A_0}}-{\theta_0\over 2}$, 
$\hat{\hat{A_x}}-{\theta_x\over 2}$ 
with $B_0$ and $B_1$ in the Appendix 1, Proposition
\ref{matrices}, case (\ref{8}), with parameters $
a={\theta_0\over 2}+{\theta_x\over 2}$, $c= \theta_0$. In particular, 
 \be
\hat{\hat{A_0}}-{\theta_0\over 2} +\hat{\hat{A_x}}-{\theta_x\over 2}
=
\pmatrix{-{\theta_0+\theta_x\over 2} & 1 \cr 0 & -{\theta_0+\theta_x\over 2} }
\label{matrinf0}
\ee
Here the values of the  
parameters satisfy the conditions  
  $a\neq 0$ and $ a\neq c$,  namely 
 $\theta_0\pm\theta_x\neq 0$. From the  matrices (\ref{8}), we obtain
$\hat{\hat{A_0}}=B_0+\theta_0/2$ and 
$\hat{\hat{A_x}}=B_1+\theta_x/2$. 
Keeping into account (\ref{matrinf0}),  by the standard theory
of fuchsian systems we have:  
$$
\Phi_0(\mu)=\left[I+O\left({1\over \mu}\right)\right]~\mu^{-{\theta_0+\theta_x\over 2}}~
\pmatrix{ 1 & \log \mu \cr 0 &1 }, ~~~\mu\to\infty.
$$
This proves the behavior of $\Psi_0(\mu)$. 
\qed

\vskip 0.3 cm

 If the monodromy of the system ({\ref{system0}) 
depends on $r$, then $r$ is a constant independent of $x$. This is the
case here.

\vskip 0.3 cm
The matching condition  
$
\Psi_{OUT}(\lambda)\sim
K_0(x) ~\Psi_0\left(\lambda/ x\right)$ becomes: 
{\small
$$
K_0(x)~ \pmatrix{1 & \log\left({\lambda\over x}\right) \cr
0 & 1 }~\sim~
G_0~\pmatrix{1 & \log \lambda 
 \cr 
0 & 1 }~~~
\Longrightarrow
~~~
K_0(x)\sim \pmatrix{1 & 0 \cr
{\theta_\infty^2-\theta_1^2\over 4~\theta_\infty~r_1} &
{1\over r_1}
}~\pmatrix{ 1 & \log x \cr 0 & 1 }.
$$
}
 From the above result, together with (\ref{saravero?0}), we  compute $
\hat{A_0}=K_0\hat{\hat{A_0}}{K_0}^{-1}$, 
$\hat{A_1}=K_0\hat{\hat{A_1}}{K_0}^{-1}$. For example, 
{\small 
$$
\hat{A_0}
= G_0~
\pmatrix{
r+{\theta_0\over 2} + {\theta_0^2-\theta_x^2\over 4}\log x
~~&
{\theta_x^2-\theta_0^2\over 4}\log^2 x~ -2\left(
r+{\theta_0\over 2}
\right) \log x ~+{4~r(r+\theta_0)\over \theta_x^2 -\theta_0^2}
\cr
\cr
{\theta_0^2-\theta_x^2 \over 4} &
{\theta_x^2-\theta_0^2\over 4}\log x - \left(
r+{\theta_0\over 2}
\right)
}
~{G_0}^{-1}.
$$
}
A similar expression holds for $\hat{A_x}$. 
The leading terms of $y(x)$ are obtained from (\ref{leadingtermaprile}) with matrix 
entries $
(\hat{A}_1)_{12}=-r_1$ and:  
$$
(\hat{A_0})_{12}=r_1~\left[ {\theta_x^2-\theta_0^2\over 4} \log^2
  x-2\left(r+{\theta_0\over 2}\right)\log x +{4~r(r+\theta_0)\over
    \theta_x^2-\theta_0^2}\right]. 
$$
The result is: 
\be
y(x)\sim x ~\left[
{\theta_x^2-\theta_0^2\over 4} \log^2 x-2\left(r+{\theta_0\over
  2}\right)\log x +{4~r(r+\theta_0)\over \theta_x^2-\theta_0^2} 
\right]
\label{STELLA}
\ee
$$
= x ~\left\{
{\theta_x^2-\theta_0^2\over 4} \log^2 x-2\left(r+{\theta_0\over
  2}\right)\log x +{4\over \theta_x^2-\theta_0^2}\left[\left(r+{\theta_0\over 2}\right)^2-{\theta_0^2 \over 4}
\right]\right\}
.
$$
The above is (\ref{intrlog1}).


\subsubsection{Case  $\theta_0\pm\theta_x=0$. Proof of (\ref{intrlog2}) }
\label{specialcasesec}

We consider here the cases (\ref{9}), (\ref{10}) 
of Proposition \ref{matrices} applied to the system (\ref{systemPhi00}). 

\vskip 0.3 cm
\noindent
{\it Case (\ref{9})}  is the case $
\sigma=0$,  $\theta_0=-\theta_x$, 
 with $a=0$, $c=\theta_0$ in the system (\ref{systemPhi00}). From Proposition 
\ref{matrices} we immediately have:
$$
\hat{\hat{A_0}}=\pmatrix{{\theta_0\over 2} & r \cr
                            0 & -{\theta_0\over 2} \cr
},~~~\hat{\hat{A_x}}=\pmatrix{{\theta_x\over 2} & 1-r \cr
                            0 & -{\theta_x\over 2} \cr
}. 
$$
The behavior of $\Psi_0$ and $\Psi_{OUT}$, and the matching are the
same of  subsection \ref{secsigma0}.  We
obtain the same $K_0(x)$. Therefore: 
$$
(\hat{A_0})_{12}= r_1~(r-\theta_0~\ln x),~~~(\hat{A_1})_{12}=-r_1.
$$
This gives the leading terms: 
\be
y(x)\sim x(r-\theta_0~\ln x)=x(r+\theta_x~\ln x).
\label{CECILIA}
\ee

In the same way, we treat the other cases. 
{\it Case (\ref{9})} with $a=c$, is the case $
\sigma=0$, $\theta_0=\theta_x$. As above, we find $
y(x)\sim x(r-\theta_0~\ln x)=x(r-\theta_x~\ln x)
$. {\it Case (\ref{10})} with $a=0$, is the case $
\sigma=0$, $\theta_0=-\theta_x$. We find $
y(x)\sim  x(r+\theta_0~\ln x)=x(r-\theta_x~\ln x)
$. {\it Case (\ref{10})} with $a=c$,  is the case 
$  
\sigma=0$, $\theta_0=\theta_x$. We find  $
y(x)\sim  x(r+\theta_0~\ln x)=x(r+\theta_x~\ln x)
$.

\vskip 0.3 cm 
Both (\ref{STELLA}) and (\ref{CECILIA}) contain more than one term,
and in principle only the leading one is certainly correct. To
prove that they are all correct, we observe that  (\ref{STELLA}) and
(\ref{CECILIA}) 
can be obtained also by direct substitution of {\small $
y(x) = x (A_1 + B_1 \ln x + C_1 \ln^2 x + D_1 \ln^3 x + ...)+
x^2(A_2+B_2\ln x +...)+... $} 
into (PVI). We can recursively 
determine the coefficients by identifying 
 the same powers of $x$ and $\ln x$. As a result we obtain only the  
 five cases (\ref{fivecases}), which include  (\ref{STELLA}) and 
 (\ref{CECILIA}).

\vskip 0.3 cm
The reader can verify that conditions (\ref{condition1}), (\ref{condition0}) are satisfied.


\section{Monodromy Data associated to the solution (\ref{intrlog1}) } 
\label{sectionmonodromiagen}

In this section, we compute the monodromy data for the solution (\ref{intrlog1}) 
 in the generic case $\theta_\nu\not\in{\bf Z}$ for any 
$\nu=0,x,1,\infty$. 
 We need some notations. Let $\gamma_E$ denote the Euler's constant. Let:  
$$
\psi_E(x)={d\ln \Gamma(x)\over dx},~~~~~x\neq 0,-1,-2,-3,....
$$
In particular, $\psi_E(1)=-\gamma_E$.

\bpr 
\label{monodromiagen}
Let $\theta_0,\theta_x,\theta_1,\theta_\infty\not\in{\bf Z}$. The monodromy 
group  associated to  (\ref{intrlog1}) is generated by:
$$
M_0= E C_{0\infty}^{(*)}~\exp\{i\pi\theta_0\sigma_3\}~ 
\left[ E C_{0\infty}^{(*)}\right]^{-1},
$$
$$
M_x=  E C_{0\infty}^{(*)}~{C_{01}^{(*)}}^{-1} ~
\exp\{i\pi\theta_x\sigma_3\}~C_{01}^{(*)}~\left[ E C_{0\infty}^{(*)}\right]^{-1},
$$
$$
M_1= B C_{01}^{-1} ~\exp\{i\pi\theta_1\sigma_3\} ~C_{01}B^{-1}.
$$
The matrices above are: 
$$
 E= \pmatrix{{4q\over \theta_x^2-\theta_0^2} & {4\over \theta_0^2-\theta_x^2} 
\cr
\cr
 {4\over \theta_x^2-\theta_0^2} & 0}
$$
 $$
q =-4i\pi\epsilon+~~~~~~~~~~~~~~~~~~~~~~~~~~~~~~~~~~~~~~~
~~~~~~~~~~~~~~~~~~~~~~~~~~~~~~~~~~~~~~~~~~~~~~~~~~~~~~~~~
$$
$$
+{1\over \theta_0^2-\theta_x^2}\left\{4r + 
2(\theta_0-\theta_x) +(\theta_x^2-\theta_0^2)\left[
\psi\left(-{\theta_0\over 2}-{\theta_x\over 2}\right)+
\psi\left(
{\theta_x\over 2}-\
{\theta_0\over 2}+1\right)+2\gamma_E
\right]\right\},
$$
where $\epsilon=\pm 1$.
$$
C_{0\infty}^{(*)}= 
\pmatrix{-{e^{i\pi\epsilon\left({\theta_0\over 2}+{\theta_x\over 2}\right)}
\Gamma(1+\theta_0)\over \Gamma\left({\theta_0\over 2}+{\theta_x\over 2}\right)
\Gamma\left({\theta_0\over 2}-{\theta_x\over 2}\right)}
&
-{e^{i\pi\epsilon\left({\theta_x\over 2}-{\theta_0\over 2}\right)}\Gamma(1-\theta_0)\over \Gamma\left(-{\theta_0\over 2}-{\theta_x\over 2}\right)
\Gamma\left({\theta_x\over 2}-{\theta_0\over 2}\right)}
\cr
\cr
{e^{i\pi\epsilon\left({\theta_0\over 2}+{\theta_x\over 2}\right)}
~\pi\sin\pi\theta_0~\Gamma(1+\theta_0)\over 
\sin\pi\left({\theta_0\over 2}-{\theta_x\over 2}\right)
\sin\pi\left({\theta_0\over 2}+{\theta_x\over 2}\right)
\Gamma\left({\theta_0\over 2}+{\theta_x\over 2}\right)
\Gamma\left({\theta_0\over 2}-{\theta_x\over 2}\right)}
&
0}
$$
\vskip 0.2 cm 
$$
C_{01}^{(*)}= 
\pmatrix{
{\Gamma(-\theta_x)\Gamma(1+\theta_0)\over \left({\theta_0\over 2}-{\theta_x\over 2}\right)
\Gamma\left({\theta_0\over 2}-{\theta_x\over 2}\right)^2
}
& 
{\Gamma(-\theta_x)\Gamma(1-\theta_0)\over \left(-{\theta_0\over 2}-{\theta_x\over 2}\right)
\Gamma\left(-{\theta_0\over 2}-{\theta_x\over 2}\right)^2
}
\cr
{\Gamma(\theta_x)\Gamma(1+\theta_0)\over \left({\theta_0\over 2}+{\theta_x\over 2}\right)
\Gamma\left({\theta_0\over 2}+{\theta_x\over 2}\right)^2
}
& 
{\Gamma(\theta_x)\Gamma(1-\theta_0)\over \left(-{\theta_0\over 2}+{\theta_x\over 2}\right)
\Gamma\left(-{\theta_0\over 2}+{\theta_x\over 2}\right)^2}
}
,
$$
\vskip 0.2 cm 
$$
C_{01}= \pmatrix{
{\Gamma(-\theta_1)
\over 
\Gamma\left(
1-{\theta_\infty\over 2}-{\theta_1\over 2}
\right)
\Gamma\left(
{\theta_\infty\over 2}-{\theta_1\over 2}
\right)}
&
-{
\Gamma\left(
1+{\theta_1\over 2}-{\theta_\infty\over 2}
\right)
\Gamma\left(
{\theta_\infty\over 2}+{\theta_1\over 2}
\right)
\over 
\Gamma(1+\theta_1)}
\cr
\cr
{\Gamma(\theta_1)
\over 
\Gamma\left(
1+{\theta_1\over 2}-{\theta_\infty\over 2}
\right)
\Gamma\left(
{\theta_\infty\over 2}+{\theta_1\over 2}
\right)}
&
0 
},
$$
\vskip 0.2 cm
$$
B= \pmatrix{ 1 & \omega 
\cr
0 & 1 },
~~~~~\omega:=  \psi_E\left({\theta_\infty\over 2}+{\theta_1\over 2}\right)-
\psi_E\left(
{\theta_1\over 2}-\
{\theta_\infty\over 2}+1\right)+2\gamma_E,
$$
With the above choice, we have: 
$$
M_1M_xM_0= C_{OUT} ~\exp\{-i\pi \theta_\infty\sigma_3\}~
C_{OUT}^{-1},
$$
where:
$$
C_{OUT}= B C_{0\infty}^{-1}D^{-1},~~~~~
D= \pmatrix{ 1 & 0 \cr \cr 0 & {1-\theta_\infty \over r_1}},
~~~~~
C_{0\infty}= 
\pmatrix{1 & -{\pi ~ e^{-i{\pi\over 2} (\theta_1+\theta_\infty)}
\over \sin{\pi \over 2} (\theta_1+\theta_\infty)}
\cr
\cr
1 
& 
-{\pi ~ e^{-i{\pi\over 2} (\theta_1-\theta_\infty)}
\over \sin{\pi \over 2} (\theta_1-\theta_\infty)}
}
$$
Let $\epsilon=-1$. Let $s(z):=\sin({\pi\over 2}z)$. The parametrization of tr$(M_1M_x)$ and tr$(M_0M_1)$ in terms of $q$ is:
$$
{\rm tr}(M_0M_x)=2
$$
$$
{\rm tr}(M_0M_1)={\bf a}q^2+({\bf b}-2{\bf a}\omega)q+({\bf c}-{\bf b}\omega +{\bf a}\omega^2)
$$
$$
{\rm tr}(M_1M_x)={\bf A}q^2+({\bf B}-2{\bf A}\omega)q+({\bf C}-{\bf B}\omega +{\bf A}\omega^2)
$$
where 
$$ 
{\bf a}= {4\over \pi^2} s(\theta_0+\theta_x)s(\theta_0-\theta_x)s(\theta_\infty+\theta_1)s(\theta_\infty-\theta_1)
$$
$$
{\bf b}={4\over \pi} (\sin\pi\theta_1 ~s(\theta_0-\theta_x)s(\theta_0+\theta_x) + \sin\pi\theta_0~s(\theta_\infty+\theta_1)s(\theta_\infty-\theta_1))
$$
$$
{\bf c}= 2\cos\pi(\theta_0-\theta_1)
$$
$$
{\bf A}={\bf a}
$$
$$
{\bf B}= {1\over 2\pi i} \Bigl[
2\cos\pi(\theta_0+\theta_1)+4\cos\pi\theta_x \cos\pi\theta_\infty - 4 e^{i\pi\theta_1}\cos\pi\theta_x - 4 e^{-i\pi\theta_0}\cos\pi\theta_\infty + 3 e^{i\pi(\theta_1-\theta_0)} - e^{i\pi(\theta_0-\theta_1)}
\Bigr]
$$
$$
{\bf C}= 2 e^{i\pi\theta_1}\cos\pi\theta_x+2e^{-i\pi\theta_0}\cos\pi\theta_\infty -2 e^{i\pi(\theta_1-\theta_0)}
$$
The parametrization of $q$, and thus or $r$, in terms of the monodromy dat is: 
$$
q= \omega +{C-c+{\rm tr}(M_0M_1)-{\rm tr}(M_1M_x)\over b-B}
$$
Since tr$(M_0M_1)$ and tr$(M_1M_x)$ 
depend on $\epsilon$ only through 
$q$, different choices of $\epsilon$ just change the branch of  
(\ref{intrlog1}), because they change $4r/(\theta_0^2-\theta_x^2)$ of $8\pi i$. 
\epr


\subsection{Derivation of Proposition \ref{monodromiagen}} 
\label{proba}
 The matching $\Psi_{OUT}\leftrightarrow \Psi_{IN}$ has been realized by: 
$$
\Psi_{OUT}(x,\lambda)
=[G_0 + O(\lambda)]~\pmatrix{1 & \log \lambda \cr
0 & 1},~~~~~~~
G_0= \pmatrix{ 1 & 0 \cr
                     {{\theta_\infty}^2-{\theta_1}^2\over 4 \theta_\infty~r_1}
&
{1\over r_1}
}.
$$
$$
\Psi_{IN}(x,\lambda)= K_0(x)\Psi_0\left({\lambda\over x}\right),~~~~~
\Psi_0(\mu)=
\left[I+O\left({1\over \mu}\right)\right]~
\pmatrix{ 1 & \log \mu \cr 0 &1 }, ~~~\mu\to\infty.
$$

\vskip 0.5 cm
\noindent
{\bf MATCHING $\Psi\leftrightarrow \Psi_{OUT}$.}
\vskip 0.2 cm
\noindent
The correct choice of $\Psi_{OUT}^{Match}$ must match with: 
$$
\Psi= \left[I+O\left({1\over \lambda}
\right)\right]\lambda^{-{\theta_\infty\over
    2}\sigma_3},~~~\lambda\to\infty. 
$$
System (\ref{systemPhi1}) is (\ref{1}) of Appendix 1, with:  
$$
a={\theta_\infty\over 2}+{\theta_1\over 2},~~~b=-{\theta_\infty\over 2}+{\theta_1\over 2},~~~ c=0. 
$$    
If we write: 
$$\Psi_{OUT}= (\lambda-1)^{\theta_1\over 2}
\pmatrix{
\varphi_1 & \varphi_2
\cr
\xi_1 & \xi_2
},
$$
then $\varphi_1$ and $\varphi_2$ are independent solutions of the hypergeometric equation (\ref{hypergeom1}):
$$
\lambda(1-\lambda)~ {d^2 \varphi \over d\lambda^2} +\bigl(1+c-(a+[b+1]+1)~\lambda
\bigr)~ {d\varphi\over d\lambda} 
-a(b+1)~\varphi=0,
$$
while $\xi_i$ are given by (\ref{xi-hypergeom1}):  
$$
\xi_i= {1\over r}\left[
\lambda(1-\lambda)~{d\varphi_i\over d\lambda} ~-a\left(
\lambda+{b-c \over a-b}
\right)~\varphi_i
\right],~~~i=1,2.
$$
We need a complete set of solutions at $\lambda=0,1,\infty$. 

We 
explain some preliminary facts. 
Let us consider a Gauss hypergeometric equation in
standard form:
\be
z~(1-z)~ {d^2 \varphi \over dz^2} +\bigl[\gamma_0-(\alpha_0+\beta_0+1)~z \bigr]~ {d\varphi\over dz}
-\alpha_0\beta_0~\varphi=0
\label{standardGAUSS}
\ee
($\alpha_0,\beta_0,\gamma_0$ here are not   the coefficients of (PVI)! We
are just using the same symbols only here). We refer to the paper by N.E. Norlund 
\cite{Norlund} in order to choose three sets of 
 two independent solutions which can be easily expanded in series at $z=0, 1, \infty$ respectively. Solutions with logarithmic or polynomial behaviors at
 $z=0$ may occur when $\gamma_0\in {\bf Z}$. The role of $\gamma_0$  at $z=1$ and $z=\infty$ is played by $\alpha_0+\beta_0-\gamma_0+1$ and $\alpha_0-\beta_0+1$ respectively.  Therefore, solutions with logarithmic or polynomial 
behaviors at
 $z=1$ may occur when  $\alpha_0+\beta_0-\gamma_0+1\in {\bf Z}$, 
at $z=\infty$ when $\alpha_0-\beta_0+1\in{\bf Z}$. 
Some more words must be said about the choice of independent solutions. We consider the point $z=0$.  

For $\gamma_0\not\in {\bf Z}$, we choose the following two independent 
solutions:  
$$
\varphi_1(z)=F(\alpha_0,\beta_0,\gamma_0;~z),
~~~~~\varphi_2(z)= z^{1-\gamma_0}F(\alpha^{\prime},\beta^{\prime},\gamma^{\prime};~z).
$$
Here $F$ is the standard hypergeometric function and 
$\alpha^{\prime}=\alpha_0-\gamma_0+1$, $\beta^{\prime}=\beta_0-\gamma_0+1$, 
$\gamma^{\prime}=2-\gamma_0$.  

If $\gamma_0=0,-1,-2,...$,
then:
$$
\varphi_1(z)=f(\alpha_0,\beta_0,\gamma_0;~z), 
~~~\varphi_2(z)= z^{1-\gamma_0}F(\alpha^{\prime},\beta^{\prime},\gamma^{\prime};~z),
~~~~~\hbox{ if } \alpha_0 \hbox{ or } \beta_0=0,-1,...,\gamma.$$
$$
\varphi_1(z)=z^{1-\gamma_0}
{\cal G}(\alpha^{\prime},\beta^{\prime},\gamma^{\prime};~z), 
~~~\varphi_2(z)= z^{1-\gamma_0}F(\alpha^{\prime},\beta^{\prime},\gamma^{\prime};~z),
~~~~~\hbox{ if } \alpha_0 \hbox{ and } \beta_0 \neq 0,-1,...,\gamma.$$
Here $f$ is the truncation of $F$ at the order $z^{-\gamma}$. ${\cal G}$ is one of the functions $g$, $g_1$, $g_0$ or $G$ with logarithmic behavior, 
introduced in  \cite{Norlund}, section 2. They are listed in Appendix 3.  

If $\gamma_0=2,3,...$, then:
$$
\varphi_1(z)=F(\alpha_0,\beta_0,\gamma_0;~z), 
~~~\varphi_2(z)= z^{1-\gamma_0}f(\alpha^{\prime},\beta^{\prime},\gamma^{\prime};~z),
~~~~~\hbox{ if } \alpha_0 \hbox{ or } \beta_0=1,2,...,\gamma-1.$$
$$
\varphi_1(z)=
F(\alpha_0,\beta_0,\gamma_0;~z), 
~~~\varphi_2(z)={\cal G}(\alpha_0,\beta_0,\gamma_0;~z) ,
~~~~~\hbox{ if } \alpha_0 \hbox{ and } \beta_0 \neq 1,2,...,\gamma-1.$$

If $\gamma_0=1$, then: 
$$
\varphi_1(z)=
F(\alpha_0,\beta_0,\gamma_0;~z), 
~~~\varphi_2(z)={\cal G}(\alpha_0,\beta_0,\gamma_0;~z).
$$
The point $z=1$ is treated in the same way, with the substitution:
$$
\alpha_0\mapsto \alpha_0,~~~\beta_0\mapsto \beta_0,
~~~
\gamma_0\mapsto \alpha_0+\beta_0-\gamma_0+1;~~~\varphi\mapsto\varphi,~~~z\mapsto 1-z.$$
 The point $z=\infty$ is treated in the same way, with the substitution:
$$
\alpha_0\mapsto \alpha_0,~~~\beta_0\mapsto \alpha_0-\gamma_0+1,
~~~
\gamma_0\mapsto \alpha_0-\beta_0+1;~~~
\varphi\mapsto z^{-\alpha_0}\varphi,~~~z\mapsto {1\over z}.$$

 In our case: 
$$
\alpha_0=a={\theta_\infty\over 2}+{\theta_1\over 2},~~~\beta_0=b+1={\theta_1\over 2}-{\theta_\infty\over 2}+1,~~~
\gamma_0=c+1=1,~~~~~~~z=\lambda.
$$
Because $\gamma_0=1$,  we have a logarithmic solution at $\lambda=0$. 
As for $\lambda=1$, $\alpha_0+\beta_0-\gamma_0+1=1+\theta_1$ and for 
$\lambda=\infty$,   $\alpha_0-\beta_0+1= \theta_\infty$. We suppose $\theta_1$ and $\theta_\infty\not\in{\bf Z}$.  
We choose the following set of independent solutions at $\lambda=0,1,\infty$ 
respectively (the upper label indicates the singularity):
$$
\left\{\matrix{ \varphi_1^{(0)}= F(\alpha_0,\beta_0,\gamma_0;\lambda),
\cr
\varphi_2^{(0)}= g(\alpha_0,\beta_0,\gamma_0;\lambda);
}
\right.
$$
\vskip 0.2 cm 
$$
\left\{\matrix{ 
 \varphi_1^{(1)}= F(\alpha_0,\beta_0,\alpha_0+\beta_0-\gamma_0+1;1-\lambda),
\cr
\varphi_2^{(1)}= (1-\lambda)^{\gamma_0-\alpha_0-\beta_0}F(\gamma_0-\alpha_0,
\gamma_0-\beta_0,\gamma_0-\alpha_0-\beta_0+1;1-\lambda);
}
\right.
$$
\vskip 0.2 cm 
$$
\left\{\matrix{ 
 \varphi_1^{(\infty)}=\lambda^{-\alpha_0} F(\alpha_0,\alpha_0-\gamma_0+1,1+\alpha_0-\beta_0;\lambda^{-1}),
\cr
\varphi_2^{(\infty)}= \lambda^{-\beta_0}F(\beta_0,
\beta_0-\gamma_0+1,1-\alpha_0+\beta_0;\lambda^{-1});
}
\right.
$$

Let:
$$
\Psi_{OUT}^{(i)}
=(\lambda-1)^{{\theta_1\over 2}}
\pmatrix{
\varphi_1^{(i)} & \varphi_2^{(i)}
\cr
\xi_1^{(i)} & \xi_2^{(i)}
},~~~i=0,1,\infty. 
$$
From Norlund, 3.(1) and 3.(2) we get:
$$
\Psi_{OUT}^{(0)}=\Psi_{OUT}^{(1)} C_{01}, ~~~|\arg\lambda|<\pi,
~~~|\arg(1-\lambda)|<\pi,
$$
where $C_{01}$
is  written  in Proposition \ref{monodromiagen}. 
From Norlund, 10.(1) and 10.(3) we obtain:  
$$
\Psi_{OUT}^{(0)}=\Psi_{OUT}^{(\infty)} C_{0\infty}, ~~~0<\arg z <\pi,
$$
where 
$C_{0\infty}$ is written  in Proposition \ref{monodromiagen}.

\vskip 0.2 cm 
\noindent
{\it $\bullet$ Note about the computation:} In order to apply the 
formulae of Norlund, 10.(1) and 10.(3) we have to transform $g$ into $g_1$, using the formula (see Norlund, formula (24)): 
\be
g(\alpha,\beta,\gamma;z)= g_1(\alpha,\beta,\gamma;z)-{\pi \over \sin \pi \alpha}
e^{i\pi \epsilon\alpha} F(\alpha,\beta,\gamma;z),
\label{norlund?}
\ee
where $\epsilon$ is an integer introduced as follows. $g(\alpha,\beta,\gamma;z)$ is defined for $|$arg$(z)|<\pi$, while  $g_1(\alpha,\beta,\gamma;z)$  is defined for $|$arg$(-z)|<\pi$. Moreover, $-z=e^{i\epsilon\pi}z$. In  $g(\alpha,\beta,\gamma;z)$, $\ln(z)$ is negative for $0<z<1$ (namely, arg$(z)=0$), while in $g_1(\alpha,\beta,\gamma;z)$, $\ln (-z)$ is negative for $-1<z<0$. Namely, for $-1<z<0$, we have 
arg$(z)=-\pi\epsilon$. 
Formula (\ref{norlund?}) holds true for $0<\arg z<\pi$ when
 $\epsilon=-1$, and for $-\pi<\arg z< 0$ when $\epsilon=1$.

 In the formulae of Norlund, 10.(1) and 10.(3) it is required that 
$|$agr$(-z)|<\pi$, namely $|$arg$(e^{i\epsilon\pi}z)|$ $<\pi$. 
This limitation must be 
 restricted to  $0<\arg z<\pi$ when
 $\epsilon=-1$, and for $-\pi<\arg z< 0$ when $\epsilon=1$ in order to apply (\ref{norlund?}). 

In our computations we have chosen  $0<\arg z<\pi$ (i.e. $\epsilon=-1$), because this is the choice which gives the  order $M_1 M_x M_0 
=\exp\{-i\pi \theta_\infty \sigma_3\}$. The choice $-\pi<\arg z< 0$ ($\epsilon=1$) gives   $M_x M_1 M_0 
=\exp\{-i\pi \theta_\infty \sigma_3\}$.

\vskip 0.3 cm 
 We expand $\varphi_1^{(0)}$, $\varphi_2^{(0)}$ in series at $\lambda=0$ and 
we get: 
$$
\Psi_{OUT}^{(0)}= G_0\bigl[I +O(\lambda)  \bigr]
\pmatrix{1 & \ln \lambda
\cr
0 & 1} ~B e^{i{\pi \over 2} \theta_1},~~~\lambda\to 0,$$
where $B$ is written  in Proposition \ref{monodromiagen}.   
 Namely:
$$
\Psi_{OUT}^{(0)}=\Psi_{OUT}~B e^{i{\pi \over 2} \theta_1}
$$
 We expand $\varphi_1^{(\infty)}$, $\varphi_2^{(\infty)}$ in series 
at $\lambda=\infty$, 
 obtaining: 
$$
 \Psi_{OUT}^{(\infty)}= \left[I + O\left({1\over \lambda}\right)\right]
\lambda^{-{\theta_\infty\over 2}\sigma_3}~D, ~~~\lambda\to \infty, 
$$
where $D$ is  written  in Proposition \ref{monodromiagen}. Namely,
$$
\Psi_{OUT}^{(\infty)}=\Psi_{OUT}^{Match}~D.
$$
Combining the above results we get: 
$$
\left.
\matrix{
 \Psi_{OUT}^{Match} &=&\Psi_{OUT}^{(\infty)}D^{-1}&
\cr
\cr
&=& \Psi_{OUT}^{(0)}C_{0\infty}^{-1}D^{-1}&
\cr
\cr
&=&
 \Psi_{OUT} ~B C_{0\infty}^{-1}D^{-1} 
e^{i{\pi \over 2} \theta_1}& \equiv  \Psi_{OUT}C_{OUT}. 
}
\right.
$$
The matrix $B C_{0\infty}^{-1}D^{-1} 
e^{i{\pi \over 2} \theta_1}$ is $C_{OUT}$. It differs from the matrix $C_{OUT}$ of proposition \ref{monodromiagen} by the factor 
$e^{i{\pi \over 2} \theta_1}$, which simplifies in the formulae. 
We also have: 
$$
 \Psi_{OUT}^{Match}= \Psi_{OUT}^{(1)} C_{01} C_{0\infty}^{-1}D^{-1}. 
$$
Finally, it is an elementary computation to see that 
$$
\Psi_{OUT}^{(1)}= (\lambda-1)^{\theta_1\over 2} 
\pmatrix{
\varphi_1^{(1)} & \varphi_2^{(1)}
\cr
\xi_1^{(1)}  & \xi_2^{(1)} 
} \mapsto \Psi_{OUT}^{(1)} e^{i\pi \theta_1\sigma_3}, ~~~\hbox{ when } 
\lambda-1\mapsto (\lambda -1) e^{2\pi i}.
$$
Thus, a choice for the matrix $M_1$ of (\ref{SYSTEM}) is 
$$
\left.
\matrix{
 M_1\equiv M_1^{OUT}&=& D C_{0\infty}C_{01}^{-1} ~e^{i\pi \theta_1\sigma_3}~ C_{01} 
C_{0\infty}^{-1} D^{-1},
\cr
\cr
&=&
C_{OUT}^{-1}\bigl[ B C_{01}^{-1} ~e^{i\pi \theta_1\sigma_3}~C_{01} B^{-1}
\bigr] C_{OUT}.
 }
\right. $$

\vskip 0.5 cm
\noindent
{\bf MATCHING $\Psi \leftrightarrow \Psi_{IN}$}
\vskip 0.3 cm 

 The system:  
$$
\Phi_0:= \mu^{-{\theta_0\over 2}} (\mu-1)^{-{\theta_x\over 2}} ~\Psi_0,~~~~~
{d \Phi_0\over d\mu}= \left[ {\hat{\hat{A_0}}-{\theta_0\over 2}\over 
\mu} + {\hat{\hat{A_x}}-{\theta_x\over 2}\over \mu-1} \right] ~\Phi_0. 
\label{systemPhi0}
$$
is (\ref{8}) of Appendix 1, with: 
$$
a={\theta_0\over 2}+{\theta_x\over 2}, ~~~c=\theta_0.
$$
 The equation for  $\xi$ is in Gauss hypergeometic form 
(\ref{nuovaiper1}): 
\be
\mu(\mu-1){d^2\xi\over d\mu^2}+\bigr(1+c-2(a+1)\mu\bigl){d\xi\over d\mu} 
-a(a+1)\xi=0,
\label{nuovaipermu}
\ee
while $\varphi$ is given by (\ref{nuovaiper2}):
$$
\varphi(\mu)={1\over a(a-c)}
\left[
\mu(\mu-1){d\xi\over d\mu}+(a\mu-c-r)\xi
\right].
$$
In the standard form 
\be
\mu~(1-\mu)~ {d^2 \xi \over d\mu^2} +\bigl[\gamma_0-(\alpha_0+\beta_0+1)~\mu \bigr]~ {d\xi\over d\mu}
-\alpha_0\beta_0~\xi=0,
\label{standardGAUSS1}
\ee
we have:  
$$
\alpha_0=a={\theta_0\over 2}+{\theta_x\over 2},~~~ \beta_0=a+1={\theta_0\over 2}+{\theta_x\over 2}+1,~~~\gamma_0=c+1=\theta_0+1;~~~~~z=\mu. 
$$ 
Therefore $\gamma_0=1+\theta_0$, $\alpha_0+\beta_0-\gamma_0+1=1+\theta_x$, 
$\alpha_0-\beta_0+1=0$, and  (\ref{nuovaipermu}) has no logarithmic solutions at $\mu=0,1$ if $\theta_0,\theta_1\not\in{\bf Z}$. On the other hand, at 
$\mu=\infty$ we may have a solution with  logarithmic or polynomial behavior. 

For $\theta_0,\theta_x\not\in{\bf Z}$, we choose the following independent 
solutions at $\mu=0,1,\infty$ respectively::
$$
\left\{
\matrix{ 
\xi_1^{(0)}=F(\alpha_0,\beta_0,\gamma_0;~\mu)
\cr
\xi_2^{(0)}=\mu^{1-\gamma_0} F(\alpha_0-\gamma_0+1,\beta_0-\gamma_0+1,2-\gamma_0;~\mu);
}
\right.
$$
\vskip 0.2 cm
$$
\left\{
\matrix{ 
\xi_1^{(1)}=F(\alpha_0,\beta_0,\alpha_0+\beta_0-\gamma_0+1;~1-\mu)
\cr
\xi_2^{(1)}=(1-\mu)^{\gamma_0-\alpha_0-\beta_0} F(\gamma_0-\beta_0,\gamma_0-\alpha_0,1+\gamma_0-\alpha_0-\beta_0;~1-\mu);
}
\right.
$$
\vskip 0.2 cm
$$
\left\{
\matrix{ 
\xi_1^{(\infty)}=\mu^{-\beta_0}
g_1(\beta_0,1-\gamma_0+\beta_0,1-\alpha_0+\beta_0;~\mu^{-1})
\cr
\xi_2^{(\infty)}=\mu^{-\beta_0}
 F(\beta_0,1-\gamma_0+\beta_0,1-\alpha_0+\beta_0;~\mu^{-1});
}
\right.
$$

Let us construct three fundamental matrices form the above three sets of independent solutions:
$$
\Psi_0^{(i)}:= \mu^{{\theta_0\over 2}} (\mu-1)^{\theta_x\over 2}\pmatrix{ \varphi_1^{(i)} & \varphi_2^{(i)} 
\cr
 \xi_1^{(i)} & \xi_2^{(i)}
},~~~i=0,1,\infty
$$
The connection formulae between solutions at $\mu=0$ and $1$ is a standard one, and can be found in any book on special functions:
$$
\Psi_0^{(0)}=\Psi_0^{(1)} C_{01}^{(*)}, 
~~~~~|\arg(\mu)|<\pi,~~~|\arg(1-\mu)|<\pi
$$
where $ C_{01}^{(*)}$ is given in the statement of  Proposition 
\ref{monodromiagen}. 
The connection formulae between solutions at $\mu=0$ and $\mu=\infty$ 
can be found in Norlund \cite{Norlund}, formulae 9.(1) and 9.(5) (case $m=1$). 
We get:
$$
\Psi_0^{(0)}=\Psi_0^{(\infty)}C_{0\infty}^{(*)}, ~~~~~|\arg(-\mu)|<\pi,
$$
where $ C_{0\infty}^{(*)}$ can be read in Proposition \ref{monodromiagen} and 
$-\mu=e^{-i\pi\epsilon}\mu$ (when $\mu<0$, arg$(\mu)=\pi\epsilon$).

\vskip 0.2 cm 
\noindent
{\it $\bullet$ Note about the computation:} In order to apply the
 formulae 9.(1) and 9.(5)
of Norlund, we have made use of the formula: 
$$g_1(\alpha,\beta,\gamma;z)=g_1(\beta,\alpha,\gamma;z)+{\pi \sin\pi(\beta-\alpha)
\over \sin\pi\beta~\sin\pi\alpha}~F(\alpha,\beta,\gamma;z).
$$

\vskip 0.2 cm
We expand $\xi_1^{(\infty)}$, $\xi_2^{(\infty)}$, $\varphi_1^{(\infty)}$,   
$\varphi_2^{(\infty)}$ for $\mu\to \infty$. We obtain:
$$
\Psi_0^{(\infty)}=\left[I+\left({1\over \mu}\right)\right]
\pmatrix{1 & \ln \mu 
\cr
0 & 1 }~ E,~~~~~ \mu\to \infty
$$
where $E$ can be read in Proposition \ref{monodromiagen}. Thus, 
$$
\Psi_0^{(\infty)}=\Psi_0~E,
$$
where $\Psi_0$ is the matrix used in the matching $\Psi_{OUT}\leftrightarrow 
\Psi_{IN}$.  
Expanding  $\xi_1^{(0)}$, $\xi_2^{(0)}$, $\varphi_1^{(0)}$,   
$\varphi_2^{(0)}$ for $\mu \to 0$ we get: 
$$
\Psi_0^{(0)} = (\mu-1)^{-{\theta_x\over 2}} 
\pmatrix{ {4(\theta_0+r)\over \theta_0^2-\theta_x^2} & 
{4r\over \theta_0^2-\theta_x^2}
\cr
1
& 1
}
~[1+O(\mu)]~ \mu^{{\theta_0\over 2}\sigma_3},~~~~~\mu\to 0.
$$
Expanding  $\xi_1^{(1)}$, $\xi_2^{(1)}$, $\varphi_1^{(1)}$,   
$\varphi_2^{(1)}$ for $\mu \to 1$ we get: 
$$
\Psi_0^{(1)}= \pmatrix{{2(\theta_0-\theta_x+2r)\over \theta_0^2-\theta_x^2}
&
{2(\theta_0+\theta_x+2r)\over \theta_0^2-\theta_x^2}
\cr
1 & 
1
}
[1+O(1-\mu)] 
(1-\mu)^{{\theta_x\over 2}\sigma_3},~~~~~\mu\to 1.
$$
The above imply that:
$$
\Psi_0^{(0)}\mapsto \Psi_0^{(0)}e^{i\pi \theta_0\sigma_3},~~~\hbox{ for } 
\mu\mapsto \mu e^{2\pi i  },
$$
$$
\Psi_0^{(1)}\mapsto \Psi_0^{(1)}e^{i\pi \theta_x\sigma_3},~~~\hbox{ for } 
\mu-1\mapsto (\mu-1) e^{2\pi i  }.
$$
Finally, we observe that:
$$
\Psi_{IN}^{Match}=\Psi_{IN} C_{OUT},
$$
$$
\Psi_{IN}= K_0(x) \Psi_0= K_0(x)~ \Psi_0^{(\infty)}~E^{-1}=
\left\{
\matrix{
K_0(x)~ \Psi_0^{(0)} ~{C_{0\infty}^{(*)}}^{-1}E^{-1}
\cr
\cr
K_0(x)~\Psi_1^{(0)} ~C_{01}^{(*)}{C_{0\infty}^{(*)}}^{-1}E^{-1}
}
\right.
$$
As a result of the matching procedure we get:
$$
M_0\equiv M_0^{IN}= C_{OUT}^{-1} ~\bigl[EC_{0\infty}^{(*)} ~e^{i\pi\theta_0\sigma_3}~
{ C_{0\infty}^{(*)} }^{-1} E^{-1}\bigr]~ C_{OUT},
$$
$$
M_x\equiv M_1^{IN}=
  C_{OUT}^{-1} ~\bigl[EC_{0\infty}^{(*)}{C_{01}^{(*)}}^{-1}~
 e^{i\pi\theta_x\sigma_3}~
C_{01}^{(*)}{C_{0\infty}^{(*)}}^{-1} E^{-1}\bigr]~ C_{OUT}.
$$ 
\qed

\vskip 0.2 cm 
When we come to the computation of the traces, we find:
$$
\hbox{tr}(M_0M_1)={\bf a}q^2+({\bf b}-2{\bf a} \omega)q+({\bf c}-{\bf b} \omega+{\bf a}\omega^2),
$$
$$
\hbox{tr}(M_1M_x)={\bf A}q^2+({\bf B}-2{\bf A} \omega)q+({\bf C}-{\bf B}
 \omega+{\bf A}\omega^2),
$$
where {\bf a}, {\bf b}, {\bf c} , {\bf A}, {\bf B}, {\bf C} are given in the statement of Proposition \ref{monodromiagen}.
The above form for the system which determines $q$ (and therefore $r$)  implies that: 
$$
q= \omega +\bigl\{ \hbox{ solution of the system for $\omega=0$}\bigr\}.
$$
Moreover: 
$$
\bigl\{\hbox{ solution of the system for $\omega=0$}\bigr\}
=
{ {\bf a}\bigl({\bf C}-\hbox{tr}(M_1M_x)\bigr)-{\bf A}\bigl({\bf c}-\hbox{tr}(M_0M_1)\bigr)\over {\bf A}{\bf b}-{\bf a}{\bf B}}
$$
$$
\equiv {  
{\bf b}\bigl({\bf C}-\hbox{tr}(M_1M_x)\bigr)-{\bf B}
\bigl({\bf c}-\hbox{tr}(M_0M_1)\bigr)
\over
{\bf a}\bigl({\bf C}-\hbox{tr}(M_1M_x)\bigr)-
{\bf A}\bigl({\bf c}-\hbox{tr}(M_0M_1)\bigr)
}
$$
Keeping into account that ${\bf a}={\bf A}$, the above formula gives the result of Proposition \ref{monodromiagen}. 


\section{Monodromy Data associated to the Solution (\ref{intrlog2})}
\label{monodromiagen1}
 
\bpr
\label{propmonodromiagen1}~~
{\rm \bf [1].}  The monodromy group associated to the solution (\ref{intrlog2}):
$$
y(x)\sim x(r+\theta_0~\ln x), 
$$
 is generated by:
$$
M_0=E ~\exp\{-i\pi\theta_0\sigma_3\}~E^{-1},~~~~~M_x=EU^{-1} ~\exp\{i\pi\theta_x\sigma_3\}~UE^{-1},
$$
$$
M_1=B C_{01}^{-1}~\exp\{i\pi\theta_1\sigma_3\}~C_{01}B^{-1};
$$
where $B$, $C_{01}$ are given in  Proposition \ref{monodromiagen}
 and:
$$
E:=\pmatrix{e^{-i{\pi\over 2}\theta_0} & {r\over \theta_0}-\Psi_E(\theta_0+1)-\gamma_E-i\pi 
\cr\cr
0 & e^{i{\pi\over 2}\theta_0}} ,~~~~~
U:=\pmatrix{1 & -\Gamma(\theta_0+1)\Gamma(-\theta_0)
 \cr\cr
0 & 1}.
$$
Conversely, the parameter $r$ is:
$$
{r\over \theta_0}=-{\pi\over 4}~{\hbox{\rm tr}(M_0M_1) \over  \sin \pi \theta_0 \sin{\pi\over 2}(\theta_\infty+\theta_1)\sin{\pi\over 2}(\theta_\infty-\theta_1)}
~+(\Psi_E(\theta_0+1)+i\pi+\gamma_E)+
$$
\be
+ {\pi \over 2}~ {\cos\pi(\theta_0+\theta_1)\over \sin\pi\theta_0
\sin{\pi\over 2}(\theta_\infty+\theta_1)\sin{\pi\over 2}(\theta_\infty-\theta_1)}
-{\omega\over 2}~{\bigl[
\cos\pi(\theta_0+\theta_1)-\cos\pi(\theta_0-\theta_1)
\bigr]\over \sin\pi\theta_0\sin\pi\theta_1}.
\label{numerostar}
\ee
$\omega$ is given in Proposition \ref{monodromiagen}.

\vskip 0.2 cm 
{\rm \bf [2].}  The monodromy group and $r$ for  the solution (\ref{intrlog2}):
$$
y(x)\sim x(r-\theta_0~\ln x), 
$$
are obtained from the results in {\rm \bf [1]},  with 
 the substitution $\theta_0\mapsto -\theta_0$.

\epr

\vskip 0.2 cm 
\noindent
{\it Proof:} For the matching $\Psi_{OUT}\leftrightarrow \Psi_{IN}$ and $\Psi \leftrightarrow \Psi_{OUT}$,  we proceed as in  the proof of 
Proposition \ref{monodromiagen}. 

\vskip 0.2 cm 
\noindent
{\bf MATCHING $\Psi\leftrightarrow \Psi_{IN}$}

\vskip 0.1 cm 
Consider the case $\theta_0=\theta_x$. For this case, the system for $\Phi_0$ 
can be chosen to be (\ref{9}) or (\ref{10}), with $a=c=\theta_0$. Here we 
refer to  
system (\ref{10}). Therefore, a fundamental solution is (see 
Proposition \ref{ipergeom}):
$$
\Psi_0^{(0)}:=
\mu^{\theta_0\over 2}(\mu-1)^{\theta_0\over 2}\Phi_0=
$$
$$
=
e^{i{\pi\over 2}\theta_0 }\pmatrix{ \mu^{-{\theta_0\over 2}}(1-\mu)^{\theta_0\over 2}
&
{\pi\over \theta_0}(1-\mu)^{-{\theta_0\over 2}}\mu^{\theta_0\over 2}
-
{1\over \theta_0+1}\mu^{{\theta_0\over 2}+1}(1-\mu)^{\theta_0\over 2}~F(1+\theta_0,1+\theta_0,2+\theta_0;\mu)
\cr
\cr
0 & \mu^{\theta_0\over 2}(1-\mu)^{-{\theta_0\over 2}}
}.
$$
Here, the branch is: $(\mu-1)=e^{i\pi}(1-\mu)$. When $\mu\to\infty$, we write the hypergeometric function as follows, using the connection formula 
9.(1) in Norlund \cite{Norlund}:
$$
F(1+\theta_0,1+\theta_0,2+\theta_0;\mu)
= e^{i\pi\theta_0}(\theta_0+1) \mu^{-1-\theta_0} g_1\left(
0,1+\theta_0,1;{1\over \mu}
\right), ~~~0<\arg\mu<2\pi.
$$
Here, we have used the branch $-\mu = e^{-i\pi}\mu$. The function $g_1$ is: 
$$ 
g_1\left(
0,1+\theta_0,1;{1\over \mu}
\right)= \Psi_E(1+\theta_0) +\gamma_E +i\pi -\ln\mu +
\sum_{\nu=1}^\infty {(1+\theta_0)_\nu\over \nu~\nu!}\mu^{-\nu},~~~~~\mu\to\infty.
$$
From the above, we obtain:
$$
\Psi_0^{(0)}=\left[1+\left({1\over \mu}\right)\right]
\pmatrix{ 1 & \ln \mu \cr 0 & 1 } E e^{i{\pi\over 2}\theta_0}
\equiv \Psi_0 ~ E e^{i{\pi\over 2}\theta_0}.
$$
Here, $\Psi_0$ is the matrix used in the matching $\Psi_{OUT}\leftrightarrow 
\Psi_{IN}$ and $E$ is in the statement of the proposition.  
When $\mu\to 1$, we use the connection formula:
$$
 F(1+\theta_0,1+\theta_0,2+\theta_0;\mu)=
$$
$$
=\Gamma(-\theta_0)\Gamma(2+\theta_0) F(1+\theta_0,1+\theta_0,1+\theta_0;1-\mu)
+
{\Gamma(\theta_0)\Gamma(2+\theta_0)\over \Gamma(1+\theta_0)^2}
(1-\mu)^{-\theta_0} F(1,1,1-\theta_0;1-\mu).
$$
Therefore, 
$$
\Psi_0^{(0)}= e^{i{\pi\over 2}\theta_0}(I+O(1-\mu))
\pmatrix{1 & {r\over \theta_0}-{\Gamma(\theta_0)\Gamma(\theta_0+2)\over 
(\theta_0+1)\Gamma(\theta_0+1)^2}
\cr
\cr
0 & 1}~(1-\mu)^{{\theta_0\over 2}\sigma_3}U,~~~~~\mu\to 1.
$$
Finally, when $\mu \to 0$,we have:
$$
\Psi_0^{(0)}=
e^{i{\pi\over 2}\theta_0} (1+O(\mu))\pmatrix{1 &  r/\theta_0
\cr 
0 & 1
}~\mu^{-{\theta_0\over 2}\sigma_3}.
$$
Let $C_{OUT}$ be the same matrix introduced in the proof of Proposition 
\ref{monodromiagen}. We have:
$$
\Psi_{IN}^{Match}=\Psi_{IN}C_{OUT} = K_0(x) \Psi_0 C_{OUT}= K_0(x) \Psi_0^{(0)}
E^{-1} C_{OUT}. 
$$
This implies that:
$$
M_x= C_{OUT}^{-1}~EU^{-1} ~\exp\{i\pi \theta_x\sigma_3\}~UE^{-1}~C_{OUT},
$$
$$
M_0=  C_{OUT}^{-1}~E ~\exp\{-i\pi \theta_0\sigma_3\}~E^{-1}~C_{OUT}.
$$
The matrix $C_{OUT}$ has been simplified in the statement of the proposition. 

\vskip 0.2 cm 
The proof for $\theta_0=-\theta_x$ is analogous (for example,
 it is the case (\ref{9}) with 
$a=0$, $c=\theta_0$).  
\qed


\section{Monodromy Data  for the Non-generic Case 
(\ref{nongenintro})} 
\label{sectionmonodromiaCHAZYmy}

We consider the non-generic case 
$$
\theta_0=2p,~~~p\in{\bf Z}, ~~~\theta_0\neq 0,~~~~~
\theta_1=\theta_x=0,~~~\theta_\infty=1.
$$
In this case, the solutions (\ref{intrlog1}) becomes (\ref{nongenintro}). 
We show here that the solutions (\ref{nongenintro}) are not in one to one
 correspondence with  a set of monodromy data. Namely, to a given set of {\it monodromy data}, as defined in Proposition \ref{pro1}, there corresponds a one 
parameter family (\ref{nongenintro}),  
where $r$ is a free parameter (i.e. $r$ is not 
a function of the traces of the product of the monodromy matrices). 

We miss the one-to-one correspondence because the conditions in 
Proposition \ref{pro1} are not realized. Namely, the matrix $R_0$ 
associated to (\ref{nongenintro}) is:
$$
R_0=0, ~~~~~~~\hbox{ while } \theta_0\in{\bf Z} \hbox{ and } \theta_0\neq 0.
$$   
This fact is contained in the following Proposition.

\vskip 0.2 cm
\bpr
\label{monodromiaCHAZYmy}
The monodromy group associated to (\ref{nongenintro}) is generated by:
$$
M_0=I,~~~~~M_x=\pmatrix{1 & 2\pi i \cr 0 & 1},
~~~~~
M_1= \pmatrix{ 1-{8 i \over \pi} \ln 2 & 
-{32 i\over \pi}(\ln 2)^2
\cr
\cr
{2 i\over \pi} & 
1+{8~i\over \pi}\ln 2
}.
$$
In particular, 
$$
\hbox{\rm tr}(M_0M_x)=\hbox{\rm tr}(M_0M_1)=2,~~~~~\hbox{\rm tr}(M_1M_x)=-2
$$
The monodromy is independent of the parameter $r$ in (\ref{nongenintro}). 
\epr

\vskip 0.2 cm 
\noindent
{\it Note:} 
With the above choice the monodrmy at infinity: $M_1M_x$ ( or $M_x M_1$) is  
  not in standard Jordan form. Namely:
$$
M^{+}_{\infty}= M_1M_x=\pmatrix{1-{8i\over \pi} \ln 2
&
-{2i\over \pi} (4 \ln 2 + i\pi)^2
\cr
\cr
{2i\over \pi }
&
-3 +{8i\over \pi} \ln 2
},~~~~~M^{-}_\infty = M_x M_1=
\pmatrix{
-3 -{8 i \over \pi} \ln 2 
&
{2 i \over \pi }(4i\ln 2 +\pi)^2
\cr
\cr
{2i\over \pi} 
&
1+{8i\over \pi} \ln 2
}
$$
They can be put in Jordan form  respectively  by the following matrices:
$$
C_{OUT}^{+} = \pmatrix{ 1-{4i\over \pi}\ln 2 
& 
-{16i\over \pi } (\ln 2)^2 r_1
\cr
\cr
{i\over \pi} 
& 
\left(
1+{4i\over \pi} \ln 2
\right)r_1
},
~~~~~C_{OUT}^{-}=  \pmatrix{ 1+{4i\over \pi}\ln 2 
& 
{16i\over \pi } (\ln 2)^2 r_1
\cr
\cr
-{i\over \pi} 
& 
\left(
1-{4i\over \pi} \ln 2
\right)r_1
},~~~~~r_1\in{\bf C}. 
$$
 We obtain:
$$
{C_{OUT}^{+}}^{-1}M^{+}_\infty C_{OUT}^{+} = \pmatrix{ -1 & 2\pi i ~r_1
\cr
0 & -1},
~~~~~
{C_{OUT}^{-}}^{-1}M^{-}_\infty C_{OUT}^{-} = \pmatrix{ -1 & 2\pi i ~r_1 
\cr
0 & -1}.
$$
On the other hand:
$$
{C_{OUT}^{+}}^{-1}M_1 C_{OUT}^{+}= {C_{OUT}^{-}}^{-1}M_1 C_{OUT}^{-}
=
\pmatrix{ 
1-{8i\over \pi} \ln 2 &
-{32 i\over \pi}(\ln 2)^2 ~r_1
\cr
\cr
{2i \over \pi ~r_1} 
& 
1+{8i\over \pi} \ln 2
},
$$
$$
{C_{OUT}^{+}}^{-1}M_xC_{OUT}^{+}=\pmatrix{-1-{8i\over \pi} \ln 2
&
{2i\over \pi} (4i \ln 2 + \pi)^2~r_1
\cr
\cr
{2i\over \pi~r_1 }
&
3 +{8i\over \pi} \ln 2
},
$$
$$
{C_{OUT}^{-}}^{-1}M_xC_{OUT}^{-}=
\pmatrix{
3 -{8 i \over \pi} \ln 2 
&
{2 i \over \pi }(4i\ln 2 -\pi)^2~r_1
\cr
\cr
{2i\over \pi~r_1} 
&
-1+{8i\over \pi} \ln 2
}
$$


\subsection{Derivation of Proposition \ref{monodromiaCHAZYmy}}

 The matching $\Psi_{OUT}\leftrightarrow \Psi_{IN}$ has been realized by 
$$
\Psi_{OUT}(x,\lambda)= 
[G_0 + O(\lambda)]~\pmatrix{1 & \log \lambda \cr
0 & 1},~~~~~~~
G_0= \pmatrix{ 1 & 0 \cr
                     {1\over 4 r_1}
&
{1\over r_1}
}.
$$
$$
\Psi_{IN}(x,\lambda)= K_0(x)\Psi_0\left({\lambda\over x}\right),~~~~~
\Psi_0(\mu)=
\left[I+O\left({1\over \mu}\right)\right]~
\pmatrix{ 1 & \log \mu \cr 0 &1 }, ~~~\mu\to\infty.
$$

\vskip 0.5 cm
\noindent
{\bf MATCHING $\Psi\leftrightarrow \Psi_{OUT}$.}
\vskip 0.2 cm
\noindent
The correct choice of $\Psi_{OUT}^{Match}$ must match with: 
$$
\Psi= \left[I+O\left({1\over \lambda}
\right)\right]\lambda^{-{1\over
    2}\sigma_3}\lambda^{R_\infty},~~~R_\infty=\pmatrix{ 0 & -r_1 
\cr
0 & 
0
},~~~~~\lambda\to\infty. 
$$
System (\ref{systemPhi1}) is (\ref{1}) of Appendix 1, with: 
$$
a={1\over 2},~~~b=-{1\over 2},~~~ c=0. 
$$    
 If we write: 
$$\Psi_{OUT}= 
\pmatrix{
\varphi_1 & \varphi_2
\cr
\xi_1 & \xi_2
},
$$
then $\varphi_1$ and $\varphi_2$ are independent solutions of the hypergeometric equation (\ref{hypergeom1}):
$$
\lambda(1-\lambda)~ {d^2 \varphi \over d\lambda^2} +\bigl(1+c-(a+[b+1]+1)~\lambda
\bigr)~ {d\varphi\over d\lambda} 
-a(b+1)~\varphi=0,
$$
and  
$$
\xi_i= {1\over r}\left[
\lambda(1-\lambda)~{d\varphi_i\over d\lambda} ~-a\left(
\lambda+{b-c \over a-b}
\right)~\varphi_i
\right],~~~i=1,2.
$$
We need a complete set of solutions at $\lambda=0,1,\infty$. 
In the standard  Gauss hypergeometric form (\ref{standardGAUSS}) we 
have $\alpha_0=\beta_0=1/2$, $\gamma_0=1$. Since $\gamma_0=1$, 
 $\alpha_0+\beta_0
-\gamma_0+1=1$ and  $\alpha_0-\beta_0+1=1$, we expect  solutions  
with logarithmic behaviors at $\lambda=0,1,\infty$. 
We choose three sets of independent solutions:
$$
\left\{
\matrix{ 
\varphi_1^{(0)}=F(\alpha_0,\beta_0,\gamma_0;\lambda)\equiv F\left(
{1\over 2},{1\over 2},1;\lambda
\right),
\cr
\varphi_1^{(0)}=g(\alpha_0,\beta_0,\gamma_0;\lambda)\equiv g\left(
{1\over 2},{1\over 2},1;\lambda
\right);
}
\right.
$$
\vskip 0.2 cm
$$
\left\{
\matrix{ 
\varphi_1^{(1)}=F(\alpha_0,\beta_0,\alpha_0+\beta_0-\gamma_0+1;1-\lambda)\equiv F\left(
{1\over 2},{1\over 2},1;1-\lambda
\right),
\cr
\varphi_1^{(1)}=g(\alpha_0,\beta_0,\alpha_0+\beta_0-\gamma_0+1;1-\lambda)\equiv g\left(
{1\over 2},{1\over 2},1;1-\lambda
\right);
}
\right.
$$
\vskip 0.2 cm
$$
\left\{
\matrix{ 
\varphi_1^{(\infty)}=\lambda^{-\beta_0}F(\beta_0,\beta_0-\gamma_0+1,\beta_0-\alpha_0+1;\lambda^{-1})\equiv \lambda^{-{1\over 2}} F\left(
{1\over 2},{1\over 2},1;{1\over \lambda}
\right),
\cr
\varphi_1^{(\infty)}=\lambda^{-\beta_0}g(\beta_0,\beta_0-\gamma_0+1,\beta_0-\alpha_0+1;\lambda^{-1})
\equiv \lambda^{-{1\over 2}}g\left(
{1\over 2},{1\over 2},1;{1\over \lambda}
\right);
}
\right.
$$

 Let 
 $$\Psi_{OUT}^{(i)}= 
\pmatrix{
\varphi_1^{(i)} & \varphi_2^{(i)}
\cr
\xi_1^{(i)} & \xi_2^{(i)}
},
$$
From Norlund, formulae 5.(1) and 5.(2) we get: 
$$
\Psi_{OUT}^{(0)}=\Psi_{OUT}^{(1)} C_{01},~~~~~C_{01}=\pmatrix{ 0 & -\pi 
\cr
-{1\over \pi} & 0}; ~~~~~|\arg \lambda|<\pi,~~|\arg(1-\lambda)|<\pi.
$$
From Norlund, formulae 12.(1) and 12.(3) we get:
$$
\Psi_{OUT}^{(0)}=\Psi_{OUT}^{(\infty)}C_{0\infty},
~~~~~
C_{0\infty}=\pmatrix{ 1 & 0 
\cr
 -{1\over \pi}e^{i{\pi \over 2}\epsilon} & 1
};
$$
$$0<\arg\lambda<\pi~~(\epsilon=1),~~~-\pi<\arg \lambda<0~~(\epsilon=
-1).
$$

\vskip 0.3 cm
\noindent 
{$\bullet$ \it Note on the computation:} In order to apply 12.(1) we need: 
$$
g_1\left( 
{1\over 2},{1\over 2},1;{1\over \lambda}
\right)
=g\left( 
{1\over 2},{1\over 2},1;{1\over \lambda}
\right)+\pi e^{i{\pi\over 2}\epsilon} F\left( 
{1\over 2},{1\over 2},1;{1\over \lambda}
\right)
$$
$$
0<\arg\lambda<\pi~~(\epsilon=1),~~~-\pi<\arg \lambda<0~~(\epsilon=
-1).
$$
$\epsilon$ appears in the computations when we express:
 $-\lambda= e^{-i\pi\epsilon}\lambda$. 

 \vskip 0.2 cm 
We expand the solutions for $\lambda\to0$ and we get:
$$
\Psi_{OUT}^{(0)}= G_0(1+O(\lambda))\pmatrix{ 1 & \ln \lambda \cr
                                             0 & 1}~B,
~~~~~B=\pmatrix{ 1 & - 4 \ln 2 \cr 0 & 1},~~~~~\lambda\to 0.
$$
Namely,
$$
\Psi_{OUT}^{(0)} = \Psi_{OUT} B.
$$
Then expansion when $\lambda\to\infty$ yields::
$$
\Psi_{OUT}^{(\infty)}= \left[I+O\left({1\over \lambda}\right)\right]
~\lambda^{-{1\over 2}\sigma_3}\lambda^{R_\infty}~D
,~~~~~\lambda\to\infty;
$$
$$
D=\pmatrix{ 1 & -\ln 16 
\cr
0 & {1\over r_1}
}
,~~~~~R_\infty=\pmatrix{0 & -r_1 \cr 0 & 0}.
$$
Namely, 
$$
\Psi_{OUT}^{(\infty)}=
 \Psi_{OUT}^{Match} D.
$$
From the above:
$$
\left.
\matrix{
\Psi_{OUT}^{Match}
& = &\Psi_{OUT}^{(\infty)}D^{-1}&
\cr
\cr
& = & \Psi_{OUT}^{(0)}C_{0\infty}^{-1}D^{-1}&
\cr
\cr
& \equiv &\Psi_{OUT}~C_{OUT}
 ,&~~~~~\hbox{ where } ~~~ C_{OUT}=BC_{0\infty}^{-1}D^{-1}.
}
\right.
$$
It is easy to see that:
$$
\Psi_{OUT}^{(1)}\mapsto \Psi_{OUT}^{(1)}\pmatrix{1 & 2\pi i \cr 0 & 1}, ~~~~~
\hbox{ when } \lambda-1\mapsto e^{2\pi i}~(\lambda-1).
$$
This, together with the connection formulae
$$
\left.
\matrix{
\Psi_{OUT}^{Match} & =\Psi_{OUT}^{(0)} C_{0\infty}^{-1} D^{-1},~~~~~~
\cr\cr
&= \Psi_{OUT}^{(1)} C_{01}C_{0\infty}^{-1} D^{-1},
}
\right.
$$
yields:
$$
\left.
\matrix{
M_1&\equiv& M_1^{OUT}&=& D C_{0\infty}C_{01}^{-1} ~ 
\pmatrix{1 & 2\pi i \cr 0 & 1}
~
C_{01}C_{0\infty}^{-1}D^{-1}
\cr
\cr
& & &=&
 C_{~OUT}^{-1} B C_{01}^{-1}~ 
\pmatrix{1 & 2\pi i \cr 0 & 1}
~
C_{01} B^{-1} C_{OUT}.
}
\right.
$$
We have two choices for $C_{OUT}$, depending on $\epsilon=\pm 1$ in $C_{0\infty}$.  These have been called $C_{OUT}^{+}$ and $C_{OUT}^{-}$ in the Note,  after Proposition \ref{monodromiaCHAZYmy}.

\vskip 0.5 cm
\noindent
{\bf MATCHING $\Psi\leftrightarrow \Psi_{IN}$} 

\vskip 0.3 cm 

 The system:  
$$
\Phi_0:= \mu^{-p}  ~\Psi_0,~~~~~
{d \Phi_0\over d\mu}= \left[ {\hat{\hat{A_0}}-p\over 
\mu} + {\hat{\hat{A_x}}\over \mu-1} \right] ~\Phi_0. 
$$
is (\ref{8}) of Appendix 1, with: 
$$
a=p, ~~~c=2p.
$$
 The equation for  $\xi$ is in Gauss hypergeometic form 
(\ref{nuovaiper1}): 
\be
\mu(\mu-1){d^2\xi\over d\mu^2}+\bigr(1+c-2(a+1)\mu\bigl){d\xi\over d\mu} 
-a(a+1)\xi=0,
\label{nuovaipermuchazy}
\ee
$$
\varphi(\mu)={1\over a(a-c)}
\left[
\mu(\mu-1){d\xi\over d\mu}+(a\mu-c-r)\xi
\right].
$$
In the standard form (\ref{standardGAUSS1}), we have:  
$$
\alpha_0=p,~~~ \beta_0=1+p,~~~\gamma_0=1+2p;~~~~~z=\mu. 
$$ 
Therefore $\gamma_0=1+2p$, $\alpha_0+\beta_0-\gamma_0+1=1$, 
$\alpha_0-\beta_0+1=0$, and  (\ref{nuovaipermuchazy}) may 
have solutions with  logarithmic or polynomial behaviors  at $\mu=0,1,\infty$.

 The choice of three sets of  independent solutions requires a distinction of sub cases $p>0$ and $p<0$. As before, 
we denote:
$$
\Psi_0^{(i)}= \mu^{p}\Phi_0^{(i)} ,~~~~~\Phi_0^{(i)}=\pmatrix{\varphi_1^{(i)}
& \varphi_2^{(i)}
\cr
\xi_1^{(i)}
& \xi_2^{(i)}
},~~~i=0,1,\infty.
$$
\vskip 0.3 cm 
\noindent 
* {\it CASE $p>0$.} We choose: 
$$
\left\{
\matrix{
\xi_1^{(0)}=F(\alpha_0,\beta_0,\gamma_0;\mu),
\cr
\xi_2^{(0)}= \mu^{1-\gamma_0} f(\alpha_0-\gamma_0+1,\beta_0-\gamma_0+1,2-\gamma_0;\mu);
}
\right.
$$
\vskip 0.2 cm 
$$
\left\{
\matrix{
\xi_1^{(1)}=F(\alpha_0,\beta_0,\alpha_0+\beta_0-\gamma_0+1;1-\mu),
\cr
\xi_2^{(1)}= g(\alpha_0,\beta_0,\alpha_0+\beta_0-\gamma_0+1;1-\mu);
}
\right.
$$
\vskip 0.2 cm 
$$
\left\{
\matrix{
\xi_1^{(\infty)}=\mu^{-\beta_0}F(\beta_0,\beta_0-\gamma_0+1,\beta_0-\alpha_0+1;\mu^{-1}),
\cr
\xi_2^{(\infty)}=\mu^{-\beta_0} g_1(\beta_0-\gamma_0+1,\beta_0,\beta_0-\alpha_0+1;\mu^{-1});
}
\right.
$$
\vskip 0.2 cm 
 From Norlund, formulae 5.(1), 5.(7) we get:
$$
\Psi_0^{(0)}= \Psi_0^{(1)} C_{01}^{(*)},~~~|\arg(1-\lambda)|<\pi,~~~~~C_{01}^{(*)}=\pmatrix{ 
0 & {p\Gamma(p)^2\over \Gamma(2p)}
\cr
-{2\Gamma(2p)\over \Gamma(p)^2} & 0 
}.
$$
From Norlund, formulae 12.(1), 12.(6) we get:
$$
\Psi_0^{(0)}=\Psi_0^{(\infty)}C_{0\infty}^{(*)},~~~|\arg(-\mu)|<\pi,
~~~~~
C_{0\infty}^{(*)}=(-1)^{p+1}\pmatrix{0 & {p^2\Gamma(p)^2\over \Gamma(2p)}
\cr
{2p~\Gamma(2p)\over \Gamma(p)^2} & 0}
,
$$
where $-\mu=e^{-i\pi \eta}\mu$, $\eta=\pm 1$.

\vskip 0.2 cm 
We compute the behavior of $\varphi_i^{(\infty)}$, $\xi_i^{(\infty)}$  
($i=1,2$)
for $\mu\to \infty$. In the computation, $\ln(-1/\mu)$ appears in $g_1$. We write $-1/\mu= e^{i\pi\eta}/\mu$, $\arg \mu=\eta \pi$ when $-\infty<\mu<0$. 
The final result (after expanding in series):
$$
\Psi_0^{(\infty)}(\mu)=
\left[I+O\left({1\over \mu}\right)\right]
\pmatrix{1 & \ln \mu 
\cr
0 & 1}~E, ~~~\mu\to \infty,~~~~~
E=
 \pmatrix{p^{-2} & Q_>~p^{-2} 
\cr
0 & -p^{-2}
},
$$
$$
Q_>=
\psi_E(p)+\psi_E(p+1)+2\gamma_E +i\pi \eta-{p+r\over p^2}
 .
$$
Namely, 
$$
\Psi_0^{(\infty)}=\Psi_0 E,
$$
where $\Psi_0$ is the matrix for the matching 
$\Psi_{OUT}
\leftrightarrow
\Psi_{IN}$. 
Expanding $\varphi_i^{(0)}$, $\xi_i^{(0)}$ for $\mu \to 0$ we get:
$$
\Psi_0^{(0)}= \pmatrix{{r+2p\over p^2} & {r\over p^2}
\cr
1 & 1 }\bigl[I+O(\mu)\bigr]\mu^{p\sigma_3},~~~\mu\to 0.
$$
Expanding $\varphi_i^{(1)}$, $\xi_i^{(1)}$ for $\mu \to 1$ we get:
$$
\Psi_0^{(1)}= \pmatrix{
{p+r\over p^2} 
&
{p+r\over p^2} (\psi_E(p)+\psi_E(p+1)+2\gamma_E)-{1\over p^2}
\cr
1
&
\psi_E(p)+\psi_E(p+1)+2\gamma_E
 }\bigl[I+O(1-\mu)\bigr]\pmatrix{1 & \ln(1-\mu) 
\cr
0 & 1},~~~\mu\to 1.
$$

\vskip 0.3 cm 
\noindent 
* {\it CASE $p<0$.} We choose: 
$$
\left\{
\matrix{
\xi_1^{(0)}=f(\alpha_0,\beta_0,\gamma_0;\mu),
\cr
\xi_2^{(0)}= \mu^{1-\gamma_0} F(\alpha_0-\gamma_0+1,\beta_0-\gamma_0+1,2-\gamma_0;\mu);
}
\right.
$$
\vskip 0.2 cm 
$$
\left\{
\matrix{
\xi_1^{(1)}=F(\alpha_0,\beta_0,\alpha_0+\beta_0-\gamma_0+1;1-\mu),
\cr
\xi_2^{(1)}= g_0(\alpha_0,\beta_0,\alpha_0+\beta_0-\gamma_0+1;1-\mu);
}
\right.
$$
\vskip 0.2 cm 
$$
\left\{
\matrix{
\xi_1^{(\infty)}=\mu^{-\beta_0}F(\beta_0,\beta_0-\gamma_0+1,\beta_0-\alpha_0+1;\mu^{-1}),
\cr
\xi_2^{(\infty)}=\mu^{-\beta_0} g_1(\beta_0,\beta_0-\gamma_0+1,\beta_0-\alpha_0+1;\mu^{-1});
}
\right.
$$

From Norlund, formulae 8.(6), 8.(11) we compute:
$$
\Psi_0^{(0)}= \Psi_0^{(1)} C_{01}^{(*)},~~~|\arg(1-\lambda)|<\pi,
~~~~~
C_{01}^{(*)}=\pmatrix{ -{p\Gamma(-p)^2\over \Gamma(-2p)} & 0 
\cr
0 & -{2\Gamma(-2p)\over \Gamma(-p^2)}
}.
$$
From Norlund, formulae 13.(1), 13.(6) we compute:
$$
\Psi_0^{(0)}=\Psi_0^{(\infty)}C_{0\infty}^{(*)},~~~|\arg(-\mu)|<\pi,
~~~~~
C_{0\infty}^{(*)}=(1)^{p+1}\pmatrix{ {p^2\Gamma(-p)^2\over \Gamma(-2p)} & 0
\cr
0 & -{2p~\Gamma(-2p)\over \Gamma(-p)^2}
}.
$$

\vskip 0.2 cm 
We compute the behavior of $\varphi_i^{(\infty)}$, $\xi_i^{(\infty)}$  
($i=1,2$)
for $\mu\to \infty$. In the computation, $\ln(-1/\mu)$ appears in $g_1$. We write $-1/\mu= e^{i\pi\eta}/\mu$, $\arg \mu=\eta \pi$ when $-\infty<\mu<0$. 
The final result (expanding in series):
$$
\Psi_0^{(\infty)}(\mu)=
\left[I+O\left({1\over \mu}\right)\right]
\pmatrix{1 & \ln \mu 
\cr
0 & 1}~E, ~~~\mu\to \infty,~~~~~
E=
 \pmatrix{p^{-2} & Q_<~p^{-2} 
\cr
0 & -p^{-2}
},
$$
$$
Q_<=\psi_E(-p)+\psi_E(-p+1)+2\gamma_E +i\pi \eta-{p+r\over p^2}
.
$$
Namely, 
$$
\Psi_0^{(\infty)}=\Psi_0 E,
$$
where $\Psi_0$ is the matrix for the matching 
$\Psi_{OUT}
\leftrightarrow
\Psi_{IN}$. 
Expanding $\varphi_i^{(0)}$, $\xi_i^{(0)}$ for $\mu \to 0$ we get:
$$
\Psi_0^{(0)}= \pmatrix{{r+2p\over p^2} & {r\over p^2}
\cr
1 & 1 }\bigl[I+O(\mu)\bigr]\mu^{p\sigma_3},~~~\mu\to 0.
$$
Expanding $\varphi_i^{(1)}$, $\xi_i^{(1)}$ for $\mu \to 1$ we get:
$$
\Psi_0^{(1)}= \pmatrix{
{p+r\over p^2} 
&
{p+r\over p^2} (\psi_E(-p)+\psi_E(1-p)+2\gamma_E)-{1\over p^2}
\cr
1
&
\psi_E(-p)+\psi_E(1-p)+2\gamma_E
 }\bigl[I+O(1-\mu)\bigr]\pmatrix{1 & \ln(1-\mu) 
\cr
0 & 1},~~~\mu\to 1.
$$

\vskip 0.3 cm 
\noindent
{\bf *}   
Both for $p>0$ and $p<0$ we have: 
$$
\left.
\matrix{
\Psi_{IN}= K_0(x) \Psi_0 &=& K_0(x) \Psi_0^{(\infty)} E^{-1},~~~~~~~~~~~~
\cr
\cr
&=& K_0(x) \Psi_0^{(0)} {C_{0\infty}^{(*)}}^{-1}E^{-1},~~~~~
\cr
\cr
&=& K_0(x) \Psi_0^{(1)}C_{01}^{(*)}  {C_{0\infty}^{(*)}}^{-1}E^{-1};
}
\right. 
$$
together with $\Psi_{IN}^{Match}= \Psi_{IN} C_{OUT}$. 
We conclude that the monodromy of (\ref{SYSTEM}) is: 
$$
M_0\equiv M_0^{IN}=I,~~~~~
M_x\equiv M_1^{IN}= C_{OUT}^{-1}~\bigl[EC_{0\infty}^{(*)}{C_{01}^{(*)}}^{-1}~
\pmatrix{ 1 & 2\pi i 
\cr
0 & 1}       
~C_{01}^{(*)} {C_{0\infty}^{(*)}}^{-1}E^{-1}\bigr]~C_{OUT}.
$$
The connection matrices $E$, $C_{0\infty}^{(*)}$, $C_{01}^{(*)}$ have different form for $p>0$ and for $p<0$. 
We also have two choices 
for $C_{OUT}$, depending on $\epsilon=\pm 1$ in $C_{0\infty}$.  These have 
been called $C_{OUT}^{+}$ and $C_{OUT}^{-}$ in the comments just after 
Proposition \ref{monodromiaCHAZYmy}.  Multiplying by $C_{OUT}$ and 
$C_{OUT}^{-1}$ to the left and right respectively we get three generators for 
the monodromy group: $$
M_0=I,~~~~~M_1= BC_{01}^{-1}~\pmatrix{ 1 & 2\pi i 
\cr
0 & 1}~ C_{01} B^{-1},~~~~~
M_x=EC_{0\infty}^{(*)}{C_{01}^{(*)}}^{-1}~
\pmatrix{ 1 & 2\pi i 
\cr
0 & 1}
~C_{01}^{(*)} {C_{0\infty}^{(*)}}^{-1}E^{-1}.
$$
With this choice,  we obtain the matrices of the Proposition \ref{monodromiaCHAZYmy}. 
We observe that
$$
C_{OUT}^{-1}M_1M_xM_0C_{OUT}=\pmatrix{-1 & 2\pi i \cr 0 & -1},~~~~~\epsilon=1;
$$ 
$$
C_{OUT}^{-1}M_xM_1M_0C_{OUT}=\pmatrix{-1 & 2\pi i \cr 0 & -1},~~~~~\epsilon=-1.
$$
$$
\hbox{tr}(M_0M_x)=\hbox{tr}(M_0M_1)=2,~~~~~\hbox{tr}(M_1M_x)=-2.
$$
\qed


\section{Logarithmic Behaviors at $x=1$ and $x=\infty$ -- Symmetries and their Action on the Monodromy Data -- Connection Problem}
\label{monodinsim}

In this section we compute the logarithmic asymptotic behaviors at $x=1,\infty$. This is easily done by applying the action of some Backlund transformations  of (PVI) on 
(\ref{intrlog1}) and (\ref{intrlog2}). 
They act as birational transformations on $y(x)$ and $x$, and as permutations on the $\theta_\nu$`s, $\nu=0,x,1,\infty$.  
  In order to know the monodromy data which are associated to the solutions of (PVI) obtained from  (\ref{intrlog1}) and (\ref{intrlog2}) by the Backlund transformations, we also compute their action on the monodromy data.   

\vskip 0.2 cm 
 The birational transformations are described in \cite{Okamoto}; some of them  
form  a representation of the permutation group and are generated by: 
$$
\sigma^1:~~~\theta_1^{\prime}=\theta_0,~~\theta_0^{\prime}=\theta_1;~~~~~~~~
\theta_x^{\prime}=\theta_x,~~\theta_\infty^{\prime}=\theta_\infty;~~~~~~~~~~y^{\prime}(x^\prime)=1-y(x),~~~x=1-x^\prime. 
$$
$$
\sigma^2:~~~\theta_0^{\prime}=\theta_\infty-1,~~\theta_\infty^{\prime}=\theta_0+1;~~~~~~~~
\theta_1^{\prime}=\theta_1,~~\theta_x^{\prime}=\theta_x;~~~~~~~~~~
y^{\prime}(x^\prime)={1\over y(x)},~~~x={1\over x^{\prime}}.
$$
$$
\sigma^3:~~~\theta_x^\prime=\theta_1,~~\theta_1^\prime=\theta_x;~~~~~~~
\theta_0^\prime=\theta_0,~~\theta_\infty^\prime=\theta_\infty;~~~~~~~~~~
y^\prime(x^\prime)={1\over x}y(x),~~~x={1\over x^\prime}.
$$
It is convenient to consider also:
\be
\label{sym1}
\theta_0^{\prime}=\theta_x,~~\theta_x^{\prime}=\theta_0;~~~~~~~~
\theta_1^{\prime}=\theta_1,~~\theta_\infty^{\prime}=\theta_\infty;~~~~~~~~~~y^{\prime}(x^\prime)= {x-y(x)\over x-1},~~~x={x^{\prime}\over x^{\prime}-1};
\ee
\be
\label{sym2}
\theta_0^\prime=\theta_\infty-1,~~\theta_x^\prime=\theta_1,~~\theta_1^\prime=\theta_x,~~\theta_\infty^\prime=\theta_0+1;~~~~~~~y^\prime(x^\prime)={x\over y(x)},~~~x=x^\prime.
\ee
\be
\label{sym3}
\theta_x^\prime=\theta_1,~~\theta_1^\prime=\theta_\infty-1,~~
\theta_\infty^\prime=\theta_x+1;~~~~~~~\theta_0^\prime=\theta_0;~~~~~y^\prime(x^\prime)={y(x)\over y(x)-x},~~~x={x^\prime-1\over x^\prime}.
\ee
The transformantion (\ref{sym1}) is the composition $\sigma^1\cdot  \sigma^3 
\cdot \sigma^1$. (\ref{sym2}) is $\sigma^2\cdot\sigma^3$. 
 (\ref{sym3}) is the composition of $\sigma^2$, 
(\ref{sym1}), (\ref{sym2}). For brevity, we will call the Backlund transformations with the name ``symmetries''. 
%


\subsection{Action on the Transcendent. Formulae 
(\ref{ass1}), (\ref{ass2}), (\ref{Ass1}), (\ref{Ass2}) }
\label{monodinsimy}

The symmetry $\sigma^3$, acting on  the transcendent (\ref{intrlog1}),  gives the behavior: 
$$
y^\prime(x^\prime)\sim {{\theta^\prime_0}^2\over {\theta^\prime_0}^2-{\theta^\prime_1}^2}+
{{\theta^\prime_1}^2-{\theta^\prime_0}^2\over 2} 
\left[
\ln {1\over x^\prime} + {4 r + 2\theta^\prime_0\over {\theta^\prime_0}^2-{\theta^\prime_1}^2}
\right]^2,~~~x^\prime\to\infty;
$$
We prove below that $\sigma^3$  maps tr$(M_0M_x)$ to tr$(M^\prime_0M^\prime_1)$, where 
$M_\nu^\prime$, $\nu=0,x,1,\infty$ are th emonodromy matrices for the system (\ref{SYSTEM}) associated to $y^\prime(x^\prime)$, with respect to the same basis of loops (see below).  Therefore $\hbox{\rm tr}(M^\prime_0M^\prime_1)=2$.
\vskip 0.2 cm 

The symmetry $\sigma^1$, acting on  the transcendent (\ref{intrlog1}), gives the behavior:  
$$
y^\prime(x^\prime)\sim 1 -(1-x^\prime) \left\{{{\theta^\prime_1}^2\over {\theta^\prime_1}^2-{\theta^\prime_x}^2} +
{{\theta^\prime_x}^2-{\theta^\prime_1}^2\over 4}
\left[
\ln (1-x^\prime) + {4 r + 2\theta^\prime_1 \over {\theta^\prime_1}^2-{\theta^\prime_x}^2}
\right]^2
\right\},~~~x^\prime\to 1.
$$
As it is proved below, $\sigma^1$ maps tr$(M_0M_x)$ to tr$(M^\prime_1M^\prime_x)$ and thus $\hbox{\rm tr}(M^\prime_1M^\prime_x)=2$.

\vskip 0.2 cm 
The action of (\ref{sym2}) gives the behavior: 
$$
y^\prime(x^\prime)\sim{1\over {{\theta^\prime_1}^2-(\theta^\prime_\infty-1)^2\over 4} 
\left[
\ln x^\prime + {4r+2\theta^\prime_\infty-2\over (\theta^\prime_\infty-1)^2-{\theta^\prime_1}^2}
\right]^2+{(\theta^\prime_\infty-1)^2\over (\theta^\prime_\infty-1)^2-{\theta^\prime_1}^2}},~~~~~x^\prime\to 0,
$$
Namely, 
$$
y^\prime(x^\prime)={4\over [{\theta_1^\prime}^2-(\theta^\prime_\infty-1)^2]\ln^2 x^\prime } 
\left[
1+{8r+4\theta^\prime_\infty-4\over {\theta^\prime_1}^2-(\theta^\prime_\infty-1)^2}{1\over \ln x^\prime} 
+
O\left({1\over \ln^2x^\prime}\right)
\right],~~~~~x^\prime\to 0.
$$
\vskip 0.2 cm 
The symmetry (\ref{sym3}) gives:
$$y^\prime(x^\prime)\sim 1+{1\over {{\theta_1^\prime}^2-{\theta_0^\prime}^2\over 4}\left[
\ln(x^\prime-1)+{4r+2\theta^\prime_0\over {\theta_0^\prime}^2-{\theta_1^\prime}^2}
\right]^2 +{{\theta_0^\prime}^2\over {\theta_0^\prime}^2-{\theta_1^\prime}^2}
},~~~~~x^\prime\to 1.
$$
Namely:
$$
y^\prime(x^\prime)= 1+{4\over ({\theta_1^\prime}^2-{\theta_0^\prime}^2)\ln^2(x^\prime-1)}
\left[
1-{8r+4\theta^\prime_0\over {\theta_0^\prime}^2-{\theta_1^\prime}^2}{1\over \ln(x^\prime-1)}+O\left({1\over \ln^2(x^\prime-1)}\right)
\right],~~~~~x^\prime\to 1
$$

\vskip 0.2 cm
The symmetry $\sigma^2$ yields:
$$
y^\prime(x^\prime)\sim{x^\prime\over 
{{\theta_x^\prime}^2-(\theta^\prime_\infty-1)^2\over 4}\left[
\ln {1\over x^\prime} +
{4r+2\theta^\prime_\infty-2\over (\theta^\prime_\infty-1)^2-{\theta^\prime_x}^2
}\right]^2+{(\theta^\prime_\infty-1)^2\over (\theta^\prime_\infty-1)^2
-{\theta_x^\prime}^2}
},~~~~~x^\prime\to\infty.
$$
Namely,
$$
y(x)={4~x^\prime\over [(\theta^\prime_\infty-1)^2-{\theta^\prime_x}^2]\ln^2x^\prime}
\left[
1-{8r+4(\theta^\prime_\infty-1)\over {\theta^\prime_x}^2-(\theta^\prime_\infty-1)^2}{1\over \ln x^\prime}+
O\left(1\over \ln^2 x^\prime
\right)
\right],~~~~~x^\prime\to\infty.
$$

\vskip 0.2 cm
 We study the action of the symmetries on (\ref{intrlog2}). If we apply $\sigma^1$ we find:
$$
y^\prime(x^\prime)\sim 1-(1-x^\prime)
\bigl(
r\pm \theta_1^\prime\ln(1-x^\prime)
\bigr),~~~~~x^\prime \to 1,~~~~~\theta_1^\prime=\pm \theta_x^\prime.
$$
The action of $\sigma^3$ gives:
$$
y^\prime(x^\prime)\sim r\pm\theta_0^\prime \ln x^\prime,
~~~~~x^\prime\to\infty,~~~~~
\theta_0^\prime=\pm \theta_1^\prime.
$$
The action of (\ref{sym2}) gives:
$$
y^\prime(x^\prime)\sim{1\over r\pm(\theta_\infty^\prime-1)\ln x^\prime},~~~~~
x\to 0,~~~~~\theta_\infty^\prime-1=\pm \theta_1^\prime.
$$
The action of (\ref{sym3}) gives:
$$
y^\prime(x^\prime)\sim 1+{1\over r\pm\theta_0^\prime\ln\left({x^\prime-1\over x^\prime}\right)},~~~~~x^\prime\to 1,~~~~~\theta_\infty^\prime-1=\pm\theta_0^\prime.
$$
Namely:
$$
y^\prime(x^\prime)= 1\pm{1\over \theta_0^\prime\ln(x^\prime-1)}
\left[
1\mp {r\over \theta_0^\prime\ln(x^\prime-1)}+O\left({1\over \ln^2(x^\prime-1)}\right)
\right],~~~~~x^\prime\to 1,~~~~~\theta_\infty^\prime-1=\pm\theta_0^\prime.
$$
The action of $\sigma^2$ gives:
$$
y^\prime(x^\prime)\sim {x^\prime\over r\pm (\theta_\infty-1)\ln x^\prime},~~~~~x^\prime\to \infty,~~~~~\theta_\infty^\prime-1=\pm\theta_x^\prime.
$$
Namely:
$$
y^\prime(x^\prime)
=
\pm{x^\prime\over (\theta_\infty^\prime-1)\ln x^\prime}\left[
1\mp{r\over (\theta_\infty^\prime-1)\ln x^\prime}+O\left(
{1\over \ln^2 x^\prime}
\right)
\right],~~~~~\theta_\infty^\prime-1=\pm\theta_x^\prime.
$$

When we drop the index $~\prime~$ from the  above formulae, we get  the asymptotic behaviors (\ref{ass1}), (\ref{ass2}), (\ref{Ass1}), (\ref{Ass2}).

\subsection{Action of (\ref{sym2}) on the Monodromy Data}
\label{cioco}

It is proved in \cite{DM1} that the action on the monodromy data is: 
$$
(\hbox{tr}(M_0M_x),\hbox{tr}(M_0M_1),\hbox{tr}(M_xM_1))\mapsto (-\hbox{tr}(M_0M_x),-\hbox{tr}(M_0M_1),\hbox{tr}(M_xM_1))
.$$
This proves that the behaviors ${1\over \ln^2 x}$ and ${1\over \ln
  x}$, obtained via  (\ref{sym2}) from the behavior computed with the matching procedure
when tr$(M_0M_x)=2$, are associated to tr$(M_0M_x)=-2$.


\subsection{Action of $\sigma^1$ and $\sigma^3$ on the Monodromy Data}
\label{monodinsimM}

To compute the action of the symmetries on the monodromy of system (\ref{SYSTEM}), it is important that we choose the same base of loops in the $\lambda$-plane that we used to 
parameterize a transcendent in terms of the monodromy data. 
Therefore, we consider an ordered base of loops in  the ``$\lambda$-plane'' 
${\bf C}\backslash\{0,x,1\}$ as we did in Sub-Section \ref{MonodromyPasqua}, figure \ref{figure1}. 

 Consider the system associated to $y(x)$: 
\be
\label{SYSTEMbis}
{d\Psi\over d\lambda}=\left[ {A_0\over \lambda}+{A_x \over \lambda-x}+{A_1
\over
\lambda-1}\right]\Psi,
\ee
The monodromy matrices of a fundamental solution $\Psi(\lambda)$  w.r.t. the chosen base of loops are denoted  $M_0$, $M_x$, $M_1$. The loop at infinity will be $\gamma_\infty=\gamma_0\gamma_x\gamma_1$, so $M_\infty=M_1M_xM_0$. 
   We 
need to construct the system associated to $y^\prime(x^\prime)$: 
\be
\label{SYSTEMprime}
{d\Psi^\prime\over d\lambda^\prime}=\left[ {A^\prime_0\over \lambda^\prime}+{A^\prime_{x^\prime} \over \lambda^\prime-x^\prime}+{A^\prime_1
\over
\lambda^\prime-1}\right]\Psi^\prime,
\ee
We will determine the relation between (\ref{SYSTEMbis}) and 
(\ref{SYSTEMprime}), between a fundamental solutions $\Psi(\lambda)$ and a 
fundamental solution 
$\Psi^\prime(\lambda^\prime)$ and between  their respective monodromy matrices 
$M_0$, $M_x$, $M_1$ and $M^\prime_0$, $M^\prime_{x^\prime}$, $M_1^\prime$.  The monodromy $M^\prime_0$, $M^\prime_{x^\prime}$, $M_1^\prime$ are understood to be referred to the order $1,2,3 = 0,x^\prime,1$. In order to do this, we will construct 
$A^\prime_j(x^\prime,y^\prime(x^\prime),dy^\prime/dx^\prime)$, $j=0,x^\prime,1$ and we will see how they are  related to the matrices $A_j(x,y(x),dy/dx)$. 

The explicit formulas to write $A_j(x,y(x),dy/dx)$ can be found at page 443-445 of \cite{JMU}: 
$$
(A_0)_{12}=-k {y\over x},~~~ (A_1)_{12}=k {y-1\over x-1},~~~(A_x)_{12}=-k {y-x\over x(x-1)};
$$
$$
{d\over dx} \ln k= (\theta_\infty-1){y-x\over x(x-1)}~~~\Longrightarrow~~~k(x)=k_0 \exp\left\{(\theta_\infty-1)\int^x {y(s)-s\over s(s-1)}ds \right\},~~~k_0\in{\bf C}.
$$
 \vskip 0.2 cm
$$
(A_i)_{11}=z_i+{\theta_i\over 2},~~~i=0,x,1.
$$
$$
z_0={y\over x \theta_\infty} \Bigl\{
y(y-1)(y-x) \tilde{z}^2 +\bigl[\theta_1(y-x)+x\theta_x(y-1)-2\kappa_2(y-1)(y-x) \bigr]\tilde{z}
+
\kappa_2^2(y-x-1)-\kappa_2(\theta_1+x\theta_x)
\Bigr\},
$$
$$
z_1=-{y-1\over (x-1)\theta_\infty}\Bigl\{
y(y-1)(y-x)\tilde{z}^2 +\bigl[
(\theta_1+\theta_\infty)(y-x)+x\theta_x(y-1)-2\kappa_2(y-1)(y-x)
\bigr]\tilde{z} +\kappa_2^2 (y-x)+\Bigr.
$$
$$-
\Bigr. \kappa_2(\theta_1+x\theta_x)-\kappa_2(\kappa_2+\theta_\infty)
\Bigr\},
$$
$$
z_x={y-x\over x(x-1)\theta_\infty}\Bigl\{
y(y-1)(y-x)\tilde{z}^2+\bigl[
\theta_1(y-x)+x(\theta_x+\theta_\infty)(y-1)-2\kappa_2(y-1)(y-x)
\bigr]\tilde{z} +
\Bigr.
$$
$$
+\Bigl. 
\kappa_2^2(y-1)-\kappa_2(\theta_1+x\theta_x)-x\kappa_2(\kappa_2+\theta_\infty)
\Bigr\},
$$
\vskip 0.2 cm 
$$
\kappa_2= -\left\{{\theta_0\over 2} +{\theta_x\over 2} +{\theta_1\over 2} +{\theta_\infty\over 2} \right\},~~~
\tilde{z}= {1\over 2} {x(x-1)\over y(y-1)(y-x)}{dy\over dx}-{1\over 2}\left\{
{1\over y-x} +{\theta_0\over y} +{\theta_x\over y-x}+{\theta_1\over y-1}
\right\},
$$
\vskip 0.2 cm
$$
(A_0)_{21}= {z_0 x\over ky} (z_0+\theta_0),~~~
(A_1)_{21}=-{(x-1)z_1\over k(y-1)}(z_1+\theta_1),~~~
(A_x)_{21}=
{x(x-1)z_x\over k(y-x)}(z_x+\theta_x).
$$
We also recall that $(A_0)_{12}/\lambda+(A_x)_{12}/(\lambda-x)+(A_1)_{12}/(\lambda-1)=
{k(\lambda-y)\over \lambda(\lambda-1)(\lambda-x)}$.

\vskip 0.3 cm
\noindent
{\large \bf Symmetry $\sigma_3$:} 
We compute the matrices $A_i^\prime$, $i=0,x^\prime,1$,  through the above formulas. 
By direct computation we find:
$$
\tilde{z}^\prime = x \tilde{z},~~~z_0^\prime=z_0,~~~z_1^\prime=z_x,~~~z_x^\prime=z_1.
$$ 
Therefore we find:
$$
 A_0^\prime=K^{-1}A_0K,~~~A_1^\prime=K^{-1}A_x K,~~~A_{x\prime}^\prime=K^{-1}A_1K;~~~~~~~
K:= \pmatrix{{k\over x k^\prime} & 0 \cr 0 & 1}.
$$
We also note that $d(\ln k^\prime)/dx^\prime= d(\ln k)/dx+(\theta_\infty-1)/x$, thus: $k^\prime=k x^{\theta_\infty-1}$. Anyway, the specific form of $k/k^\prime$ is not important here. What is important is that the matrix $K$ is {\it diagonal}. Then we can write  
$$
{d\Psi^\prime\over d\lambda^\prime}=\left[ {A^\prime_0\over \lambda^\prime}+{A^\prime_{x^\prime} \over \lambda^\prime-x^\prime}+{A^\prime_1
\over
\lambda^\prime-1}\right]\Psi^\prime
= K^{-1}\left[ {A_0\over \lambda^\prime}+{A_1 \over \lambda^\prime-x^\prime}+{A_x
\over
\lambda^\prime-1}\right]K\Psi^\prime,
$$
With the change of variables:
$$
\lambda^\prime = {\lambda \over x}, ~~~x^\prime={1\over x},
$$
we get:
$$
{d\Psi^\prime\over d\lambda}= K^{-1}\left[ {A_0\over \lambda}+{A_1 \over \lambda-1}+{A_x
\over
\lambda-x}\right]K\Psi^\prime.
$$
With the gauge:
$$
 \Psi=K\Psi^\prime,
$$
We finally get (\ref{SYSTEMbis}): 
\be
\label{sysmod}
{d\Psi\over d\lambda}= \left[ {A_0\over \lambda}+{A_1 \over \lambda-1}+{A_x
\over
\lambda-x}\right]\Psi.
\ee
It is important to note that the gauge is {\it diagonal}, a fact that ensures 
that, for the gauge-transformed system, the solution $\lambda$ of the equation obtained by setting the  matrix element $(1,2)$ equal to zero  defines the same $y(x)$.   
 We conclude that the systems (\ref{SYSTEMbis}) and (\ref{SYSTEMprime}) are related by a {\it diagonal} gauge transformation and the exchange of the point 
$x$ and $1$. In other words, we can take as (\ref{SYSTEMprime}) the 
system:  
\be
\label{sysmodprime}
{d\Psi\over d\lambda^\prime}= \left[ {A_0\over \lambda^\prime}+{A_1 \over \lambda^\prime-x^\prime}+{A_x
\over
\lambda^\prime-1}\right]\Psi,
\ee
where $\Psi(\lambda)$ is also a fundamental matrix solution of (\ref{SYSTEMbis}). The equation defining $y^\prime(x^\prime)$ is: 
$$
 \left[ {A_0\over \lambda^\prime}+{A_1 \over \lambda^\prime-x^\prime}+{A_x
\over
\lambda^\prime-1}\right]_{1,2}=0~~~\Longrightarrow~~~\lambda^\prime=y^\prime(x^\prime),
$$ 
while:  
$$
 \left[ {A_0\over \lambda}+{A_1 \over \lambda-1}+{A_x
\over
\lambda-x}\right]_{1,2}=0~~~\Longrightarrow~~~\lambda=y(x),
$$ 
Therefore, (\ref{SYSTEMprime}) can be obtained from  (\ref{SYSTEMbis}) simply by a change of variables $\lambda^\prime=\lambda/x$, $x=1/x^\prime$. The result is that the   points $\lambda=x,1$ are exchanged to $\lambda^\prime=1,x^\prime$.

\vskip 0.2 cm
We compute the monodromy of (\ref{sysmodprime}) in terms of the monodromy of (\ref{sysmod}). For the latter, we have fixed in the beginning of the section a ordered base of loops $\gamma_0$, $\gamma_x$, $\gamma_1$. But for (\ref{sysmodprime}), the  points $1,x^\prime$ are exchanged. The  loops $\tilde{\gamma}_0$,  $\tilde{\gamma}_1$  $\tilde{\gamma}_{x^\prime}$ of figure \ref{figure2}  
\begin{figure}
\epsfxsize=12cm
\centerline{\epsffile{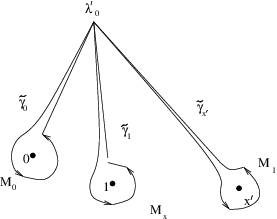}}
\caption{}
\label{figure2}
\end{figure}
correspond to the order 1,2,3. 
 Their monodromy matrices are:  
$$
M_{\tilde{\gamma}_0}=M_0,~~~M_{\tilde{\gamma}_1}=M_x,~~~M_{\tilde{\gamma}_{x^\prime}}=M_1.
$$
We need a new basis of loops such that the order $1,2,3$ be $0,x^\prime,1$. Let us  denote these loops $\gamma_0^\prime,\gamma_{x^\prime}^\prime,\gamma_1^\prime$ of  figure \ref{figure4}.
\begin{figure}
\epsfxsize=12cm
\centerline{\epsffile{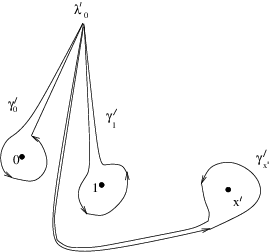}}
\caption{}
\label{figure4}
\end{figure}
For the basis in  figure \ref{figure4} we easily see that:  
$$
\gamma_0^\prime=\tilde{\gamma}_0,
~~~~~\gamma_{x^\prime}^\prime=\tilde{\gamma}_1~\tilde{\gamma}_x\tilde{\gamma}_1^{-1},
~~~~~\gamma_1^\prime=\tilde{\gamma}_1.
$$

Let $M_0^\prime, M_{x^\prime}^\prime, M_1^{\prime}$  be the monodromy matrices for the orderded loops 
$\gamma_0^\prime, \gamma_{x^\prime}^\prime, \gamma_1^{\prime}$.
 Therefore we have: 
$$
M_0^\prime=M_{\tilde{\gamma}_0}= M_0,
$$
$$
M_{x^\prime}^\prime=M_{\tilde{\gamma}_1}^{-1}M_{\tilde{\gamma}_{x^\prime}}M_{\tilde{\gamma}_1}\equiv M_x^{-1}M_1M_x,
$$
$$
M_1^\prime=M_{\tilde{\gamma}_1}\equiv M_x. 
$$
 From the above results we compute the traces:
$$\left.\matrix{
\hbox{\rm tr}(M_0^{\prime}M_{x^\prime}^\prime)=&-~\hbox{\rm tr}(M_0M_1)&
-~\hbox{\rm tr}(M_0M_x)\hbox{\rm tr}(M_1M_x)&
+~4\bigl(
\cos(\pi \theta_\infty)\cos(\pi\theta_x)+\cos(\pi\theta_0)\cos(\pi \theta_1)
\bigr),
\cr
\cr
\hbox{\rm tr}(M_0^{\prime}M_1^{\prime})=&
\hbox{\rm tr}(M_0M_x),
\cr
\cr
\hbox{\rm tr}(M_1^{\prime}M_{x^\prime}^\prime)=& \hbox{\rm tr}(M_1M_x).
}
\right.
$$
The above follow from the identity: 
$$ 
 \hbox{tr}(AB)= \hbox{tr}(A) \hbox{tr}(B) - \hbox{tr} (AB^{-1}), 
~~~~A,B \hbox{ $2\times 2$ matrices }, ~~~ \det(B)=1
$$ 
and from:
$$
\hbox{tr}(M_1M_xM_0)= e^{ i \pi \theta_{\infty} }+ 
 e^{ -i \pi \theta_{\infty} },
~~~~\hbox{tr}(M_i) = e^{ i \pi \theta_i }+ 
 e^{ -i \pi \theta_i },~~i=0,x,1
$$

\vskip 0.3 cm 
\noindent
{\large \bf Symmetry $\sigma_1$:} We repeat the  computation  $A_0^\prime$, $A_{x^\prime}^\prime,A_1^\prime$  as above. As a result we find that 
the system (\ref{SYSTEMprime}) is -- up to diagonal conjugation: 
\be
{d\Psi\over d\lambda^\prime}= \left[ {A_0\over \lambda^\prime-1}+{A_1 \over \lambda^\prime}+{A_x
\over
\lambda^\prime-x}\right]\Psi,~~~\lambda^\prime=1-\lambda,~~~x^\prime=1-x,
\label{sysmodprime1}
\ee
where $\Psi(\lambda)$ is also a fundamental matrix of (\ref{SYSTEMbis}). 
In other words, (\ref{sysmodprime1}) can be  obtained from (\ref{SYSTEMbis}) by the change of variables $\lambda^\prime=\lambda-1$, $x=1-x^\prime$. 
The relation between the two systems is simply that the points $\lambda=0,1$ are exchanged to $\lambda^\prime= 1,0$.  
The base $\gamma_0,\gamma_x\gamma_1$ becomes the basis  $\tilde{\gamma}_1,\tilde{\gamma}_{x^\prime},\tilde{\gamma}_0$, in figure \ref{figure5}. 
\begin{figure}
\epsfxsize=12cm
\centerline{\epsffile{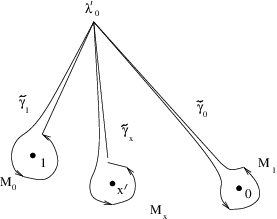}}
\caption{}
\label{figure5}
\end{figure}
The monodromy matrices are: 
$$
M_{\tilde{\gamma}_1}=M_0,~~~M_{\tilde{\gamma}_{x^\prime}}=M_x,~~~M_{\tilde{\gamma}_0}=M_1.
$$
We introduce the ordered basis  $\gamma_0^\prime,\gamma_{x^\prime}^\prime,\gamma_1^\prime$ of figure \ref{figure6}  
\begin{figure}
\epsfxsize=12cm
\centerline{\epsffile{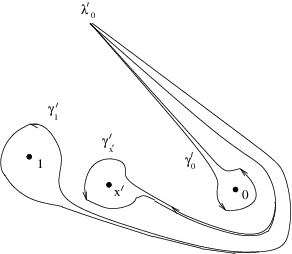}}
\caption{}
\label{figure6}
\end{figure}
and we easily compute:
$$
\gamma_0^\prime=\tilde{\gamma}_0,~~~\gamma_{x^\prime}^\prime=\tilde{\gamma}_0^{-1}
\tilde{\gamma}_{x^\prime}\tilde{\gamma}_0,~~~\gamma_1^\prime=\tilde{\gamma}_0^{-1}
\tilde{\gamma}_{x^\prime}^{-1}\tilde{\gamma}_1\tilde{\gamma}_{x^\prime}\tilde{\gamma}_0.
$$
Therefore:
$$
M_0^\prime=M_1,~~~M_{x^\prime}^\prime= M_1 M_x M_1^{-1},~~~
M_1^\prime=M_1M_xM_0M_x^{-1}M_1^{-1};
$$ 
and:
$$
\left. \matrix{
\hbox{\rm tr}(M_0^\prime M_{x^\prime}^\prime)=&\hbox{\rm tr}(M_1M_x),
\cr
\cr
\hbox{\rm tr}(M_0^\prime M_1^\prime)=&
-~
\hbox{\rm tr}(M_0M_1)&-
\hbox{\rm tr}(M_1M_x)\hbox{\rm tr}(M_0M_x)~+~4\bigl(
\cos(\pi\theta_\infty)\cos(\pi\theta_x)+\cos(\pi\theta_1)\cos(\pi\theta_0)
\bigr)
\cr
\cr
\hbox{\rm tr}(M_1^\prime M_{x^\prime}^\prime)=&\hbox{\rm tr}(M_0M_x). 
}
\right.
$$


\subsection{Connection Problem} 
\label{CCPR}

When we act with a Backlund transformation  on  $y(x)$ for $x\to 0$, we obtain the asymptotic behavior for $x^\prime\to$~(the image of $x=0$). 
 $r$ in (\ref{intrlog1}) is expressed in terms of the monodromy data. 
Let us write the dependence on the monodromy data in a synthetic way as follows:  
$$
y(x)=y(x;\Theta; {\bf TR}_{MM}),
$$ 
where $\Theta=\theta_0,\theta_x,\theta_1,\theta_\infty$; ${\bf TR}_{MM}=
\hbox{\rm tr}(M_0M_x), \hbox{\rm tr}(M_0M_1),\hbox{\rm tr}(M_1M_x)$.   

When we act with a symmetry on the above transcendent, we get: 
$$
y^{\prime}\bigl(~x^\prime;\Theta(\Theta^\prime);{\bf TR}_{MM}\bigl(
{\bf TR}_{M^\prime M^\prime}\bigr)~\bigr).
$$
Here $\Theta(\Theta^\prime)$ stands for the $\theta_\nu$'s expressed in terms 
of the  $\theta_\nu^\prime$'s, and ${\bf TR}_{MM}\bigl(
{\bf TR}_{M^\prime M^\prime }\bigr)$ stands for the  traces of the products of the $M_j$'s as functions of the traces of the products of the $M_j^\prime$'s. For example: 
\vskip 0.2 cm 
For $\sigma_3$: 
$$\left.\matrix{
2\equiv\hbox{\rm tr}(M_0M_x)=&
\hbox{\rm tr}(M_0^{\prime}M_1^{\prime}),
\cr
\cr
\hbox{\rm tr}(M_0M_1)=&-~\hbox{\rm tr}(M_0^\prime M_{x^\prime}^\prime)&
-~\hbox{\rm tr}(M_0^\prime M_1^\prime)\hbox{\rm tr}(M_1^\prime M_{x^\prime}^\prime)&
+~4\bigl(
\cos(\pi \theta_\infty^\prime)\cos(\pi\theta_1^\prime)+\cos(\pi\theta_0^\prime)\cos(\pi \theta_x^\prime)
\bigr),
\cr
\cr
 \hbox{\rm tr}(M_1M_x)=&\hbox{\rm tr}(M_1^{\prime}M_{x^\prime}^\prime).
}
\right.
$$

\vskip 0.2 cm
For $\sigma^1$:
$$
\left.
\matrix{
2\equiv\hbox{\rm tr}(M_0M_x)=&\hbox{\rm tr}(M_1^\prime M_{x^\prime}^\prime), 
\cr
\cr
\hbox{\rm tr}(M_0M_1)=&
-~\hbox{\rm tr}(M_0^\prime M_1^\prime)
&-
\hbox{\rm tr}(M_1^\prime M_{x^\prime}^\prime)\hbox{\rm tr}(M_0^\prime M_{x^\prime}^\prime)~+~4\bigl(
\cos(\pi\theta_\infty^\prime)\cos(\pi\theta_x^\prime)+\cos(\pi\theta_1^\prime)\cos(\pi\theta_0^\prime)
\bigr)
\cr
\cr
\hbox{\rm tr}(M_1M_x)=&\hbox{\rm tr}(M_0^\prime M_{x^\prime}^\prime),
}
\right.
$$

  In order to   obtain the formulas which express $r$ in terms of the monodromy data 
for the solutions (\ref{ass1})  or (\ref{Ass1}) when $x\to \infty$ and $x\to 1$, we  substitute in  (\ref{rrr}) of Proposition  
\ref{monodromiagen} or in (\ref{numerostar}) of Proposition 
(\ref{propmonodromiagen1}) the $\theta_\nu$'s as functions of the 
$\theta_\nu^\prime$'s  and the tr$(M_iM_j)$ as functions of the  tr$(M_i^\prime M_j^\prime)$. When this is done, we can drop the index $\prime$. 
 The above also proves that the behaviors for $x\to\infty$ in (\ref{ass1}) and   (\ref{Ass1}) are associated to tr$(M_0M_1)=2$, while the behaviors for $x\to 1$  are associated to tr$(M_1M_x)=2$.


\subsection{The case of (\ref{nongenintro}): asymptotic behavior  (\ref{imbroglio})}
\label{caseimbro}

 We apply the above results for the transformation of the traces to the case 
(\ref{nongenintro}). First of all, we observe that  
the solutions obtained from the above by the symmetry (\ref{sym2}) are: 
$$
y(x)\sim -{1\over p^2 \ln^2 x} ~\left[
1-2{p+r\over p^2} {1\over \ln x} + {4p^2+6rp+3r^2\over p^4}{1\over \ln^2x}
\right],
$$
 with:
$$
\theta_0=\theta_x=\theta_1=0,~~~\theta_\infty=2p+1.
$$
These  contain the family of  {\it Chazy solutions} studied in 
\cite{M} (for $\mu=-1/2$ in \cite{M}),  namely:
$$
y(x)\sim-{1\over \ln^2 x}  ~\left[
1-{2+2r\over \ln x} + {4+6r+3r^2\over \ln^2x}
\right],~~~~~\theta_0=\theta_x=\theta_1=0,~~~\theta_\infty=3 ~~~(p=1).
$$

The symmetry $\sigma^3$ transforms (\ref{nongenintro}) into : 
\be
 y^\prime(x^\prime)\sim 1-p^2\left(\ln{1\over x^\prime} +{r+p\over p^2}\right)^2, ~~~~~x^\prime\to\infty,
\label{QUADRATO}
\ee
$$
\bigl(\hbox{\rm tr}(M_0M_x),\hbox{\rm tr}(M_0M_1),\hbox{\rm tr}(M_1M_x)
\bigr)~~\longmapsto ~~\bigl(\hbox{\rm tr}(M_0^\prime M_x^\prime),
\hbox{\rm tr}(M_0^\prime M_1^\prime),\hbox{\rm tr}(M_1^\prime M_x^\prime)
\bigr)\equiv(2,2,-2),
$$
$$
(\theta_0,\theta_x,\theta_1,\theta_\infty)=(2p,0,0,1)~~~\longmapsto~~~
(\theta_0^\prime,\theta_x^\prime,\theta_1^\prime,\theta_\infty^\prime)=(2p,0,0,1).
$$
Therefore, the transformed solution 
is again associated to the same monodromy data of (\ref{nongenintro}). 

\vskip 0.2 cm 
Now we apply (\ref{sym3}). We obtain: 
\be
y^\prime(x^\prime)=1-{1\over p^2\left(
\ln(1-x)+{r+p\over p^2}
\right)^2
},~~~~~x^\prime\to 1
\label{TRIANGOLO}
\ee
$$
\bigl(\hbox{\rm tr}(M_0M_x),\hbox{\rm tr}(M_0M_1),\hbox{\rm tr}(M_1M_x)
\bigr)~~\longmapsto~~ \bigl(\hbox{\rm tr}(M_0^\prime M_x^\prime),
\hbox{\rm tr}(M_0^\prime M_1^\prime),\hbox{\rm tr}(M_1^\prime M_x^\prime)
\bigr)\equiv(2,2,-2),
$$
$$
(\theta_0,\theta_x,\theta_1,\theta_\infty)=(2p,0,0,1)~~~\longmapsto~~~
(\theta_0^\prime,\theta_x^\prime,\theta_1^\prime,\theta_\infty^\prime)=(2p,0,0,1).
$$
The transformation of the traces by the action of (\ref{sym3}) 
 will be proved in the second paper. 
The transformed solution 
is again associated to the same monodromy data of (\ref{nongenintro}).

Actually, a transcendents (\ref{nongenintro}) has a  behaviors (\ref{TRIANGOLO})  at $x=1$ and a behavior (\ref{QUADRATO}) at  $x=\infty$. 
Namely, it is the transcendent  (\ref{imbroglio}). The parameters $r$ appearing in (\ref{nongenintro}), (\ref{QUADRATO}) and  (\ref{TRIANGOLO}) are not the same. Their relation will be determined below.

\vskip 0.2 cm 
The rigorous proof of  
(\ref{imbroglio}) is as follows. For $\theta_0=\theta_x=\theta_1
=0$ and $\theta_\infty=2p+1$, $p\in{\bf Z}$,  (PVI) was completely studied 
in \cite{M}. There are two classes of solutions: 

\vskip 0.1 cm
(1) Chazy solutions for any 
$p\neq 0$. The Chazy solutions for a given $p\neq 0$ can be 
obtained applying a birational transformation to the Chazy solutions for $p=1$. 

(2) Picard solutions for any $p$.  The Picard solutions for a given $p\neq 0$ can be 
obtained applying a birational transformation to the Picard solutions for 
$p=0$. 

\vskip 0.1 cm

The symmetry (\ref{sym2}) transforms the Chazy solutions of (PVI) with $
\theta_0=\theta_x=\theta_1=0$, $ \theta_\infty=2p+1$, $p=1$,    
to  the solution: 
\be
\label{miachazy}
y(x)= {8 x ~\omega \omega^{\prime}\bigl(2(x-1)\omega^{\prime}+\omega)\bigr)
(2x\omega^{\prime}+\omega)
\over
\bigl[
(2x\omega^\prime+\omega)^2-4x{\omega^\prime}^2
\bigr]^2
},
\ee
associated to 
$$
\theta_0=2p,~~~p=1,~~~ \theta_x=\theta_1=0,~~~ \theta_\infty=1.
$$  
Here,   
$$\omega=\omega_1+\nu\omega_2, ~~~\nu\in{\bf C},~~~~~\omega^{\prime}=d\omega/dx. 
$$
 The $\omega_i$, $i=1,2$ are two independent solutions of the hypergeometric equation $x(x-1)\omega^{\prime\prime}+(1-2x)\omega^\prime-1/4\omega=0$, namely:
$$
\omega_1= F\left({1\over 2},{1\over 2},{1\over 2};x\right),
~~~~~
\omega_2=g\left({1\over 2},{1\over 2},{1\over 2};x\right).
$$
Any other case $p\in {\bf Z}$, $p\neq 0$, can be obtained by a birational 
transformation  of (\ref{miachazy}), as it is already proved in \cite{M} 
for the Chazy solutions.  
If we expand (\ref{miachazy}) for $x\to 0$ we obtain (\ref{nongenintro}), with:
$$
\nu=1/(4\ln 2-1+\rho_0),~~~~~\rho_0\equiv {r+p\over p^2}. 
$$
Thanks to the representation (\ref{miachazy}), we  can compute the parameters in (\ref{imbroglio}): 
$$
\rho_\infty= {\pi(4\ln 2-1+\rho_0)\over \pi -i (4\ln 2-1+\rho_0)}-2\ln 2+1,
~~~\rho_1={\pi^2\over 4\ln 2 -1+\rho_0}-\ln 2 +1.
$$
This is done by expanding $\omega_1$, $\omega_2$ for $x\to 1$, $x\to\infty$. 
In order to do this, we use the connection formulae in Norlund \cite{Norlund}.
From 5.(1) and 5.(2), we get:
$$
\omega_1=-{1\over \pi} g\left({1\over 2},{1\over 2},{1\over 2};1-x\right),
~~~~~
\omega_1=-\pi F\left({1\over 2},{1\over 2},{1\over 2};1-x\right);
$$
From 12.(1), 12.(3) we get:
$$
\omega_1= {x^{-{1\over 2}}\over \pi} \left[
\pi  F\left({1\over 2},{1\over 2},{1\over 2};{1\over x}\right)-ig\left({1\over 2},{1\over 2},{1\over 2};{1\over x}\right)
\right],
$$
$$
\omega_2= x^{-{1\over 2}}g\left({1\over 2},{1\over 2},{1\over 2};{1\over x}
\right).
$$

It is not possible to compute the relation between 
$\rho_0$, $\rho_\infty$ and $\rho_1$ by the method of monodromy
 preserving deformations, due to the lack of one to one correspondence 
between a solution (the parameter $r$, i.e. $\rho_0$) and the monodromy data.

\vskip 0.2 cm

\noindent
{\it Note 1:} 
The pure braid group (Appendix 2) acts as follows: 
$$
\beta_i\cdot\beta_i:~
\bigl(\hbox{\rm tr}(M_0M_x),\hbox{\rm tr}(M_0M_1),\hbox{\rm tr}(M_1M_x)
\bigr)=(2,2,-2)~\longmapsto ~(2,2,-2),~~~~~i=1,2.
$$
It  leaves $\bigl(\hbox{\rm tr}(M_0M_x),\hbox{\rm tr}(M_0M_1),\hbox{\rm tr}(M_1M_x)
\bigr)$ invariant, thus the log-behaviors at $x=0,1,\infty$ are preserved in  
the analytic continuation of  (\ref{imbroglio}).

\vskip 0.2 cm 
\noindent
{\it Note 2:} 
 The symmetry $\sigma^1$ transforms: 
$$
\bigl(\hbox{\rm tr}(M_0M_x),\hbox{\rm tr}(M_0M_1),\hbox{\rm tr}(M_1M_x)
\bigr)\mapsto \bigl(\hbox{\rm tr}(M_0^\prime M_x^\prime),
\hbox{\rm tr}(M_0^\prime M_1^\prime),\hbox{\rm tr}(M_1^\prime M_x^\prime)
\bigr)\equiv(-2,2,2),
$$
$$
(\theta_0,\theta_x,\theta_1,\theta_\infty)=(2p,0,0,1)\mapsto
(\theta_0^\prime,\theta_x^\prime,\theta_1^\prime,\theta_\infty^\prime)=(0,0,2p,1).
$$
Therefore, the  solution:  
$$
y^\prime(x^\prime) \sim 1-(1-x^\prime)\left[
-p^2\left(
\ln(1-x)+{r+p\over p^2}
\right)^2+1
\right],~~~~~x^\prime\to 1
$$
is not associate to the same monodromy data of (\ref{nongenintro}).


\section{Appendix 1}

\bpr 
\label{matrices}
Let $B_0$, $B_1$ be $2\times 2$ matrices such that 
$$
\hbox{Eigenvalues }(B_0)=0,-c,~~~\hbox{Eigenvalues }(B_1)=0,c-a-b.
$$
and $B_0+B_1$ is either  diagonalizable: 
$$ 
B_0+B_1=\pmatrix{ -a & 0 \cr 0 & -b} ~~\hbox{ (it may happen that }a=b),
$$
or it is a  Jordan form: 
$$ 
B_0+B_1=\pmatrix{ -a & 1 \cr 0 & -a}.
$$
Then, $B_0$ and $B_1$  can be computed as in the following cases. 
Let $r$, $s$  be any complex numbers. 
\vskip 0.2 cm
\noindent 
{\bf 1) Diagonalizable case.} 

\noindent
Case $a\neq b$:
\be
B_0:= \pmatrix{ {a(b-c)\over a-b} & r 
 \cr 
    {ab(a-c)(c-b)\over r(a-b)^2} & {b (c-a)\over a-b}
},~~
B_1= \pmatrix{ {a(c-a)\over a-b} & -r 
\cr
               -(B_0)_{21} & {b(b-c)\over a-b}
},~~~r\neq 0
\label{1}
\ee
\vskip 0.2cm
\noindent
Case $a=b$. We have two sub-cases:
\be
 \hbox{ If } a=b=c:~~~B_0=\pmatrix{ -c-s & r \cr
                                    -{s(c+s)\over r} & s},~~
                      B_1=\pmatrix{s & -r \cr
                                   {s(c+s)\over r} & -c-s}.
\label{6}
\ee
\be
\hbox{ If } a=b=0:~~~ B_0=\pmatrix{-c-s & r \cr
                                   -{s(c+s)\over r} & s},
   ~~~~~                   B_1=-B_0.
\label{7}
\ee
 The transpose matrices of all the above cases are also possible.

\vskip 0.2 cm
\noindent
{\bf 2) Jordan case.}  

\noindent
For $a\neq 0$ and $a\neq c$ we have:
\be
B_0=\pmatrix{ r & {r(r+c)\over a(a-c)} \cr 
             a(c-a) & -c-r },~~
B_1=\pmatrix{-a-r & 1-{r(r+c)\over a(a-c)} \cr
             a(a-c) & c-a+r }.
\label{8}
\ee
For $a=0$, or $a=c$, we have two possibilities:
\be
 B_0=\pmatrix{0 & r \cr 0 & -c},~~ 
                    B_1=\pmatrix{-a & 1-r \cr 0 & -a+c }; 
\label{9}
\ee
or
\be
B_0=\pmatrix{-c & r \cr 0   & 0} ,~~~B_1=\pmatrix{ c-a & 1-r \cr 0 & -a}
\label{10}
\ee
\epr


\bpr 
Let $B_0$ and $B_1$ be as in Proposition \ref{matrices}. 
The linear system:  
$$
{d\over dz} \pmatrix{ \varphi \cr \xi} = \left[
{B_0\over z} +{B_1\over z-1}
\right]~\pmatrix{ \varphi \cr \xi}
$$
may be reduced  to a Gauss hyper-geometric equation, in the
following cases. 

\vskip 0.2 cm
\noindent
Diagonalizable case (i.e. from (\ref{1}) to (\ref{7})):
\be
z(1-z)~ {d^2 \varphi \over dz^2} +\bigl(1+c-(a+[b+1]+1)~z \bigr)~ {d\varphi\over dz}
-a(b+1)~\varphi=0.
\label{hypergeom1}
\ee
The component $\xi$ is obtained by
 the following equalities,
 according to the different cases of Proposition \ref{matrices}.

\vskip 0.2 cm
\noindent 
Cases (\ref{1}):
\be
\xi= {1\over r}\left[
z(1-z)~{d\varphi\over dz} ~-a\left(
z+{b-c \over a-b}
\right)~\varphi
\right]
\label{xi-hypergeom1}
\ee

\vskip 0.2 cm
\noindent
Case (\ref{6}): 
$$
\xi= {1\over r} \left[
z(1-z)~{d\varphi\over dz} +(c+s-c~z)~\varphi
\right]
$$

\vskip 0.2 cm
\noindent
Case (\ref{7}):
$$
\xi={1\over r} \left[
z(1-z)~{d\varphi\over dz} + (c+s)~\varphi
\right]
$$

\vskip 0.2 cm
\noindent
Jordan case (\ref{8}): The equation for  $\xi$ is in Gauss hypergeometic form: 
\be
z(z-1){d^2\xi\over dz^2}+\bigr(1+c-2(a+1)z\bigl){d\xi\over dz} 
-a(a+1)\xi=0,
\label{nuovaiper1}
\ee
\be
\varphi(z)={1\over a(a-c)}
\left[
z(z-1){d\xi\over dz}+(az-c-r)\xi
\right].
\label{nuovaiper2}
\ee

\vskip 0.2 cm
\noindent
Jordan case (\ref{9}): The equation for $\xi$:
$$
{d\xi\over dz}= \left(
-{c\over z}+{c-a\over z-1}
\right)\xi~~\Longrightarrow~~\xi(z)=
\left\{
\matrix{
D~z^{-c}(1-z)^c,  & a=0;
\cr
\cr
D~ z^{-c}, & a=c; 
}
\right.~~~D\in{\bf Z}
$$
The equation for $\varphi$:
$$
{d\varphi\over dz}= 
\left\{
\matrix{
\left[
{r\over z}+{1-r\over z-1}
\right]~D{(1-z)^c\over z^c},& a=0;
\cr
\cr
-{c\over z-1}\varphi+\left[
{r\over z}+{1-r\over z-1}
\right]~{D\over z^c}, & a=c;
}
\right.
$$
The equation for $\varphi$ can be integrated. If $c\not\in{\bf Z}$ we obtain (by variation of parameters): 
$$\varphi(z)=
\left\{
\matrix{
E+ D\left[
-{r\over c}(1-z)^c z^{-c}+{1\over c-1}~z^{1-c}F(1-c,1-c,2-c;z)
\right], & a=0;
\cr
\cr
E (1-z)^{-c}+D\left[
-{r\over c} z^{-c}+{1\over c-1}~z^{1-c} (1-z)^{-c} F(1-c,1-c,2-c;z)
\right],& a=c;
}
\right.~~~~~D,E\in{\bf C}
$$
If $c\in{\bf Z}$, the solution contains a logarithmic term.   

\vskip 0.2 cm
\noindent
Jordan case (\ref{10}): The equation for $\xi$:
$$
{d\xi\over dz}=-{a\over z-1}\xi ~~\Longrightarrow~~\xi(z)=
\left\{
\matrix{D, & a=0;
\cr\cr
D~(1-z)^{-c}, & a=c;
}
\right.~~~D\in{\bf C}
$$
The equation for $\varphi$:
$$
{d\varphi\over dz}=
\left\{
\matrix{
\left(
-{c\over z}+{c\over z-1}
\right)\varphi+\left(
{r\over z}+{1-r\over z-1}
\right)D, & a=0;
\cr
\cr
-{c\over z}\varphi+\left(
{r\over z}+{1-r\over z-1}
\right){D\over (1-z)^a}, & a=c;
}
\right.
$$
The equation for $\varphi$ can be integrated. If $c\not\in{\bf Z}$ we obtain (by variation of parameters): 
$$\varphi(z)=
\left\{
\matrix{
E (1-z)^c z^{-c}+D\left[
{r\over c} -{1\over c+1}~z(1-z)^c F(1+c,1+c,2+c;z)
\right], & a=0;
\cr
\cr
E z^{-c}+D\left[{r\over c} (1-z)^{-c}-{1\over c+1}~z F(1+c,1+c,2+c;z)
\right],
& a=c;
}
\right.
~~~E,D,\in{\bf C}
$$
If $c\in{\bf Z}$, the solution contains a logarithmic term.  
 
\label{ipergeom}
\epr

\section{Appendix 2: Action of the Braid Group and Analytic Continuation}

The subject of this Appendix is well known. Let us denote a branch of a 
transcendent,  
in one to one correspondence with the monodromy data $\theta_0,\theta_x,\theta_1,\theta_\infty$; 
$\hbox{\rm tr}(M_0M_x),\hbox{\rm tr}(M_0M_1),\hbox{\rm tr}(M_1M_x))$, 
with  the following notation: 
$$
y(x;\theta_0,\theta_x,\theta_1,\theta_\infty;
\hbox{\rm tr}(M_0M_x),\hbox{\rm tr}(M_0M_1),\hbox{\rm tr}(M_1M_x)),
$$
Its analytic continuation,  when $x$ goes around a loop around one of the 
singular points $x=0,1,\infty$,  is obtained by an action of the pure braid 
group on the monodromy data. This means that the  new branch is:
$$
y(x;\theta_0,\theta_x,\theta_1,\theta_\infty;
\hbox{\rm tr}(M_0^\beta M_x^\beta),\hbox{\rm tr}(M_0^\beta M_1^\beta ),\hbox{\rm tr}(M_1^\beta M_x^\beta)),
$$
where $\beta$ is a pure braid, and $M_j\mapsto M_j^\beta$ is its action.

It is convenient to replace (\ref{SYSTEM}) by 
$$
{d\Psi\over d\lambda}= \left[{A_0(u)\over \lambda-u_1}+{A_x(u)\over \lambda-u_2}+{A_1(u)\over \lambda-u_3}\right]\Psi,
$$
where we have restored three parameters of isomonodromy deformation $u_1,u_2,u_3$.  The ordered basis of loops $\gamma_1,\gamma_2,\gamma_3$ is in figure \ref{figure7}. 
\begin{figure}
\epsfxsize=12cm
\centerline{\epsffile{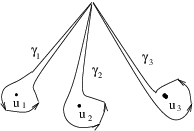}}
\caption{}
\label{figure7}
\end{figure}
The monodromy matrices which correspond to the loops are  $M_0,M_x,M_1$. 

When $x$ goes around a loop around $x=0$, the monodromy data of the
 system (\ref{SYSTEM}) change by the action of the pure braid $\beta_1 \cdot \beta_1$, where $\beta_1$ is the elementary braid which exchanges $u_1$ and $u_2$, namely which continuously deforms $(u_1,u_2,u_3)\mapsto(u_1^\prime,u_2^\prime,u_3^\prime):=(u_2,u_1,u_3)$. The basis $\gamma_1,\gamma_2,\gamma_3$ is deformed, but it is still denoted by  $\gamma_1,\gamma_2,\gamma_3$ in figure  
\ref{figure8}. The monodromy matrices remain unchanged, because the deformation is monodromy preserving. The monodromy matrices obtained by the action of the braid are the monodromy matrices  for:  
 $$
{d\Psi\over d\lambda}= \left[{A_0(u^\prime)\over \lambda-u^\prime_1}+{A_x(u^\prime)\over \lambda-u^\prime_2}+{A_1(u^\prime)\over \lambda-u^\prime_3}\right]\Psi,
$$
w.r.t to the basis $\gamma_1^\prime,\gamma_2^\prime,\gamma_3^\prime$ of figure 
\ref{figure8}. 

\begin{figure}
\epsfxsize=12cm
\centerline{\epsffile{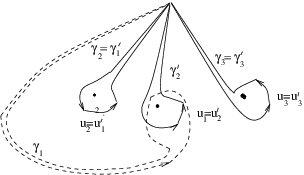}}
\caption{}
\label{figure8}
\end{figure}
We have:  
$$
\gamma_1^\prime=\gamma_2,~~~\gamma_2^\prime=\gamma_2^{-1}\gamma_1\gamma_2,
~~~
\gamma_3^\prime=\gamma_3.
$$
Therefore:
$$
M_0^{\beta_1}= M_x,~~~M_x^{\beta_1}=M_xM_0M_x^{-1},~~~M_1^{\beta_1}=M_1.
$$
If follows that:
$$
M_0^{\beta_1\cdot\beta_1}= M_xM_0M_x^{-1},
$$
$$
M_x^{\beta_1\cdot\beta_1}= M_xM_0M_xM_0^{-1}M_x^{-1},
$$
$$
M_1^{\beta_1\cdot\beta_1}=M_1;
$$
\vskip 0.2 cm 
$$
\hbox{\rm tr}( M_0^{\beta_1\cdot\beta_1}M_x^{\beta_1\cdot\beta_1})=
\hbox{\rm tr}(M_0M_x)
$$
$$
\hbox{\rm tr}( M_0^{\beta_1\cdot\beta_1}M_1^{\beta_1\cdot\beta_1})=
-\hbox{\rm tr}( M_0M_1)-\hbox{\rm tr}(M_1M_x)\hbox{\rm tr}(M_0M_x)
+
4\bigl(
\cos(\pi\theta_\infty)\cos(\pi\theta_x)+\cos(\pi\theta_1)\cos(\pi\theta_0)
\bigr),
$$
\vskip 0.2 cm
$$
\hbox{\rm tr}( M_1^{\beta_1\cdot\beta_1}M_x^{\beta_1\cdot\beta_1})=
\hbox{\rm tr}(M_1M_x)
\bigr[
\hbox{\rm tr}(M_0M_x)^2-1
\bigl]
+
\hbox{\rm tr}(M_0M_x)\hbox{\rm tr}(M_0M_1)+
$$
$$
-4
\bigl[
\cos(\pi\theta_\infty)\cos(\pi\theta_x)+\cos(\pi\theta_1)\cos(\pi\theta_0)
\bigr]\hbox{\rm tr}(M_0M_x)+4\bigl[
\cos(\pi\theta_\infty)\cos(\pi\theta_0)+\cos(\pi\theta_1)\cos(\pi\theta_x)
\bigr].
$$
We observe that tr$(M_0M_x)$ is unchanged. This means that the log-behavior at $x=0$ is preserved when $x$ goes around a small loop around $x=0$.

\vskip 0.2 cm 

When $x$ goes around a loop around $x=1$, the monodromy data of the
 system (\ref{SYSTEM}) change by the action of the pure braid $\beta_2 \cdot \beta_2$, where $\beta_2$ is the elementary braid which exchanges $u_2$ and $u_3$, namely which continuously deforms $(u_1,u_2,u_3)\mapsto(u_1^\prime,u_2^\prime,u_3^\prime):=(u_1,u_3,u_2)$. The basis $\gamma_1,\gamma_2,\gamma_3$ is deformed, and we still denote it $\gamma_1,\gamma_2,\gamma_3$ in figure 
\ref{figure9}.
 The monodromy matrices remain unchanged.  
The monodromy matrices obtained by the action of the braid group are the monodromy matrices  
 w.r.t to the basis $\gamma_1^\prime,\gamma_2^\prime,\gamma_3^\prime$ of 
 figure \ref{figure9}. 
\begin{figure}
\epsfxsize=12cm
\centerline{\epsffile{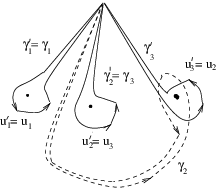}}
\caption{}
\label{figure9}
\end{figure}
We have:  
$$
\gamma_1^\prime=\gamma_1,~~~
\gamma_2^\prime=\gamma_3,~~~\gamma_3^\prime=\gamma_3^{-1}\gamma_2\gamma_3.
$$
$$
M_0^{\beta_2}=M_0,~~~M_x^{\beta_2}=M_1,~~~M_1^{\beta_2}=M_1M_xM_1^{-1}. 
$$
Therefore:
$$
\hbox{\rm tr}( M_0^{\beta_2\cdot\beta_2}M_x^{\beta_2\cdot\beta_2})=
-\hbox{\rm tr}(M_0M_x)-\hbox{\rm tr}(M_0M_1)\hbox{\rm tr}(M_1M_x)
$$
$$+4\bigl(
\cos(\pi\theta_\infty)\cos(\pi\theta_1)+\cos(\pi\theta_0)\cos(\pi\theta_x)
\bigr),
$$
\vskip 0.1 cm    
$$
\hbox{\rm tr}( M_0^{\beta_2\cdot\beta_2}M_1^{\beta_2\cdot\beta_2})=
\hbox{\rm tr}(M_0M_1)
\bigr[
\hbox{\rm tr}(M_1M_x)^2-1
\bigl]
+
\hbox{\rm tr}(M_0M_x)\hbox{\rm tr}(M_1M_x)+
$$
$$
-4
\bigl[
\cos(\pi\theta_\infty)\cos(\pi\theta_1)+\cos(\pi\theta_0)\cos(\pi\theta_x)
\bigr]\hbox{\rm tr}(M_1M_x)+4\bigl[
\cos(\pi\theta_\infty)\cos(\pi\theta_x)+\cos(\pi\theta_0)\cos(\pi\theta_1)
\bigr],
$$
\vskip 0.1 cm 
$$
\hbox{\rm tr}( M_1^{\beta_2\cdot\beta_2}M_x^{\beta_2\cdot\beta_2})=
\hbox{\rm tr}( M_1M_x).
$$
We observe that tr$(M_1M_x)$ is unchanged. This means that the log-behavior at $x=1$ is preserved when $x$ goes around a small loop around $x=1$. 

\vskip 0.2 cm 
 Any pure braid can be obtained by the two generators $\beta_1\cdot\beta_1$, $\beta_2\cdot\beta_2$ introduced above.

\section{Appendix 3: Functions introduced in \cite{Norlund}} 
$$(a)_n:=a(a+1)(a+2)...(a+n-1),~~~~~(a)_{-n}:={1\over 
(a-1)(a-2)(a-3)...(a-n)}.
$$
\vskip 0.2 cm 
$$
F(a,b,c;z)=\sum_{n=0}^\infty{(a)_n(b)_n\over n!(c)_n}z^n.
$$
\vskip 0.2 cm
$$
G(a,b,c;z)=\sum_{n=1}^{c-1}(-1)^{n-1}(n-1)!
{(a)_{-n}(b)_{-n}\over (c)_{-n}}z^{-n}
+
$$
$$
+\sum_{n=0}^\infty{(a)_n(b)_n\over n!(c)_n}\bigl([
\psi_E(a+n)-\psi_E(a)+\bigr.$$
$$
\bigl.+\psi_E(b+n)-\psi_E(b)-\psi_E(c+n)+\psi_E(c)-
\psi_E(1+n)+\psi_E(1)]+\ln z
\bigr)z^n.
$$
\vskip 0.2 cm 
$$
g(a,b,c;z)=\sum_{n=1}^{c-1}(-1)^{n-1}(n-1)!
{(a)_{-n}(b)_{-n}\over (c)_{-n}}z^{-n}+
$$
$$
+\sum_{n=0}^\infty{(a)_n(b)_n\over n!(c)_n}[\psi_E(a+n)+\psi_E(b+n)-\psi_E(c+n)-\psi_E(1+n)+\ln z]z^n.
$$
\vskip 0.2 cm 
$$
g_1(a,b,c;z)=\sum_{n=1}^{c-1}(-1)^{n-1}(n-1)!
{(a)_{-n}(b)_{-n}\over (c)_{-n}}z^{-n}+
$$
$$
+\sum_{n=0}^\infty{(a)_n(b)_n\over n!(c)_n}[\psi_E(1-a-n)+\psi_E(b+n)-\psi_E(c+n)-\psi_E(1+n)+\ln z]z^n.
$$
\vskip 0.2 cm 
$$
g_0(a,b,c;z)=\sum_{n=1}^{c-1}(-1)^{n-1}(n-1)!
{(a)_{-n}(b)_{-n}\over (c)_{-n}}z^{-n}+
$$
$$
+\sum_{n=0}^\infty{(a)_n(b)_n\over n!(c)_n}[\psi_E(1-a-n)+\psi_E(1-b-n)-\psi_E(c+n)-\psi_E(1+n)+\ln z]z^n.
$$


\end{document}